\documentclass{amsart}
\usepackage{latexsym}
\usepackage{amssymb}
\usepackage{amsmath}
\usepackage[english]{babel}
\usepackage[all]{xy}
\usepackage{epsfig}
\usepackage{amsthm}
\usepackage{tikz}
\usepackage{eurosym}
\usetikzlibrary{arrows,decorations.pathmorphing,backgrounds,positioning,fit,petri,decorations}
\usetikzlibrary{calc,intersections,through,backgrounds,mindmap,patterns,fadings}
\usetikzlibrary{decorations.text}
\usetikzlibrary{decorations.fractals}
\usetikzlibrary{fadings}
\usetikzlibrary{shadings}
\usetikzlibrary{shadows}
\usetikzlibrary{shapes.geometric}
\usetikzlibrary{shapes.callouts}
\usetikzlibrary{shapes.misc}
\usetikzlibrary{spy}
\usetikzlibrary{topaths}
\usepackage{setspace}
\usepackage{yfonts}
\usepackage{lscape}
\usepackage{natbib}
\usepackage{url}
\usepackage{alltt}
\usepackage{graphicx}
\usepackage{mathrsfs}
\usepackage{wrapfig}
\usepackage{kantlipsum}
\usepackage{xcolor}

\graphicspath{{immagini//}}

\usepackage[all]{xy}
\usepackage{setspace}

\newcommand{\R}{\mathbb{R}}

\renewcommand{\theenumii}{\arabic{enumii}}
\def\s#1.{\stepcounter{enumii}\theenumi.\theenumii\ \textsl{#1.}\\}

\newcommand\mb{\medbreak}
\newcommand\n{\noindent}
\newcommand\ip{\raise1pt\hbox{\large$\lrcorner$}\,}

\newcommand{\ba}{\begin{array}}\newcommand{\ea}{\end{array}}
\def\lie#1({\mathfrak{#1}(}

\newcommand{\nit}[1]{\mb\n\textit{#1.}}

\renewcommand{\ge}{\geqslant}
\renewcommand{\le}{\leqslant}

\def\,{\kern2pt}
\def\.{\,\cdot\,}

\numberwithin{equation}{section}

\newtheorem{teor}{Theorem}
\newtheorem{prop}[teor]{Proposition}

\newtheorem{corol}[teor]{Corollary}
\newtheorem{defi}[teor]{Definition}
\newtheorem{rem}[teor]{Remark}

\begin{document}

\title{{A gauge theory of complex adaptive systems}}

\bigskip\author{{Gueorgui M. Mihaylov$^1$ and Sergio L. Cacciatori$^2$}}
\thanks{$^1$HALEON $\&$ Department of Mathematics, King's College London}
\thanks{$^2$Department of Science and High Technology (DiSAT), and Como Lake centre for AstroPhysics (CLAP), Universit\`a dell’Insubria,
Via Valleggio 11, IT-22100 Como, Italy,\\ and INFN sezione di Milano, via Celoria 16,
IT-20133 Milano, Italy}

\begin{abstract}\noindent We introduce a geometric construction of a gauge field theory of a complex adaptive system. It is based on a suitable simplicial formulation of a discrete geometry that manifests relevant properties valid in the classical differentiable case. Bundles' non-triviality naturally arises from local collective interactions between agents. Key elements of the theory of principal and associated bundles, such as local obstructions for triviality and characteristic classes, are opportunely defined in this context. Complexity is modelled as the result of local and topological obstructions for the triviality of these geometric structures.  
\end{abstract}

\maketitle

\bigbreak

\null\hfill \emph{``Everything is nothing with a twist.''}

\hfill \emph{Kurt Vonnegut}

\section{Introduction}

In recent years, we have witnessed a strong and continually increasing interest in the description, analysis, and applied and theoretical approaches to modelling complex systems. Various implications of concepts such as emergence, system irreducibility, and collective behaviour have been explored by different scientific disciplines, philosophy, mathematics, physics, engineering, system theory, etc. Emergent properties are those phenomena that manifest at a high or global level of organisation of a system and cannot be directly described and deduced from the characteristics of its subsystems, elementary constituents, or ``agents''. Interesting classical examples of emergent phenomena are conductivity and superconductivity, magnetic susceptibility, phase transitions, shock wave propagation, emergence of nonlinear dynamics in physical and biological systems, chemical properties of large biomolecules, etc. Rapid technological development of data infrastructures and the availability of adequate computational resources have enabled unprecedented visibility on large and complex real-world systems and opportunities to capture and model their behaviour. Relevant characteristics of complex industrial, transportation, or engineering systems, such as reliability, safety, overall efficiency, and resilience, can be efficiently modelled as emergent phenomena. We adopt a broad definition of a complex system as a system that manifests emergent properties and emergent collective behaviour.\smallbreak 

This article is part of a long research effort focused on the applications of geometric and topological methods for the description of emergent phenomena in complex adaptive systems and for the ``quantification'' of complexity itself.  This article collects in a systematic way ideas presented by the authors over the last years in a series of seminars (at Politecnico di Torino, University of Modena and Reggio Emilia, ISI Foundation Torino, University of Cambridge), conference lectures (Winter School of Mathematics and Physics and a Joint Meeting of the London Mathematical Society and the Institute of Mathematics and its Applications), conference courses (ICTD-CMMNO 2016 Congress, Silesian University of Technology) and tutorials (European Conference of the PHM Society Bilbao), etc.\smallbreak

Interesting examples and applications that arise directly from industrial projects provided additional motivation for this research. For example, in a project developed jointly by the Polytechnic University of Turin and Tetra Pak, the overall performance of industrial plants has been efficiently modelled as an emergent phenomenon in a complex adaptive system \cite{MihaylovSpallanzani}. Analogous geometric approaches proved to be very helpful in finding efficient industrial-scale solutions to complex multiple-vehicle routing problems and complex supply chain applications.\smallbreak

\subsection{Frustration and complexity}

Geometric frustration is the main mechanism that drives emergent phenomena in many complex systems.  The term frustration was introduced in \cite{Toulouse1977} initially in the context of magnetic systems. It refers to the impossibility of finding a configuration that simultaneously minimises the energy of all the pairwise interactions between elementary constituents/agents in the system. This impossibility is determined by the geometry and/or the topology of the system. The states of multiple agents cannot be simultaneously synchronised (or consistently interpreted) due to local contradictions, competing tendencies, and conflicting dynamical behaviour. Geometric frustration and topological frustration are obstructions to the consistent propagation of local order. Even simple interactions between agents determine a large class of structural realisations at a system level.\smallbreak 

Geometric frustration determines highly degenerate/``rugged'' energy landscapes, with numerous nearly equivalent ground states and complex, non-trivial dynamical paths toward equilibrium. The rugged energy landscape and the existence of highly degenerate ground states determine long-range correlations and apparent/emergent organised collective behaviour of the systems.\smallbreak

In fact, in systems affected by geometric frustration, local modifications of the state of the agents cause a non-local response, i.e. other regions of the system might reconfigure in order to conserve the energy level. The collective response of the system appears as globally entangled fluctuations caused by long-range correlations. When frustration impedes local minimisation of energy, the system tends to compromise globally, or in other words, collective coherence patterns are provoked by local contradictions.\smallbreak

Furthermore, in a complex interconnected energetic landscape, the set of degenerate low-energy states often forms a connected and organised hierarchical structure/manifold. The affinity between states can be understood in terms of the overlap of the states of multiple agents and can be efficiently measured by ultrametric relations (see, for example \cite{Grossman1989}). States with the same energy can be similar or very different. The transitions between states and the patterns towards equilibrium states are not random but follow collective trajectories ultimately determined by the geometry and the topology of the system. Furthermore, systems tend to evolve into states that have more accessible neighbouring states. This entropic selection of patterns manifests itself as an emergent order in the evolution of the system. Collective order emerges from an apparent disorder; the systems in their totality behave differently with respect to the sum of their parts and often manifest nonlinear dynamics.\smallbreak 

Geometric frustration creates degenerate and overlapped ground states. This leads to the proliferation of the so-called metastable states of disordered systems. In the theory of spin glasses, metastability refers to the system's tendency to become confined to one of many thermodynamically stable ergodic components (pure states). Metastable states/ensembles are stable under thermal fluctuations, hierarchically organised and inaccessible to each other dynamically due to high energy barriers. A formal measure of the complexity of a system is defined as the logarithm of the number of metastable states with a given free energy divided by the number of elementary constituents. In this context, a defining property of a complex system in statistical mechanics (we refer the reader to \cite{Parisi2002}) is the coexistence of a high number (potentially infinite) phases.\smallbreak 

Similarly, topological frustration arises when global topological properties of a system, such as non-trivial homology, periodic boundary conditions, genus, etc., prevent all local interactions from being simultaneously minimised. For example, topological frustration is generated by frustrated loops and is not directly related to quenched disorder or local geometric properties.	In particular, topological frustration can be formally linked to non-trivial homotopy of the configuration space or the underlying manifold or reflects the inability to extend globally some form of local order without singularities or defects.\smallbreak

The idea of modelling the interactions between elementary constituents/agents in a system in terms of some effective interaction field has been successfully applied in various contexts. For example, the Dynamical Mean Field Theory approach to spin glasses provides a useful simplification and approximation of the set of 1-1 interactions based on averaging the interactions over the surrounding agents under the assumption of self-consistency \cite{Turkowski2021}. In infinite-range/fully connected spin glass models (e.g. Sherrington-Kirkpatrick model \cite{Panchenko2013}), frustration is a consequence of the quenched disorder and not from geometry. In these mean field models, frustration is modelled statistically and not locally or geometrically. Other effective field models built on finite-dimensional lattices (e.g. the Edwards-Anderson model, Bethe lattices, Kagome lattices \cite{Edwards1975, Schmidt2015, Mezard2000}) introduce geometric constraints.\smallbreak

We recall some relevant examples to illustrate the above considerations on the role of geometric frustration.\smallbreak

Spin glasses are magnetic systems characterised by a random/disordered (``quenched'') mixture of ferromagnetic and antiferromagnetic interactions embedded within frustrated spin configurations on a lattice.  Frustration may arise from either long-range antiferromagnetic interactions or from the intrinsic geometry of the lattice. In both cases, it is impossible to minimise the energy of all antiferromagnetic bonds simultaneously without causing contradictions. Competing interactions prevent global energy minimisation. The resulting emergent properties of spin glasses, such as magnetic susceptibility, phase transitions, and vitrification, are classic examples of the crucial role of geometric frustration.\smallbreak 

In many large-scale physical or artificial systems with finite interaction configurations, even simple competing forces can give rise to complex, "exotic" collective behaviours. Frustration plays an important role in modelling various structural aspects of biomolecules, particularly in understanding how the molecular structure is linked to biological functions, physical and chemical properties of condensed matter, etc.\smallbreak

Examples of the effects of geometric or vertex frustration can be found in \cite{Gilbert2014, Saccone2019, Hofhuis2021, Makarova2021, Abakumov2012, Rojas2012, Tsirlin2017, Lang2014}.   \smallbreak

For studies on the effects of topological frustration, we refer the reader to \cite{Filippi2024, Drisko2017, Mari2022, Gosavi2006, Odavic2023}. \smallbreak

\subsection{The main idea}

In differential geometry and differential topology, there are standard methods to quantify the obstructions that impede extending pointwise geometric constructions to their local analogues (defined on open sets) and extending local phenomena to global ones (for example, integrable complex, symplectic and other G-structures). The theory of fibre bundles provides a set of tools to define the obstructions for a point-wise phenomenon to be extended and globalised. Local geometric obstructions to integrability are captured by tensor objects such as curvature and torsion \cite{Sternberg83, Salamon}, and global topological obstructions are captured by non-trivial characteristic classes \cite{Milnor74}. In various fundamental theories in physics, precisely the geometric obstructions to the triviality of geometric structures are exploited to define the dynamics of field theories (a curvature tensor in General Relativity and the Yang-Mills theory, a torsion tensor in teleparallel gravity, etc.).\smallbreak

The fundamental role played by frustration and the above considerations motivate our research for a gauge theory of complex adaptive systems. A successful gauge theory in this context should provide a formal framework in which geometric and topological obstructions are harmonically interrelated (similar to the Chern-Weil theory of characteristic classes). It should capture the role played by geometric and topological obstructions in the description of the dynamics of a system. Conceptualising and highlighting the field aspects of the theory (material fields that describe the states of the agents and gauge fields that describe the interactions) would allow us to understand and interpret the dynamics of the system in terms of collective degrees of freedom. To be applicable to a larger class of systems, such a gauge theory should be built with minimal assumptions/requirements on the underlying manifold. We expect that a rigorous gauge theory of complex adaptive systems would provide a very useful set of tools capable of modelling and quantifying complexity.\smallbreak 

The first fundamental ingredient of such a rigorous gauge theory is the construction of a non-trivial fibre bundle associated with a discrete set of agents. The main goal of this article is to provide a consistent construction of non-trivial principal bundles, in which the non-triviality naturally arises from local multi-object interactions between the agents. We will define, formalise, and prove relevant elements of this construction, which are essential for the definition and computational exploitation of the gauge theory formalism, and prove their relevant properties. We emphasise the analogies in formulae, transformation rules and relevant properties with the standard differentiable gauge theories.\smallbreak

Our geometric construction combines the following elements:\smallbreak

\noindent 1. System formalism, i.e. modelling a system as a set of agents and their interactions or relations. In particular, we adopt a standard way of describing a system in which each agent can be described by variables of two types, ``position'' and ``internal state''. The internal state can be deterministic or, in analogy with the fitness network model, we can consider deterministic ``spatial'' coordinates and random ``internal'' state variables.  The states of the agents can be modelled by a multidimensional random variable extracted from an ideally continuous population, a real n-dimensional vector space with a continuous density of probability distribution. \smallbreak  

 \emph{Example.} A system of airports can be described by geographic positions and data on daily passenger traffic. \smallbreak
 
\noindent 2. The states of the agents are interpreted as a field on a suitable manifold. We expect that large-scale or global phenomena in a complex system can be efficiently described by a set of collective degrees of freedom. \smallbreak

\emph{Example.} Musical harmony satisfies this definition of an emergent phenomenon. Indeed, the global behaviour of a vibrating string is efficiently modelled in terms of collective normal modes (harmonic oscillators in correspondence to the eigenfunctions of the Laplacian) rather than the motion of (infinitesimal) portions of the string. 
The vibrating string can be broadly interpreted in line with the definition of complex systems mentioned above. If one interprets stationary sinusoidal waves as specific states of organisation of the infinitesimal elements/segments of the vibrating string, then a physical state is indeed an overlap of infinite ``pure phases''. The dynamical behaviour of the system is efficiently described by collective harmonic oscillators.\smallbreak 

\noindent 3. Collective and geometry-dependent (e.g. path-dependent) interaction gauge field, that more directly accounts for geometric frustration as a micro mechanism with minimal assumptions on the underlying discrete manifold.\smallbreak

\noindent 4. Importance of the topology of the configuration space. Topological data analysis leverages topological structures defined on discrete data sets by associated simplicial complexes. Standard algorithms based on metric affinities or other relations allow the construction of these simplicial complexes (\v{C}ech complexes and Vietoris–Rips complexes) \cite{Kaczynski2004, Virk2022, Schenk2021, Dey2022, Oudot2015}. Simplicial complexes can also be associated with point clouds and graphs based on criteria that reflect collective multi-agent interactions (for example, neurons in a neural network firing together). We will consider a simplicial complex defined by the deterministic coordinates of the agents. In addition, the concept of topological complexity refers to a topological invariant that characterises the space of continuous paths on a path-connected topological space $X$ \cite{Cohen2016}. It is defined in a similar way to the Lusternik-Schliemann category of $X$ and represents the Schwarz sectional genus of the fibration of the continuous paths in $X$ over the space of endpoints $X\times X$ \cite{Schwarz66}. The concept of topological complexity is related to the motion planning problem for mechanical systems. \cite{Farber2003}. Generalisations to higher topological orders have been developed in \cite{Rudyak2010, Ternero2018, Gonzalez2018}.\smallbreak

Various constructions proposed in the research literature on systems theory, graph theory, graph-based models, and topological data analysis, incorporate one or more of the elements mentioned above to describe complex behaviour of systems (\cite{Orus2019, Hubisch2024, Barthel2022, Bodnar2022}, just to cite a few examples). Our goal is to integrate all these components into a coherent and unified framework.\smallbreak

As a helpful intuition to keep in mind throughout the paper, we outline a version of a network fitness model that combines geometric (topological) and statistical features referred to as spatial and internal degrees of freedom. We assume that each agent in the network can ``observe'' the internal (probability) spaces of all the other agents, but there is no a priori way to identify vectors sampled from the internal spaces of distinct agents. Agents can agree on the interpretation of some quantities, for example, a set of moments of the probability distribution $f$, but they cannot identify state vectors. Since the internal spaces can be identified up to a family of transformations that preserve the invariant quantities (moments), these assumptions give rise to a classical gauge theory (see Appendix A for more details). We consider a probability space $(\Omega,f)$ where an action of a group $G$ is defined. Relevant examples are vector spaces of random variables with continuous probability densities and a Lie group action, or a discrete set with a discrete probability distribution and the pair groupoid. An interesting consequence of the above assumptions is that the only way to compare the internal states of distinct agents in the network is by using a parallel transport defined by a suitable connection (a gauge field). Referring to the fitness model intuition, this implies that the probability of the existence of a directed link from one agent to another depends on how the first ``sees'' the internal state of the second via parallel transport on a manifold. The links will be determined by the collective states of the gauge field.  The coupled dynamics of the gauge field and the ``material field'' that describes the states of the vertices needs to be determined by maximising a suitable action functional (for example, a probability-valued functional).\smallbreak

The construction introduced in this paper includes a classical Lie group $G$ acting on $V$ and identified as the stabiliser of a prescribed tensor. We will provide consistent constructions of a principal $G$-bundle and a vector bundle with fibre $V$ associated with it over suitable discrete manifolds.\smallbreak

An important characteristic of our construction is that it combines some aspects of a theory with some aspects of a model. Some properties arise naturally or can be proved from a minimal set of axioms in our mathematical construction. Other elements of the structure are conventional, and there are multiple possible consistent choices. These aspects will often be highlighted and explained in the paper, and some alternative feasible choices will be suggested.\smallbreak   

Finally, the ambition of the article is to be as self-contained as possible and to adopt a style and level of detail that will hopefully make it accessible to specialists in different areas. The examples and some of the constructions introduced in the paper (statistically motivated candidates of gauge groups, toy examples of gauge-invariant actions, etc.) aim to show that our theoretical framework is non-empty and non-trivial and have no ambition to be strictly realistic or exhaustive. In future work, numerical simulations, more advanced examples, and realistic applications of our construction will be developed.

\subsection{Further considerations and motivation for our approach}

The considerations on emergence and complexity can be applied in the context of the so-called Geometric Deep Learning. Deep neural network architectures are often extremely complex adaptive systems, and the relevant characteristics of the learning process are collective and emergent. Unsurprisingly, various tools, ideas, and theoretical approaches from statistical mechanics, condensed matter physics, etc. have been and are currently applied in the attempt to model, describe, and analyse deep NN architectures \cite{Mezard2009, Agliari2020}. Graph neural networks are designed to process the states of agents and the states of connections by implementing various mechanisms of information exchange and propagation. Many deep NN architectures can be interpreted or mapped as GNNs \cite{Bronstein}). A series of recent publications \cite{Bronstein, AslanPlatt} have recognised the importance of local and global symmetries and introduced the gauge theory formalism and the description of graph neural networks. Certain groups arise as natural candidates for structural groups in the gauge formulation of Geometric Deep Learning. Certain features (in the Machine Learning sense of the concept) are formalised as sections of associated bundles, and all operations are expressed as convolutions with respect to equivariant integral kernels, in remarkable analogy with the 2016 paper \cite{MihaylovSpallanzani}, which introduces a similar approach to real-world complex (engineering) systems.\medbreak

In \cite{MihaylovPost}, the author proposes a more radical geometric approach to Geometric Deep Learning to fully leverage the potential of a gauge theory in this space. Connections can play the role of interaction fields, local and topological obstructions can play a dynamical role, the GNN evolves a section of a suitable associated bundle, etc. A key ingredient that enables this view is a consistent definition of a non-trivial principal bundle over a discrete set of agents, naturally deduced from the local multi-agent interactions.

\tableofcontents 

%%%%%%%%%%%%%%%%%%%%%%%%%%%%%%%%

\section{Discrete k-paths}

In a series of publications, Desburn, Hirani, Marsden, Leok, etc. \cite{Hirani1, Hirani2}, develop a discrete geometry driven by strong topological motivations. In this construction, a discrete differentiable manifold is by definition a simplicial complex, and the differential n-forms are real n-cochains on the complex. The De Rham cohomology is identified by definition with the real simplicial cohomology, where the differential is the coboundary operator. We wish to generalise this simplicial construction to a theory which contains discrete versions of vector-valued and group-valued skew-symmetric (1,k)-tensors. We start with the following:

\begin{defi} A discrete differentiable manifold $M$ is a simplicial complex.
\end{defi}

There is a standard way to endow a simplicial complex with a topological structure. The simplicial complex is considered a partially ordered set with respect to the inclusion relation. An application of Alexandrov's topology defines a subset of a simplicial complex to be closed if it is a simplicial complex itself. So, according to this point-set topology, the open sets are the complements of a simplicial (sub)complex. For example, eliminating a (compact) face of a simplex, we obtain an open set.\smallbreak

Consider the set of $n!$ possible orderings of the vertices $(A_0,A_1,A_2,\ldots,A_n)$ of an (n-1)-simplex. The symmetric group acts transitively on this set. The subgroup of even permutations divides the set of orderings into two orbits called orientations of the n-simplex.\smallbreak

\noindent\textbf{Notation:} In this paper, we denote an n-simplex by a sequence $A_0A_1\ldots A_n$ of its vertices, which is indicative of its orientation. In addition, $\overline{\sigma}$ and $\underline{\sigma}$ will denote the same simplex with the opposite orientation.

\begin{defi} Consider a simplicial complex $\Sigma$. A 1-path is an ordered sequence of oriented 1-simplices.\end{defi}

The set $P_1$ of 1-paths with the composition of ordered sequences into an ordered sequence is the monoid freely generated by the set of 1-simplices  in $\Sigma$. This composition is obviously associative, and the identity element is the empty set. The orientation of 1-simplices  is sufficient to induce a group structure on $P_1$.\smallbreak

The inverse of a path is the same path taken with the opposite orientation, i.e. we set:
\begin{align*}
    (AB)\underline{(AB)}=(AB)(BA)&:=\emptyset, \\  (AB)(CD)(EF)\ldots(FE)(DC)(BA)&:=\emptyset.
\end{align*}
%$$\begin{array}{rcl}(AB)\underline{(AB)}=(AB)(BA)&:=&\emptyset, \\  (AB)(CD)(EF)....(FE)(DC)(BA)&:=&\emptyset.\end{array}$$
In other words, two adjacent oppositely oriented 1-simplices  cancel with each other. The composition operation on the set of paths is intuitive. We can extend the same definition to simplices of each order.

\begin{defi} Given a simplicial complex $\Sigma$, the set $P_k$ of ordered sequences of k-simplices, called k-paths, is a group with respect to the composition of ordered sequences into an ordered sequence. The identity element is $\emptyset$ and the inverse is defined by the following relations:
\begin{align*}
    (ABCD\ldots)\underline{(ABCD\ldots)}&:=\emptyset, \\  \overline{\sigma}_1\overline{\sigma}_2\overline{\sigma}_3\ldots\underline{\sigma}_3\underline{\sigma}_2\underline{\sigma}_1&:=\emptyset.
\end{align*}
\end{defi}
%The orientation can be conventionally extended to the set of 0-simplices. \smallbreak

We can adapt to this context the standard definition of boundary operator. \begin{equation}\label{bound} \partial_k (A_0A_1\ldots A_k)=\prod_i(-1)^i(A_0A_1\ldots\tilde{A}_i\ldots A_k).\end{equation} \noindent 
In the expression above, the i-th vertex is omitted, and we occasionally use the notation $(-1)^i$ to indicate the alternating orientation of the $(k-1)$-simplices. Since the product of k-paths is not abelian, the standard boundary operator does not produce a well-defined map between the set of k-paths and the set of (k-1)-paths. In fact, $\partial$ associates to each k-path $s$ a set of (k-1)-paths which differ by ordering of the (k-1)-faces in each k-simplex contained in $s$.  Consider for example the 2-simplex $(ABC)$, we have:
\begin{align*}
   \partial(ABC)&=(BC)(CA)(AB), \\
\partial(BCA)&=(CA)(AB)(BC), \\
\partial(CAB)&=(AB)(BC)(CA). 
\end{align*}
In the research literature, there are various prescriptions for avoiding analogous ambiguities. For example, one can introduce a priori a partial ordering on the set of vertices that defines an ordering on the boundary elements. Similar to the definition of the cup product between simplices (see below). \smallbreak

For the purposes of our construction, it is not strictly necessary to avoid the ambiguity introduced by the boundary operator. We adopt the following definition.

\begin{defi} Given a k-path $s$, we call a realisation of its boundary any (k-1)-path obtained via $\partial s$.
\end{defi}

\begin{rem} \label{PartOrd}In more intuitive terms, a geometric realisation of the boundary is a (k-1)-path in which the (k-1)-faces of each k-simplex can be reordered, but the order of the k-simplices is preserved (the geometric realisations of the boundary can change the order ``within'' but not ``between'' k-simplices).  \end{rem}

In the rest of this section, we will introduce a multiplicative construction based on k-paths of the standard simplicial homology. We denote by $B_k$ the image of the boundary operator, which is the set of all the realisations of boundaries of k-paths of order (k+1).
\begin{prop}The property of being a realisation of the boundary of a given (k+1)-path is a well-defined equivalence relation in $B_k$. \end{prop} 
\nit{Proof} The equivalence classes are composed of ordered sequences of k-simplices which differ by even permutations inside portions of the sequence, portions composed by faces of the same (k+1)-simplex. This relation is reflexive, symmetric, and transitive.\qed\smallbreak

 We will call this relation \emph{boundary ordering equivalence}. 

\begin{defi}The boundary of a k-path is the equivalence class of its boundary realisations.
\end{defi}

The realisations of the boundaries of a k-path form a (k+1)!/2-lief covering space of the set of boundaries.\smallbreak

An intuitive way to introduce the notion of a closed k-path is as follows:

\begin{defi} A k-path is called a cycle if there exists an empty realisation of its boundary.\end{defi}

Any realisation of the boundary of a closed k-path contains the (k-1)-face $\overline{\sigma}$ and the (k-1)-face $\underline{\sigma}$ in distinct adjacent k-simplices exactly the same number of times.  The operation of reordering the faces of each k-simplex of the k-path (compatible with the overall orientation) always permits putting $\overline{\sigma}$ and $\underline{\sigma}$ as adjacent simplices in the realisation of the boundary, and in such a case, they cancel with each other.  We denote by $Z_k$ the set of closed k-paths. In these terms, we can recover the standard simplicial homology theory.\smallbreak

\begin{prop} For any k-path $s$, there exists an empty realisation of $\partial_{k-1}\partial_k s$. \label{dd}\end{prop}

\nit{Proof} The boundary operator applied twice leads to a (k-2)-path in which each element is contained twice with two opposite orientations $\overline{\sigma}$ and $\underline{\sigma}$. On the (k-2)-path $\partial_{k-1}\partial_k s$, two operations can be performed to maintain it in the same boundary-equivalence class, even permutations of the faces of the k-simplices in $s$ and even permutations of the faces of the (k-1)-simplices in $\partial_k s$. It is easy to see that with these two operations, it is always possible to put each simplex and its inverse in adjacent positions. We write $$\overline{\sigma}\underline{\sigma}=\emptyset=\emptyset\emptyset=[\overline{\sigma}\underline{\sigma}][\overline{\sigma}\underline{\sigma}],$$\noindent and then, once cancelled, replace both $\overline{\sigma}$ and $\underline{\sigma}$ with the empty set $[\overline{\sigma}\underline{\sigma}]$. With this substitution, we apply iteratively the above equivalence operations and obtain an empty realisation of $\partial_{k-1}\partial_k s$.\qed\smallbreak

The relation between the set of k-cycles $Z_k$ and k-boundaries $B_k$ in a discrete manifold $M$ is the standard one. Indeed, the set $B_k$ is strictly contained in $Z_k$ and the proper inclusion captures the distinction between (k+1)-simplices and ``(k+1)-holes''. So we define the k-th homology group $H_k$ that characterises a simplicial complex as the quotient $Z_k/B_k$. This quotient recovers exactly the standard k-th simplicial homology group in terms of the equivalence classes of the boundary ordering relation.

\begin{rem} \label{LocalTopological} Paths and equivalence classes of paths will play different roles in our construction. We will use distinct realisations of boundaries in certain contexts and opportune equivalence classes of k-paths in others. In the following sections, the former will be used in the definition of local tensor-like objects (and obstructions), and the latter will be exploited in the definition of topological classes (and obstructions) in our model. The order will be crucial for 1-paths and the definition of the gauge field, but we will see that having a realisation of a form and a consistent set of transformation rules will be sufficient for all practical purposes. \end{rem}

%%%%%%%%%%%%%%%%%%%%%%%%%%%%%%%%
\section{Integral structures}

In classical differential geometry, the notion of differential structure substantially allows to perform computations. A k-differential structure on a manifold $M$ is an equivalence class of atlases of $\mathscr{C}^k$-maps from $M$ to $\R^n$. Vector fields, differential forms, and other tensors obtain their local expressions relative to a differential structure. Integration on manifolds is performed in an analogous way, i.e. exploiting volume forms and regular maps onto domains in $\R^n$. In a discrete context, the point-wise operations as a contraction of tensors should be replaced by their finite version, i.e. a result obtained via an integration on submanifolds of suitable dimension. In simple terms, we need a procedure for evaluating various objects on k-paths.

\begin{defi} Given a discrete differentiable manifold $M$ and a fixed class $\mathscr{G}$ of groups $G$, we denote by $\Omega_{G}^k$ the set of group homomorphisms $Hom(P_k,G)$. We call an integral structure the set of $\Omega_{G}^k$ for all possible values of k and each $G\in \mathscr G$, endowed with the coboundary operator $d$ defined by:

\begin{equation}\label{cobound}\langle \xi,\partial\sigma \rangle=\langle d\xi, \sigma \rangle\end{equation} 
for any $\xi\in \Omega_{G}^k$, $\sigma\in P_{k+1}$.
\end{defi}

Due to the boundary ordering ambiguity, if a group $G\in \mathscr G$ is not abelian, the coboundary operator does not induce a well-defined group homomorphism $ d:\Omega_{G}^k\longrightarrow \Omega_{G}^{k+1}$. We will return to this point.\smallbreak

The above definition is quite general, but different groups can be selected for modelling specific problems. The integral structure can be equipped with further constructions determined by specific relations between the target groups. For the purposes of this article, we will construct an integral structure that uses real vector spaces and classical Lie groups acting on them via standard representations. We will refer to this construction as \emph{discrete Lie integral structure}.\smallbreak
Vector spaces give rise to an object analogous to the classical (k,1) skew-symmetric tensor.

\begin{defi} Consider a discrete manifold $M$ and a (real) vector space $V$ (a group with respect to the sum). An element of $\Omega_{V}^k$ is called $V$-valued skew-symmetric tensor or a $V$-valued differential k-form on $M$.  \label{vector_valued} \end{defi}

A vector-valued differential form $v(\sigma)$ can be represented by a map from the set of k-simplices into $V$. In particular, a differential form with values in $V$ associates a vector $v\in V$ to $\overline{\sigma}$ and $-v$ to $\underline{\sigma}$. We will call the coboundary operator \eqref{cobound} applied on a form $\xi$ the differential of $\xi$. In particular, a 0-form is a map $v(X)$ that associates a vector to each 0-simplex. Observe that the differential of a vector-valued 0-form can be interpreted as the ``finite variation'' of $v$ between the boundary points of an oriented 1-simplex:\begin{equation}dv(XY)=v(Y)-v(X)=-dv(YX).\label{defiV0}\end{equation} Since the vector space is an abelian group concerning the summation, the ordering ambiguity of the boundary (coboundary) operator is irrelevant. The classical definitions of closed and exact $V$-valued differential form can be introduced directly in this context as the kernel and the image of the differential operator.\smallbreak

For reasons that will become clear later on, we can generalise Definition~\eqref{vector_valued}:

\begin{defi} Given a discrete manifold $M$ and vector space $V$, we call a differential form on $M$ an element of $Hom(P_k, \bigwedge (V))$ where $\bigwedge(V)$ is the exterior algebra of $V$ (an abelian group with respect to the vector sum). 
 \label{exterior_valued}\end{defi}\medbreak

Now consider a real function $f$ on $M$ and a $V$-valued 0-form $\xi$. The simplicial version of the standard rule:$$d(fv)=df\,\,v+f\,\,dv$$ on a 1-simplex with boundary $X,Y$ becomes:
\begin{align}
f(Y)v(Y)-f(X)v(X)=&\frac{1}{2}\Big[(f(Y)-f(X))v(Y)+(f(Y)-f(X))v(X) \cr
& \quad+f(Y)(v(Y)-v(X))+f(X)(v(Y)-v(X))\Big]\label{leib}
\end{align}
We decide to multiply the variation of the real function by both the values that the tensor field assumes on the boundary and vice versa. In these terms, it is easy to prove that:

\begin{prop}Denote by $\mathscr F$ the space of $\R$-valued functions on a discrete differentiable manifold $M$. The space of tensors on $M$ is an $\mathscr F$-module.\end{prop}

%The equation \eqref{leib} can be easily generalised to tensors of order $n$.$$\begin{array}{rcl} d(fv)(A_0A_1...A_n) &=& \sum (-1)^k fv(A_0A_1...\bar{A}_k...A_n)=\\
%                                         &=& \sum (-1)^k f(A_0A_1...\bar{A}_k...A_n) v(A_0A_1...\bar{A}_k...A_n)=\\
%                                         &=& \frac{1}{2}(\sum_i (-1)^k f(A_0A_1...\bar{A}_i...A_n)\sum_j v(A_0A_1...\bar{A}_j...A_n)+\\
%                                         & & \,\,\,\,\,\,\,\,\,\,\,\,\,\,\, +(-1)^{n-1}\sum_i (-1)^i v(A_0A_1...\bar{A}_i...A_n) \sum_j f(A_0A_1...\bar{A}_j...A_n))\end{array}$$

The second element of the Lie integral structure includes classical Lie groups as targets. We choose to multiply the elements in $G$ on the left. 

 \begin{defi} Given a discrete differentiable manifold and a group $G$, a $G$-valued differential k-form is an element of $\Omega_{G}^k$ i.e. a group homomorphism between $P_k$ and $G$.
\end{defi}

A k-form can be represented by a map from the set of k-simplices to $G$. It associates $g\in G$ to an oriented simplex and $g^{-1}\in G$ to the same simplex if considered with the opposite orientation.\smallbreak

When the group $G$ is not abelian, the coboundary operator is affected by the boundary ordering ambiguity, so a k-form is mapped by the differential operator into a set of (k+1)-forms. Different methods that can eliminate this ambiguity are proposed in the literature (see for example \cite{Bergner2022, Gioan2011, Adams2003}).\footnote{We avoid introducing globally defined objects, quantities, and full or partial vertex orders. Local solutions for the coboundary ambiguity can be achieved by minimising or maximising a $G$-invariant norm on the elements of $G$, or based on a suitable probability weight function induced on $G$. We will comment further on this point in the following sections. } The boundary ordering equivalence allows to capture the topology of a discrete differentiable manifold by means of a multiplicative construction of a cohomological complex. The closure and exactness of differential forms are well defined in terms of suitable equivalence classes. The proper use of an integral structure (i.e. using distinct realisations of the differential of a $G$-valued form) will be relevant in its own right (as pointed out in Remark~\eqref{LocalTopological}). \smallbreak

\begin{defi}Given a $G$-valued k-form $\omega$, we call a realisation of the differential of $\omega$ any (k+1)-form $d\omega$ in the image of the coboundary operator.\end{defi}

The boundary ordering equivalence affects differential forms.

\begin{defi}  Two k-forms are equivalent if they are realisations of the differential of the same (k-1)-form.\end{defi}

This definition enables the introduction of closed and exact $G$-valued differential forms.

\begin{defi} A $G$-valued differential k-form $\omega$ is called exact if $\omega=d\theta$ for some differential (n-1)-form $\theta$. \end{defi}

\begin{defi} A $G$-valued k-form is called closed if its differential $d\omega$ admits a realisation identically equal (on each (k+1)-simplex) to the identity element of $G$.\end{defi}

In other words, a $G$-valued k-form $\omega$ is called closed if $d\omega=Id\in G$ for some ordering of the elements in the boundary of each (n+1)-simplex. %The use of the term \emph{differential form} is justified if referred to these equivalence classes:

\begin{prop}For each k-form $\omega$, there is a realisation of $dd\omega$ which is equal to $Id\in G$. \end{prop}

\nit{Proof} This follows from Proposition~\eqref{dd}. Given a k-form $\omega$, its differential $d(d\omega)$ is computed on some (k+2)-simplex $\sigma$.  It is always possible to find a permutation (compatible with the overall orientation) of the values of $\omega$ on $\partial(\partial \sigma)$ and a permutation of the corresponding values of $d\omega$ on $\partial \sigma$ to obtain $Id$ as a product.\qed\smallbreak

\begin{rem} The support of a differential k-form in a discrete differentiable manifold consists of a set of k-simplices in the complex. The differential of a form exists where the coboundary operator makes sense. Where it is impossible to apply the coboundary operator, we conventionally set $d\omega=Id$ (the form is treated as a top-dimensional form). From this point of view, no particular regularity conditions on the simplicial complex (constant dimension, regularity of the boundary, etc.) are required to define and exploit an integral structure.
\label{reg}\end{rem}

\begin{defi} Given a discrete manifold $M$, its cohomology group $H^k(G)$ is the quotient of the closed over the exact elements of $\Omega^k_G$.\end{defi}

These definitions imply that 0-forms are $G$-valued functions on $M$. The differential of a function is:\begin{equation} d\omega (XY):= \omega(X)^{-1}\omega(Y).
\label{defi0}\end{equation} This expression can also be interpreted as a ``finite variation'' of $\omega$ along the 1-simplex XY.\smallbreak

The discrete Lie integral structure can be equipped with further structures by setting a natural ``interaction'' between the objects introduced in this section.\smallbreak

Consider a classical Lie group $G$ and a representation $\rho$ of $G$ on the vector space $V$. For each $V$-valued differential k-form $v$ and a $G$-valued differential k-form $\omega$, we can define a $V$-valued differential k-form $\omega v$ obtained by the action (via $\rho$) of $\omega$ on $v$ on each k-simplex (multiple $G$-valued forms can be composed). In these circumstances we have a well-defined $d$ operator, $$d(\omega\,v)(A_0A_1\ldots A_{k+1})=\sum(-1)^{i}\rho(\omega((A_0A_1\ldots\bar{A}_i\ldots A_{k+1}))) v((A_0A_1\ldots\bar{A}_i\ldots A_{k+1})).$$ 

Another operation is the simplex-wise product of $G$-valued k-forms, giving a k-form:$$\omega\,\theta(XYZ\ldots)=\omega(XYZ\ldots)\,\theta(XYZ\ldots).$$\medbreak

Because of its analogy with the wedge product, the so-called cup product between discrete differential forms is a more relevant operation. The cup product between a k-form and an m-form on a (k+m)-simplex $\sigma$ defines a (k+m)-form on $\sigma$. The standard definition of the cup product for $\R$-valued cochains is:\begin{equation}    
(\omega\smile\theta)\sigma=\omega(A_0\ldots A_k)\theta(A_{k}\ldots A_{k+m})
\label{CupProduct}\end{equation} \noindent where the right-hand side is the product in $\R$, and, according to standard terminology, $\omega$ is evaluated on a ``front'' face and $\theta$ is evaluated on a ``back'' face. The cup product satisfies the Leibniz rule \cite{Hatcher}:$$d(\omega\smile \theta)=d\omega\smile\theta + (-1)^k\omega \smile d\theta$$An important consequence of this property is that the cup product of an exact cochain and a closed cochain is exact or, in other words, the exact cochains form an ideal. As a consequence, the cup product on cochains induces a (topological) cup product on cohomology groups: $$H^k\smile H^m\longrightarrow H^{k+m}.$$

The cup product can be directly extended to vector-valued (additive) differential forms introduced above: $$ (\omega\smile\theta)\sigma=\omega(A_1\ldots A_k) \wedge \theta(A_{k+1}\ldots A_{k+m}),$$

\noindent where in view of Definition~\eqref{exterior_valued}, $\omega$ and $\theta$ are $\bigwedge^m R^n$-valued k-form and m-forms respectively, and the right-hand side is a product in the exterior algebra of $\R^n$. The proof replicates the one for $\R$-valued cochains in \cite{Hatcher}.\smallbreak

\begin{prop}
    
 Given a $V$-valued differential k-form $\omega$ and m-form $\theta$,$$d(\omega\smile \theta)=d\omega\smile\theta+(-1)^k\omega \smile d\theta.$$
 
\end{prop}

\nit{Proof} Consider a (k+m+1)-simplex $\sigma$ and let us initially select an order of its vertices $A_0,\ldots,A_{k+m+1}$.\bigbreak

\noindent$\begin{array}{rl} 
(d\omega\smile\theta)\sigma  & =d\omega[A_0,\ldots,A_{m+1}]\theta[A_{m+1},\ldots,A_{m+k+1}]  \\

                             & =\omega[\partial(A_0,\ldots,A_{m+1})]\theta[A_{m+1},\ldots,A_{m+k+1}]\\
                             
                             & =\sum_{i=0}^{m+1}(-1)^i\omega[A_0,\ldots,\bar{A}_i,\ldots,A_{m+1})]\theta[A_{m+1},\ldots,A_{m+k+1}]\\
                             & \\

(-1)^m(\omega\smile d\theta)\sigma &=(-1)^m(\omega[A_0,\ldots,A_{m}]d\theta[A_{m},\ldots,A_{m+k+1}])\\

                              & =(-1)^m(\omega[A_0,\ldots,A_{m}]\theta[\partial(A_{m},\ldots,\bar{A}_i,\ldots,A_{m+k+1})])\\
                              & =\sum_{i=m}^{m+k+1}(-1)^i\omega[A_0,\ldots,A_{m}]\theta[A_{m},\ldots,\bar{A}_i,\ldots,A_{m+k+1}])\\
                              
                              & \\

d(\omega\smile\theta)\sigma   & =\sum_{i=0}^{i=m+k+1}(-1)^i((\omega\smile\theta) [A_0,\ldots,\bar{A}_i,\ldots,A_{m+k+1}])\\
                              & =\sum_{i=0}^{i=m}\{(\omega\smile\theta) [A_0,\ldots,\bar{A}_i,\ldots,A_{m+k+1}] \\
                              & \ \ +\sum_{i=m+1}^{i=m+k+1}(-1)^i(\omega\smile\theta) [A_0,\ldots,\bar{A}_i,\ldots,A_{m+k+1}]\\
                              &=\sum_{i=0}^{i=m}(-1)^i\omega\ [(A_0,\ldots,\bar{A}_i,\ldots,A_{m+1})]\theta[A_{m+1},\ldots,A_{m+k+1}]\\
                              &\ \ +\sum_{i=m+1}^{i=m+k+1}(-1)^i\omega[(A_0,\ldots,A_{m}] \theta[A_m,\ldots,\bar{A}_i,\ldots,A_{m+k+1}]\end{array}$\smallbreak

The equality is valid because the following terms cancel: \smallbreak

\noindent $$ (-1)^{m+1}\omega[A_0,\ldots,A_m,\bar{A}_{m+1}]\theta[A_{m+1},\ldots,A_{m+k+1}]
+(-1)^{m}\omega[A_0,\ldots,A_m]\theta[\bar{A}_m,A_{m+1},\ldots,A_{m+k+1}]=0$$

\qed{}
\bigbreak

The above argument can be directly applied to G-valued differential forms if $G$ is abelian.

\begin{prop}
    
 Given an abelian group $G$, a $G$-valued differential k-form $\omega$ and m-form $\theta$, there exist realisations of $d\omega$, $d\theta$  and $d(\omega\smile\theta)$ such that $$d(\omega\smile \theta)=d\omega\smile\theta\,\,(\omega \smile d\theta)^{(-1)^k}$$  
 
 \noindent $(-1)^k$ denotes the inverse of the element of $G$ accordingly with orientation.\label{AbelianIdeal}\end{prop}

The validity in the Abelian case will be relevant for the construction of characteristic classes below.\smallbreak 

The generalisation of this construction to $G$-valued differential forms with non-abelian $G$ is not straightforward because $(gh)^{-1}=h^{-1}g^{-1}$ for $g,h\in G$, therefore computing the inverse in the second term on the left-hand side and on the right-hand side gives different results.   
This difficulty is not resolved by the equivalence relations induced by permutations of the vertices that we exploit in our construction. Exploring more general equivalence relations that can lead to a consistent generalisation is a potential direction for future research.\bigbreak

The following way to identify special realisations of the boundary of a k-path will be very useful later on.

\begin{defi} Given a differential k-form $\omega$ and a k-simplex $ABC\ldots $,  we call a point-based value of $d\omega$ in $A$ the differential computed by omitting first the vertex $A$ in the expression \eqref{bound} of the boundary operator. \end{defi}

\nit{Example} The point-based values in A of the differential of a 2-form on the simplex $ABCD$ are:
\begin{align*}
d\omega|_A=&\omega(ABC)^{-1}\omega(ABD)\omega(ACD)^{-1}\omega(BCD),\\ 
d\omega|_A=&\omega(ACD)^{-1}\omega(ACB)\omega(ADB)^{-1}\omega(CDB),\\ 
d\omega|_A=&\omega(ADB)^{-1}\omega(ADC)\omega(ABC)^{-1}\omega(DBC).
\end{align*}
%\smallbreak
The point-based value of a differential form reduces the boundary ordering ambiguity. Obviously, up to dimension 2, the point-based value of a differential form is unique, so in such a context, this definition substantially restores the classical pairing between vector fields and differential forms. More precisely, 1-simplices  in $M$ (the shortest parts of discrete curves in the manifold) can play the role of tangent vectors, etc. This interpretation of 1-simplices  combines the topological motivation for the introduction of the simplicial complex with classical aspects of differential geometry. 

\begin{defi}A simplicial map is a map between simplicial complexes such that the images of the vertices of a simplex always span a simplex. We say that a simplicial mapping is continuous if the preimage of each closed set (simplicial subcomplex) is a closed set. A bijective simplicial map is called a simplicial isomorphism.
\end{defi}

 We call \emph{faced} each simplex in a complex that is not a face of another simplex. In other words, it is locally the simplex with the highest dimension. We now introduce the following key concept.

\begin{defi} \label{local} Given two simplicial complexes $M$ and $N$, a local simplicial isomorphism is a simplicial map $s:M\longrightarrow N$ which maps each faced simplex in $M$ isomorphically to a faced simplex in $N$.\end{defi}

\begin{defi} We call \emph{local} any phenomenon referred to faced simplices in a simplicial complex.\label{defliLocal}\end{defi}

One may expect that the behaviour of integral structures on manifolds under simplicial mappings highlights relevant properties of the manifolds, such as the local dimension, characterises the support of differential k-forms, etc.

\begin{defi}\label{localPB}Consider a local simplicial isomorphism $s:M\longrightarrow N$ between two discrete differentiable manifolds. Given a differential k-form $\omega$ on $N$, its local pull-back $s^*\omega$ on $M$ is the k-form defined on each k-simplex $\sigma\in M$ by $s^*\omega(\sigma):=\omega(s\sigma)$.\end{defi}

Analysing in detail the behaviour of the boundary operator and the differential under simplicial mappings goes beyond the purposes of this paper, but the following result is quite obvious:
\begin{prop}Simplicial isomorphisms commute with the boundary operator and with the differential.
\end{prop}

\begin{rem} Another intermediate solution to the boundary ordering ambiguity can be based on our notion of locality. One can fix a local ordering of the vertices, i.e. an ordering valid on each faced simplex. Typically, such local orderings contradict each other on the intersections of faced simplices. The general construction presented in this paper can be simplified by a local ordering of the vertices that leads to a local elimination of the coboundary ambiguity. We will not develop this idea further in this paper. \label{locord} \end{rem}

The version of the Hodge star operator proposed by Hirani-Marsden-Leok in terms of dual mesh is more difficult to introduce in our construction. Volume integrals can be easily defined on vector-valued forms. Multiplicative integrals over 1-paths (curves) have a direct intuitive meaning. Defining a general volume integral from the k-path integration is non-trivial in the non-abelian multiplicative case.  

\section{Principal bundles}

The formalism of fibre bundles is the proper context for the description of the geometric obstructions for certain point-wise objects to be extended to structures on open sets, and certain local constructions to be promoted to global phenomena. In this section, we provide a construction of a semidiscrete principal bundle based on the concept of locality introduced below (Definition~\eqref{local}). It is a generalisation of the classical concept of a principal bundle.\smallbreak

Consider a discrete differentiable manifold $M$ and a Lie group $G$. The set of faced simplices forms a covering of the base manifold $M$. We first consider a faced n-simplex $\sigma\in M$ and a (trivial) $G$-bundle over the set of 0-simplices  in $\sigma$. In the remainder of the paper, we will consider $G$ acting on the left.  The choice of the left action is purely conventional. Our construction can be replicated in terms of the right action if we wish to better resemble the ``base change'' intuition of a $G$-structure over the frame bundle.

We call the total space of this bundle $P_{\sigma 0}$ and it contains simplices  of the following types:\smallbreak

 - points in $P_{\sigma 0}$ are 0-simplices and they project to 0-simplices  in $\sigma$;\smallbreak
 
  - type I simplices are determined by a generic set of points of the same $G$-orbit;  \smallbreak

  - type II k-simplices are those simplices  whose vertices belong to k distinct orbits;\smallbreak

  - any other simplex in $P_{\sigma 0}$ is called generic;\smallbreak

  - $\pi^{-1}\sigma$ contains type II n-faces which project isomorphically to $\sigma$.\smallbreak

Similarly to the previous sections, simplices can be combinatorially defined and exist as basic entities contractible to a single point (vertex); they are not, in general, geometrically embedded, i.e. they do not arise as loci of points. In the following, we will also discuss a geometric definition of type II simplices in the $G$-orbit. \smallbreak

For a faced n-simplex $\sigma$, the free $G$-action on the fibres of $P_{\sigma 0}$ induces a set $\epsilon_n(G)$ of well-defined transformations on each of the above types of simplices. The action $\epsilon_n(G)$ reduces to $G$ on points as in the classical case. We will use the generic notation $\epsilon(G)$ for faced simplices of various dimensions. \smallbreak

The structure of the semidiscrete principal bundle will be constructively defined starting from $P_{\sigma 0}$.

\begin{defi}\label{localdef}A semidiscrete principal bundle is locally a triple  $(P,M,\pi)$.  The total space $P$ is a family of simplicial complexes endowed with a left simply transitive local action $\epsilon(G)$ induced by a Lie group $G$. The projection $\pi:P\longrightarrow M$ maps every $G$-orbit into a 0-simplex in $M$.\end{defi}

A quotient with respect to the action $\epsilon(G)$ is isomorphic to the set of faced simplices in $M$. In this specific sense, a semidiscrete principal bundle is a fibre bundle $F\hookrightarrow P\rightarrow M$ in which the total space $P$ is locally simplicially isomorphic to a product of a discrete differentiable manifold $M$ and a Lie group $G$.

\begin{defi}\label{trivialisations} A trivialisation of the principal bundle $P$ associates to each faced simplex in $\sigma \in M$ a set of diffeomorphisms $\psi_A :F\longleftrightarrow G$ which is in bijective correspondence with the 0-simplices  $A\in\sigma$, extended by the $\epsilon(G)$-action.\end{defi}

A trivialisation is a way to identify locally the fibre over each 0-simplex with the structure group and express locally the total space as a product. 

\begin{defi}\label{section} A section of a principal bundle is a local right inverse $s:M\longrightarrow P$ of the bundle projection $\pi$.\end{defi}

We will use $s$ to indicate a section of $P$.\smallbreak

The simplicial definitions introduced so far can be interpreted in a purely combinatorial sense, but we introduce some additional local properties in the principal bundle construction to ensure the geometric consistency of the concept of local triviality. Since all simplices are contractible to a point, we require the bundle over a faced simplex $\sigma$ to retract (to be homotopically equivalent) to a fibre over a vertex $A\in\sigma$, i.e. we require the inclusion $i:\pi^{-1}(A)\subset\pi^{-1}(\sigma)$ to be a homotopy equivalence. In more detail, given a deformation retraction $\delta$ from $\sigma$ to $A$ and $\delta_p(\pi(Id)):=\Delta(Id,t)$, we require that the homotopy extension problem
$$\begin{array}{ccc}
    \pi^{-1}(\sigma)\times {0}                   &     \stackrel{i}{\longrightarrow} &  P \\
\,\,\,\downarrow^i  &                                     & \,\downarrow^\pi \\
  \pi^{-1}(\sigma)\times [0,1]   & \stackrel{\Delta}{\longrightarrow}     &  M \end{array}$$\smallbreak \noindent always has a solution, i.e., there exists $\tilde\Delta:\pi^{-1}(\sigma)\times [0,1]\longrightarrow P$, which covers $\Delta$. By construction $\tilde{\Delta}$ is a deformation retraction from $\pi^{-1}(\sigma)$ to  $\pi^{-1}(A)$.\smallbreak

The locality of Definition~\eqref{section} implies that if a simplex $\sigma\in M$ belongs to the intersection of m-faced simplices in $M$, then $s(\sigma)$ is a k-fold covering of $\sigma$, with $1\leq k\leq m$. The construction of the semidiscrete principal bundle is completed by a definition of transition maps between faced simplices (defined as open sets).\smallbreak

\begin{figure}[!htbp]
\begin{center}
\begin{tikzpicture}[>=latex,decoration={zigzag,amplitude=.5pt,segment length=2pt}]
\shade[ball color=cyan] (-1.5,0) -- (1.5,0) -- (0,2.598) -- cycle;
\draw [thick] (-1.5,0) -- (1.5,0) -- (0,2.598) -- cycle;
\shade[ball color=yellow] (3,0) -- (6,0) -- (4.5,2.598) -- cycle;
\draw [thick] (3,0) -- (6,0) -- (4.5,2.598) -- cycle;
\draw [ultra thick,->] (0.1,2.7) .. controls (1,3.3) and (3.5,3.3) .. (4.4,2.7);
\draw [ultra thick,->] (1.6,-0.1) .. controls (1.9,-0.5) and (2.5,-0.4) .. (2.9,-0.1);
\draw [ultra thick,->] (-1.4,-0.1) .. controls (1.0,-2) and (3.1,-2) .. (5.9,-0.1);
\shade [ball color=cyan] (9,-1) .. controls (6,1) and (6,4) .. (9,4) .. controls (12,4) and (12,1) .. cycle;
\draw [ultra thick] (9,-1) .. controls (6,1) and (6,4) .. (9,4) .. controls (12,4) and (12,1) .. cycle;
\shade [ball color=cyan!50!black] (9,-1) .. controls (7,0.5) and (7,2) .. (9,2) .. controls (11,2) and (11,0.5) .. cycle;
\draw [thick, densely dotted] (9,-1) .. controls (7,0.5) and (7,2) .. (9,2) .. controls (11,2) and (11,0.5) .. cycle;
\filldraw [white] (9,-1) .. controls (7.3,0.3) and (7.6,1.3) .. (8.1,1.55) .. controls (9.1,1.9) and (8.9,0.0) .. cycle;
\filldraw [white] (9,-1) .. controls (10.7,0.3) and (10.4,1.3) .. (9.9,1.55) .. controls (8.9,1.9) and (9.1,0.0) .. cycle;
\draw [thick, densely dotted] (9,-1) .. controls (7.3,0.3) and (7.6,1.3) .. (8.1,1.55) .. controls (9.1,1.9) and (8.9,0.0) .. cycle;
\draw [thick, densely dotted] (9,-1) .. controls (10.7,0.3) and (10.4,1.3) .. (9.9,1.55) .. controls (8.9,1.9) and (9.1,0.0) .. cycle;
\shade [ball color=yellow] (9,-1) circle (4pt);
\node at (0,1) {\Large$\pmb {\sigma_1}$};
\node at (4.5,1) {\Large$\pmb {\sigma_2}$};
\node at (9,3) {\Large$\pmb {\sigma_1}$};
\node at (9.5,-1.2) {\Large$\pmb {\sigma_2}$};
\end{tikzpicture}
\caption{Starting from two 2-symplices, we get a 2-parachute (cyan and dark cyan) with three 1-edges (dotted lines) and a point (yellow).}
\label{fig:parachute2}
\end{center}
\end{figure}
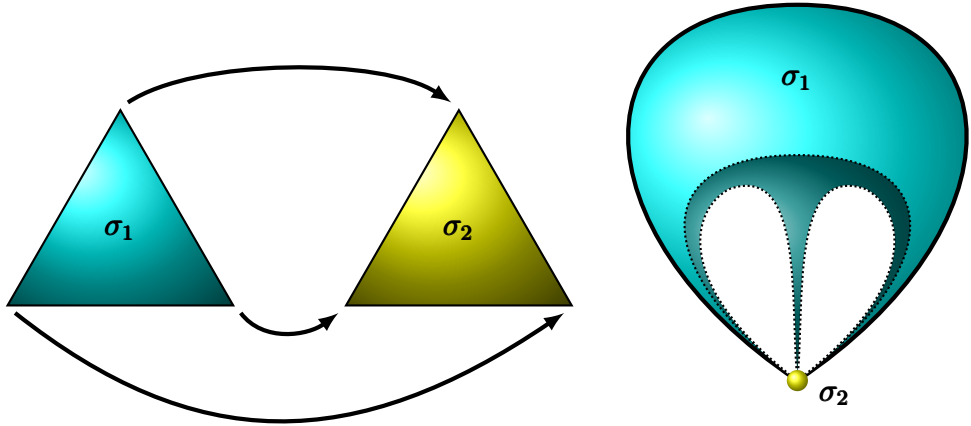

Suppose that an n-simplex $\sigma\in M$ is the common face of the highest dimension of two-faced simplices in $M$, which gives rise to two distinct simplices $\sigma_1$ and $\sigma_2$ in $s(\sigma)$. A section of the bundle can be chosen locally in a way that the vertices of $\sigma_1$ and $\sigma_2$ are identified in $\sigma\times F \in P$ by the $\epsilon(G)$-action. The simplex $\sigma_2$ can be contracted to a point, and in this way, the 1-simplices in $\sigma_1$ are equivalent to 1-spheres with one point in common. Each 2-simplex in $\sigma_1$ transforms into a 2-dimensional ``parachute'', which is equivalent to a 2-sphere with 3 holes with their boundaries sharing a common point (see Fig.~\eqref{fig:parachute2}). 

%%%%%%%%%%%%%%%%%%%%%%%%%%%%%%%%%%%%%%%%%%%%%%%%%%%%%%%%%%%%%%%%%%%%

\begin{figure}[!htbp]
\begin{center}
\begin{tikzpicture}[>=latex,decoration={zigzag,amplitude=.5pt,segment length=2pt}]
%%%%%%%%%%%%%%%%%%%%%%%%%%%%%%%%%%%%%%%%%%%%%%%%%%%%%%%%%%
\draw[thick,dashed] (-1.5,-6) -- (1.8,-4.9);
\shade[ball color=cyan,opacity=0.8] (-1.5,-6) -- (1.2,-6.2) -- (0,2.598-6) -- cycle;
\draw [thick] (-1.5,-6) -- (1.2,-6.2) -- (0,2.598-6) -- cycle;
\shade [top color=white,bottom color=cyan!40!black,opacity=0.8] (1.2,-6.2) -- (0,2.598-6) -- (1.8,-4.9) -- cycle;
\draw [thick] (1.2,-6.2) -- (0,2.598-6) -- (1.8,-4.9) -- cycle;
\draw[thick,dashed] (5+1.5,-6) -- (5-1.8,-4.9);
\shade[ball color=yellow,opacity=0.8] (5+1.5,-6) -- (5-1.2,-6.2) -- (5-0,2.598-6) -- cycle;
\draw [thick] (5+1.5,-6) -- (5-1.2,-6.2) -- (5-0,2.598-6) -- cycle;
\shade [top color=white,bottom color=yellow!40!black,opacity=0.8] (5-1.2,-6.2) -- (5-0,2.598-6) -- (5-1.8,-4.9) -- cycle;
\draw [thick] (5-1.2,-6.2) -- (5-0,2.598-6) -- (5-1.8,-4.9) -- cycle;
\draw [ultra thick,->] (0.1,2.7-6) .. controls (1,3.3-6) and (3.5,3.3-6) .. (4.9,2.7-6);
\draw [ultra thick,->] (1.85,-0.1-4.8) .. controls (2.2,-0.5-4.8) and (2.8,-0.4-4.8) .. (3.15,-0.1-4.8);
\draw [ultra thick,->] (1.3,-0.1-6.2) .. controls (1.6,-0.5-6.2) and (3.3,-0.4-6.2) .. (3.7,-0.1-6.2);
\draw [ultra thick,->] (-1.4,-0.1-6) .. controls (1.0,-2-6) and (3.1,-2-6) .. (6.5,-0.1-6);
\shade [ball color=cyan] (10,-1-6) .. controls (6,1-6) and (6,4-6) .. (10,4-6) .. controls (14,4-6) and (14,1-6) .. cycle;
\draw [ultra thick] (10,-1-6) .. controls (6,1-6) and (6,4-6) .. (10,4-6) .. controls (14,4-6) and (14,1-6) .. cycle;
\node at (0,1-6) {\Large$\pmb {\sigma_1}$};
\node at (4.8,1-6) {\Large$\pmb {\sigma_2}$};
\node at (10,3-6) {\Large$\pmb {\sigma_1}$};
\node at (10.5,-1.2-6) {\Large$\pmb {\sigma_2}$};
%%%%%%%%%%%%%%%%%%%%%%%%%%%%%
\shade [ball color=cyan!50!black] (10,-1-6) .. controls (8.8,0.3-6) and (9.1,1.3-6) .. (9.4,1.55-6) .. controls (10.1,1.9-6) and (9.9,0.0-6) .. cycle;
\draw [thick] (10,-1-6) .. controls (8.8,0.3-6) and (9.1,1.3-6) .. (9.4,1.55-6) .. controls (10.1,1.9-6) and (9.9,0.0-6) .. cycle;
\shade [ball color=cyan!50!black] (10,-1-6) .. controls (11.2,0.3-6) and (10.9,1.3-6) .. (10.6,1.55-6) .. controls (9.9,1.9-6) and (10.1,0.0-6) .. cycle;
\draw [thick] (10,-1-6) .. controls (11.2,0.3-6) and (10.9,1.3-6) .. (10.6,1.55-6) .. controls (9.9,1.9-6) and (10.1,0.0-6) .. cycle;
\shade [ball color=cyan!50!black] (10,-1-6) .. controls (8.5,-0.3-6) and (7.5,1.3-6) .. (8.4,1.55-6) .. controls (9.1,1.8-6) and (8.7,0.0-6.1) .. cycle;
\draw [thick] (10,-1-6) .. controls (8.5,-0.3-6) and (7.5,1.3-6) .. (8.4,1.55-6) .. controls (9.1,1.8-6) and (8.7,0.0-6.1) .. cycle;
\shade [ball color=cyan!50!black] (10,-1-6) .. controls (11.5,-0.3-6) and (12.5,1.3-6) .. (11.6,1.55-6) .. controls (10.9,1.8-6) and (11.3,0.0-6.1) .. cycle;
\draw [thick] (10,-1-6) .. controls (11.5,-0.3-6) and (12.5,1.3-6) .. (11.6,1.55-6) .. controls (10.9,1.8-6) and (11.3,0.0-6.1) .. cycle;
%%%%%%%%%%%%%%%%%%%%%%%%%%%%%%
\filldraw [white] (10,-1-6) .. controls (9.0,0.3-6) and (9.2,1.0-6) .. (9.3,1.1-6) .. controls (9.5,1.1-6) and (9.3,0.0-6) .. cycle;
\draw [densely dotted] (10,-1-6) .. controls (9.0,0.3-6) and (9.2,1.0-6) .. (9.3,1.1-6) .. controls (9.5,1.1-6) and (9.3,0.0-6) .. cycle;
\filldraw [white] (10,-1-6) .. controls (9.05+0.3,0.3-6) and (9.2+0.3,1.0-6) .. (9.3+0.25,1.1-6) .. controls (9.4+0.3,1.1-6) and (9.3+0.3,0.0-6) .. cycle;
\draw [densely dotted] (10,-1-6) .. controls (9.05+0.3,0.3-6) and (9.2+0.3,1.0-6) .. (9.3+0.25,1.1-6) .. controls (9.4+0.3,1.1-6) and (9.3+0.3,0.0-6) .. cycle;
\filldraw [white] (10,-1-6) .. controls (9.05+0.6,0.3-6) and (9.2+0.5,1.0-6) .. (9.3+0.5,1.1-6) .. controls (9.4+0.5,1.1-6) and (9.3+0.5,0.0-6) .. cycle;
\draw [densely dotted] (10,-1-6) .. controls (9.05+0.6,0.3-6) and (9.2+0.5,1.0-6) .. (9.3+0.5,1.1-6) .. controls (9.4+0.5,1.1-6) and (9.3+0.5,0.0-6) .. cycle;
%%%%%%%%%%%%%%%%%%%%%%%%%%%%%%
\filldraw [white] (10,-1-6) .. controls (8.6,-0.3-6) and (8.0,1.0-6) .. (8.2,1.1-6) .. controls (8.4,1.1-6) and (8.2,0.0-6.0) .. cycle;
\draw [densely dotted] (10,-1-6) .. controls (8.6,-0.3-6) and (8.0,1.0-6) .. (8.2,1.1-6) .. controls (8.4,1.1-6) and (8.2,0.0-6.0) .. cycle;
\filldraw [white] (10,-1-6) .. controls (8.6+0.25,-0.3-6) and (8.0+0.25,1.0-6) .. (8.2+0.25,1.1-6) .. controls (8.4+0.25,1.1-6) and (8.2+0.25,0.0-6.0) .. cycle;
\draw [densely dotted] (10,-1-6) .. controls (8.6+0.25,-0.3-6) and (8.0+0.25,1.0-6) .. (8.2+0.25,1.1-6) .. controls (8.4+0.25,1.1-6) and (8.2+0.25,0.0-6.0) .. cycle;
\filldraw [white] (10,-1-6) .. controls (8.6+0.5,-0.3-6) and (8.0+0.5,1.0-6) .. (8.2+0.5,1.1-6) .. controls (8.4+0.5,1.1-6) and (8.2+0.5,0.0-6.0) .. cycle;
\draw [densely dotted] (10,-1-6) .. controls (8.6+0.5,-0.3-6) and (8.0+0.5,1.0-6) .. (8.2+0.5,1.1-6) .. controls (8.4+0.5,1.1-6) and (8.2+0.5,0.0-6.0) .. cycle;
%%%%%%%%%%%%%%%%%%%%%%%%%%%%%%
%%%%%%%%%%%%%%%%%%%%%%%%%%%%%%
\filldraw [white] (10,-1-6) .. controls (11.0,0.3-6) and (10.8,1.0-6) .. (10.7,1.1-6) .. controls (10.5,1.1-6) and (10.7,0.0-6) .. cycle;
\draw [densely dotted] (10,-1-6) .. controls (11.0,0.3-6) and (10.8,1.0-6) .. (10.7,1.1-6) .. controls (10.5,1.1-6) and (10.7,0.0-6) .. cycle;
\filldraw [white] (10,-1-6) .. controls (10.95-0.3,0.3-6) and (10.8-0.3,1.0-6) .. (10.7-0.25,1.1-6) .. controls (10.6-0.3,1.1-6) and (10.7-0.3,0.0-6) .. cycle;
\draw [densely dotted] (10,-1-6) .. controls (10.95-0.3,0.3-6) and (10.8-0.3,1.0-6) .. (10.7-0.25,1.1-6) .. controls (10.6-0.3,1.1-6) and (10.7-0.3,0.0-6) .. cycle;
\filldraw [white] (10,-1-6) .. controls (10.95-0.6,0.3-6) and (10.8-0.5,1.0-6) .. (10.7-0.5,1.1-6) .. controls (10.6-0.5,1.1-6) and (10.7-0.5,0.0-6) .. cycle;
\draw [densely dotted] (10,-1-6) .. controls (10.95-0.6,0.3-6) and (10.8-0.5,1.0-6) .. (10.7-0.5,1.1-6) .. controls (10.6-0.5,1.1-6) and (10.7-0.5,0.0-6) .. cycle;
%%%%%%%%%%%%%%%%%%%%%%%%%%%%%%
\filldraw [white] (10,-1-6) .. controls (11.4,-0.3-6) and (12.0,1.0-6) .. (11.8,1.1-6) .. controls (11.6,1.1-6) and (11.8,0.0-6.0) .. cycle;
\draw [densely dotted] (10,-1-6) .. controls (11.4,-0.3-6) and (12.0,1.0-6) .. (11.8,1.1-6) .. controls (11.6,1.1-6) and (11.8,0.0-6.0) .. cycle;
\filldraw [white] (10,-1-6) .. controls (11.4-0.25,-0.3-6) and (12.0-0.25,1.0-6) .. (11.8-0.25,1.1-6) .. controls (11.6-0.25,1.1-6) and (11.8-0.25,0.0-6.0) .. cycle;
\draw [densely dotted] (10,-1-6) .. controls (11.4-0.25,-0.3-6) and (12.0-0.25,1.0-6) .. (11.8-0.25,1.1-6) .. controls (11.6-0.25,1.1-6) and (11.8-0.25,0.0-6.0) .. cycle;
\filldraw [white] (10,-1-6) .. controls (11.4-0.5,-0.3-6) and (12.0-0.5,1.0-6) .. (11.8-0.5,1.1-6) .. controls (11.6-0.5,1.1-6) and (11.8-0.5,0.0-6.0) .. cycle;
\draw [densely dotted] (10,-1-6) .. controls (11.4-0.5,-0.3-6) and (12.0-0.5,1.0-6) .. (11.8-0.5,1.1-6) .. controls (11.6-0.5,1.1-6) and (11.8-0.5,0.0-6.0) .. cycle;
%%%%%%%%%%%%%%%%%%%%%%%%%%%%%%
\shade [ball color=yellow] (10,-1-6) circle (4pt);
\end{tikzpicture}
\caption{Starting from two 3-symplices, we obtain a 3-parachute (cyan) having four 2-faces (dark cyan), three 1-edges for each face (dotted lines), and a dot (yellow).}
\label{fig:parachutes3}
\end{center}
\end{figure}
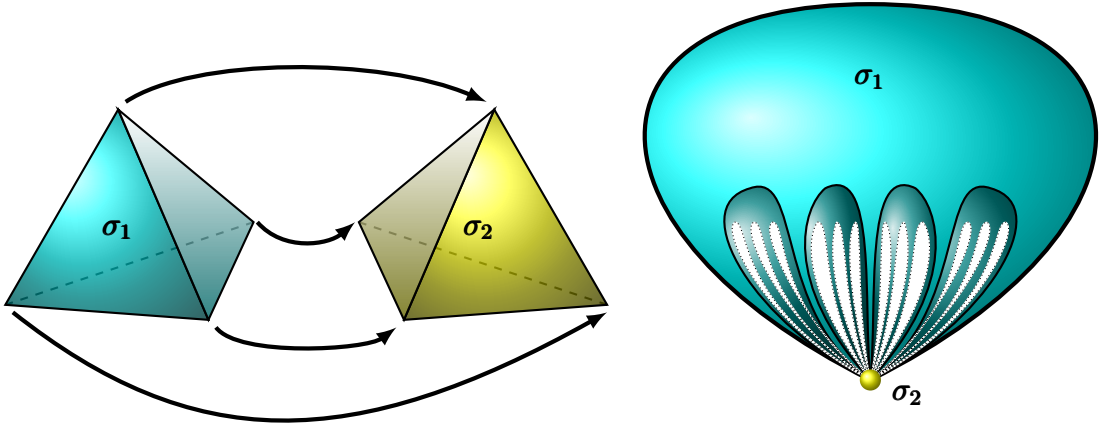

By applying these considerations recursively, each m-simplex in $\sigma_1$ gives rise to an m-dimensional "parachute" in which the vertices are collected in a point, the 1-simplexes transformed in 1-spheres, and the simplices of dimension between 2 and m-1 are transformed in ``parachutes'' of intermediate dimensions with the combinatorial relations proper to the faces of an m-simplex $\sigma$ (see Fig.~\eqref{fig:parachutes3}). In the absence of better and widely accepted terminology, we will refer to these constructions as \emph{m-parachutes}.\smallbreak

%%%%%%%%%%%%%%%%%%%%%%%%%%%%%%%%%%%%%%%%%%%%%%%%%%%%%%%%%%%%%%%%%%%%

The geometric consistency conditions introduced above guarantee that the m-parachute can be retracted from $\sigma\times F$ to $F$ and ultimately mapped to $G$ in a way that $\sigma_2$ defines a point $X_0\in G$. Observe that the dimension of $\sigma$ can be higher than the dimension of $G$.\smallbreak

One can always \emph{assign} to each 1-sphere in the m-parachute an element in $\pi_1(G)$.\smallbreak

When the identity element of $\pi_1(G)$ is assigned to a 1-sphere, the corresponding edges of $\sigma_1$ and $\sigma_2$ are called \emph{identifiable}. Or in other words, 1-parachutes that are contractible in $G$ correspond to identifiable edges. The non-identifiable edges give rise to a bouquet of 1-spheres $\mathcal{B}\subset G$. \smallbreak 

If all the 1-spheres in a 2-parachute are contractible in $G$, then the 2-parachute is a 2-sphere that contains $X_0$, so the retraction $\delta$ defines an element in $\pi_2(G)$.\smallbreak

Suppose a 2-parachute contains a non-trivial bouquet $\mathcal{B}$. The 2-parachute can be contracted to $\mathcal{B}$. In fact, a 2-sphere with n holes is topologically equivalent to a bouquet of 1-spheres. This means that such a 2-parachute does not define an element in $\pi_2(G)$ with base point $X_0$.\smallbreak

These considerations can be directly generalised to k-parachutes with $k\neq m$, while keeping in mind the combinatorics determined by the relations between faces in $\sigma$. Starting from the vertices and then considering k-faces with increasing k, we can assign trivial and non-trivial elements of $\pi_k(G)$ up to a maximum bouquet of spheres of various dimensions in each m-parachute and ultimately in $\sigma_1$. The obvious inclusions of simplices into simplices of higher order unambiguously define the maximum bouquet of spheres for each m-parachute.\smallbreak

We interpret the assigned non-trivial elements of $\pi_k(G)$ as topological obstructions for the identifiability of k-faces in $\sigma_1$ and $\sigma_2$. Intuitively, faces can be identified up to a certain dimension at which an obstruction topologically equivalent to a sphere makes them unidentifiable.\smallbreak

It can be proved (analogously to 1 and 2-dimensional cases) that faces that give rise to m-parachutes with a dimension higher than the dimension of the maximum bouquet they contain are contractible to the elements of the bouquet. This follows from the fact that the homotopy groups $\pi_m(G)$ of n-spheres are all trivial for $m<n$. \smallbreak

On the one hand, this means that m-parachutes with non-trivial maximum bouquets of spheres are not spheres and therefore they do not directly define elements in $\pi_m(G)$. On the other hand, we should account for the fact that the corresponding m-faces in $\sigma_1$ and $\sigma_2$ are not identifiable since they ``inherit'' the obstructions defined by the maximum bouquet $\mathcal{B}$ they contain. These considerations motivate the following definition. 

\begin{defi} The topological obstructions to the identifiability of m-faces in $\sigma_1$ and $\sigma_2$ are elements assigned from the relative homotopy group $\pi_m(G,\mathcal{B})$.  
\end{defi}

Key definitions and properties of relative homotopy groups are summarised in Appendix A. If all faces up to dimension m-1 are identifiable, then the maximum bouquet of spheres collapses to the base point $X_0$ and the relative homotopy group is $\pi_m(G)$. \smallbreak

If $\sigma$ is the intersection of multiple-faced simplices in $M$, then $s(\sigma)$ contains multiple m-simplices $\sigma_i$, and the topological obstructions are defined pairwise. The identifiability of faces in $\sigma_i$ and $\sigma_j$ is, in general, unrelated to the identifiability of faces in $\sigma_j$ and $\sigma_k$. This means that the maximum number of spheres in the m-parachutes defined by these distinct couples of simplices is, in general, different. Although a theory can be potentially developed in these more general terms by suitably defining the action and the composition of the topological $G$-correction of the transition maps (see below), we introduce the following restriction to the structural data in our construction.\smallbreak

The long sequence of groups and group  homomorphisms induced by the inclusions $i:(\mathcal{B},X_0)\longrightarrow (G,X_0)$ and $j:(G,X_0)\longrightarrow (G,\mathcal{B})$ is exact (see Appendix~A and \cite{Hatcher,Gray, Hu} for more details): $$\ldots\pi_{n+1}(G,\mathcal{B})\stackrel{\delta}{\longrightarrow}\pi_n(\mathcal{B},X_0)\stackrel{i_\ast}{\longrightarrow}\pi_n(G,X_0)\stackrel{j_\ast}{\longrightarrow} \pi_n(G,\mathcal{B})\stackrel{\delta}{\longrightarrow}\pi_{n-1}(\mathcal{B},X_0)\ldots$$\medbreak
 
In general, the group homomorphism $j_\ast:\pi_n(G,X_0)\longrightarrow \pi_n(G,\mathcal{B})$ is neither injective nor surjective, and its image and kernel are typically non-trivial. We impose the condition that the topological obstructions in $P$ are always chosen from the image of $j_*$ in $\pi_n(G,\mathcal{B})$. These elements correspond precisely to the kernel of the connecting homomorphism $\delta$. More importantly, the generators of $Im( j_\ast)\subset\pi_n(G,\mathcal{B})$ originate from the generators of the quotient $\pi_n(G,X_0)/\ker(j_\ast)$ (see Appendix A).
This implies that, by restricting to elements in $Im(j_\ast)$, we effectively select elements of $\pi_n(G)$ for each pair of n-simplices in $s(\sigma)$. This restriction allows us to interpret the topological obstructions in $s(\sigma)$ across multiple pairs of simplices as elements of $\pi_n(G)$ and to use the group structure. Observe that this construction accounts for the various maximal bouquets defined by different pairs of simpleices, each with its own associated kernel and corresponding constraint.\smallbreak 

Note that in the special (and very intuitive) case where common n-faces are identifiable up to order $n-1$, the maximal bouquet of spheres reduces to a point. Consequently, topological obstructions can be freely selected from $\pi_n(G)$ without restrictions.
In the remainder of this paper (unless otherwise stated), we refer to the topological obstructions included in the structural data of a bundle as elements of $\pi_n(G)$, selected from the quotient $\pi_n(G,X_0)/\ker(j_\ast)$. \smallbreak 

\begin{defi} We refer to the structural group $G$ and the set of topological obstructions in $\pi_n(G)$ assigned to each n-simplex in the intersection of faced simplices in a principal bundle $P$ as the structural data of $P$. \label{structural}
\end{defi}

Different ways of generating or assigning these structural elements naturally or by prescribing them in a data-driven way, will be discussed in a dedicated section below. This is one of the cases where ``theoretical'' and ``modelling'' aspects will be consistently combined in our construction.\smallbreak

The local Definition~\eqref{localdef} and Definition~\eqref{structural} are combined in the following construction. 

\begin{defi}\label{transitions} Consider a principal bundle $P$ with assigned set of structural data. Given an injective map $\beta_n:\pi_k(G)\longrightarrow G$ that preserves the group identity element and a k-simplex $\sigma\in M$, the transition maps $\zeta_{ij}$ between the i-th and the j-th simplices  in $s(\sigma)$ are defined as elements of $(\epsilon(G),\pi_k(G))$ such that:\smallbreak

-  For a vertex $X$, which belongs to two faced simplices $\sigma_i$ and $\sigma_j$, given a section $s$, we denote $s(X)|\sigma_i=X_i$ and analogously $X_j$. Consider a trivialisation $\psi_X$ of the bundle in $X$ (according to Definition~\ref{trivialisations}). Then $\psi_{X}(X_i)=\zeta_{ij}\psi_X(X_j)$.

\smallbreak 

- The second term acts on the first term via $(\beta_n(\mathcal{S})\epsilon(g_1\ldots g_n), \mathcal{S})$ for $\mathcal{S}\in\pi_n(G)$.

\end{defi}

The operation in the second term $\pi_n(G)$ is a product or a sum, depending on $n$; we will refer to the second term as the topological (obstruction) term. \smallbreak

The first term connects trivialisations and extends the $\epsilon(G)$-action between faced simplices. Observe that the transition maps on 0-simplex $X\in M$ satisfy the usual cocycle rule $\zeta_{ij}(X) \zeta_{jk}(X)= \zeta_{ik}(X)$, whereas the $\epsilon(G)$-action does not, so the construction is a generalisation of the classical case. As discussed above, the maximum bouquets defined by distinct pairs of simplices are not directly related; this means that the topological term in the product in general does not satisfy the cocycle rule. \smallbreak

In some circumstances, the topological terms can satisfy the cocycle rule. For example, in the case where all maximum bouquets are collapsed to points, it is possible to assign the topological obstructions starting from a subset of pairs and completing the remaining pairs by assigning the opportune sums/products in $\pi_n(G)$.\smallbreak  

\textbf{NB.} We complete the construction of a semidiscrete principal bundle by extending the local Definition~\eqref{localdef} and by replacing the local $\epsilon (G)$ action with $(\chi\epsilon(G),\pi_\bullet(G))$.  We use the notation $\pi_\bullet (G)$ to generically refer to elements of different dimensions. The ``topological corrections'' $\beta_n$ to the local group action $\epsilon(G)$ account for the topological obstructions. Clearly, the local action can be easily recovered, as the action in the local section of $P$ indices the identity in the second term.\smallbreak

\begin{defi}\label{globaldef}
A semidiscrete principal bundle is locally a triple $(P, M, \pi)$.  The total space $P$ is a family of simplicial complexes endowed with a left, simply transitive local action $(\epsilon(G), \pi_\bullet(G))$ induced by a Lie group $G$. The projection $\pi:P\longrightarrow M$ maps every $G$-orbit to a 0-simplex in $M$. The topological obstructions and the topological correction in $\zeta_{ij}$ are structural data that define the principal bundle through its transition maps.\end{defi}

\begin{rem} The differential is a local operation; this means that it is performed ``before'' transition maps are applied. Whether or not the transition maps commute with the differential depends on the assignment of the topological data and the definition of $\chi_n$ for consecutive values of $n$. We will develop further this point in Section~8 by introducing a concept of torsion suitable for semidiscrete principal bundles.\end{rem}

Classically trivial principal bundles admit smooth global sections. We adapt this concept by saying that:

\begin{defi}    
A semidiscrete principal bundle is trivial if it admits sections that are global simplicial isomorphisms, or equivalently, if all the topological terms in the transition maps $(\epsilon(G),\pi_n(G))$ are equal to the identity. \label{trivial1}
\end{defi} This means that common faces of faced simplices in $M$ are mapped to a common face and not to multiple faces in $s(M)$.  As expected, in such a case, the local property of being a product of a faced simplex and a Lie group $G$ is extended to the whole simplicial complex $M$. \smallbreak

\begin{defi}
For $n>1$, we call an obstruction form $\chi_n$ on $M$ a $\pi_n(G)$-valued n-form that associates to each n-simplex $\sigma\in M$ the sum (product for n=1) of all the topological obstructions between n-simplices in $s(\sigma)$.
\label{ObstructionForm}
\end{defi}

The role of the obstruction n-form is to capture the $\pi_n(G)$ obstructions to the triviality of the bundle. For $n>1$, the homotopy groups are abelian, so the obstruction n-forms are unambiguously defined. For $n=1$, the exact order of composition of the topological obstructions is irrelevant; therefore, any arbitrary order can be selected for the definition of the obstruction form.\smallbreak 

\begin{teor} The obstructions for the triviality of a semidiscrete principal $G$-bundle $P$ on $M$ are determined by elements in $$H^1(\pi_1(G))\times H^2(\pi_2(G))\times\ldots\times H^{n-1}(\pi_{n-1}(G))$$ \noindent where $n$ is the highest dimension of a simplex in $M$. \label{triviality}\end{teor}

\nit{Proof} We analyse how topological obstructions (and obstruction forms) behave with respect to the differential operator, and we prove that the obstruction forms $\chi_n$ are closed, but not exact. We consider an n-simplex $\sigma\in M$ in the intersection of more than one faced simplices in M. By construction, the topological obstruction in $\pi_n(G)$ cannot be associated with simplices of lower dimensions. Therefore, in general, the topological obstructions in $\pi_n(G)$ cannot be expressed as the differential of a $\pi_n(G)$-valued  (obstruction) (n-1)-form. \smallbreak

We prove that $\chi_n$ is closed for each component/summand in Definition~\eqref{ObstructionForm} associated with pairs in $s(\sigma)$. \smallbreak

If $\sigma$ is the highest order common face of two-faced simplices, then by construction there cannot be topological obstructions associated with the (n+1)-faces of the two-faced simplices. For this reason, we adopt the convention/definition that in these cases we treat $\chi$  as a top-dimension form and set $d\chi$ as equal to the identity.\smallbreak

If $\sigma$ is not the common face of the highest dimension, consider a (n+1)-simplex $\tau\in M$ such that $\sigma\subset\partial\tau$. We have to prove that $d\chi(\tau)=\Sigma(-1)^i \chi(\sigma_i)$ (with product notation for n=1) is always contractible to a point in $\tau\times F$, i.e. the differential resolves the obstruction.\smallbreak

We select a pair $\tau_i$ and $\tau_j$ in $s(\tau)$. For simplicity, let us first consider the case in which all (n-1)-faces of two simplices in $s(\sigma)$ are identifiable, and the special case in which all the n-faces of the pair are identifiable with the exception of one. We can select a section such that the n-faces are identified. In $\tau\times F$, this configuration is topologically equivalent to an (n+1)-disk with the obstruction n-sphere mapped to its boundary. The extra dimension of $\tau$ over $\sigma$ allows us to contract the topological obstruction into a point in $\tau\times F$. \smallbreak

Suppose that there are multiple topological $S^n$-obstructions over $\partial\tau$. Observe that $\tau_1$ can be contracted to a point contained in each of these n-spheres. This bouquet of n-spheres can be retracted to $F$, and the n-spheres are mapped to $\pi_n(G)$ generators assigned by the structure data of $P$. The expression $\Sigma(-1)^i \chi(\sigma_i)$ computed in $\pi_n(G)$ with a fixed base point gives one element $\mathcal{S}\in\pi_n(G)$. As a result, a generator of $\mathcal{S}$ becomes the boundary of $\tau_2$ in $\tau\times F$ and contains the base point $\tau_1$ as in the special case described above.\smallbreak

\begin{figure}[!htbp]
\begin{center}
\begin{tikzpicture}[>=latex,decoration={zigzag,amplitude=.5pt,segment length=2pt}]
\shade[ball color=black!30!white] (0,1+6) ellipse (1.3 and 1.8);
\filldraw [white] (0,2.3+6) ellipse (1.4 and 0.5);
\draw [thick, top color=white,bottom color=pink] (0,2+6) ellipse (1.08 and 0.25); 
\filldraw [red] (-2.4,1.2+6) -- (-2.1,1.2+6) -- (-2.1,1.35+6) -- (-1.9,1+6) -- (-2.1,0.65+6) -- (-2.1, 0.8+6) -- (-2.4,0.8+6) -- cycle;
%%%%%%%%%%%%%%%%%%%%%
\draw [thick, ->] (-0.1,2-0.25+6) -- (0.1,2-0.25+6); 
\draw [thick, ->] (0.1,2+0.25+6) -- (-0.1,2+0.25+6);
%%%%%%%%%%%%%%%%%%%%%
\shade[ball color=black!30!white] (-4.3,1+6) ellipse (1.3 and 1.8);
\filldraw [white] (-5.7,2+6) -- (-3.1,2+6) -- (-3.1,2.8+6) -- (-5.7,2.8+6) -- cycle;
\draw [thick, top color=white,bottom color=pink] (-4.84,2+6) ellipse (0.54 and 0.15); 
\draw [thick, top color=white,bottom color=pink] (-3.76,2+6) ellipse (0.54 and 0.15); 
%%%%%%%%%%%%%%%%%%%%%
\draw [thick, ->] (-4.94,2-0.15+6) -- (-4.74,2-0.15+6); 
\draw [thick, ->] (-4.74,2+0.15+6) -- (-4.94,2+0.15+6);
\draw [thick, ->] (-3.86,2-0.15+6) -- (-3.66,2-0.15+6); 
\draw [thick, ->] (-3.66,2+0.15+6) -- (-3.86,2+0.15+6); 
%%%%%
%%%%%
\shade[ball color=black!30!white] (0,1) ellipse (1.3 and 1.8);
\filldraw [white] (-1.4,2) -- (1.4,2) -- (1.4,2.8) -- (-1.4,2.8) -- cycle;
\filldraw [red] (-2.4,1.2) -- (-2.1,1.2) -- (-2.1,1.35) -- (-1.9,1) -- (-2.1,0.65) -- (-2.1, 0.8) -- (-2.4,0.8) -- cycle;
\shade [top color=white,bottom color=pink] (0,1.5) .. controls (-0.4,1.3) and (-1.08,1.7) .. (-1.08,2) arc (180:0:0.54 and 0.15) -- cycle; 
\filldraw [white] (0.54,2) ellipse (0.54 and 0.15); 
\shade [top color=white,bottom color=pink] (1.08,2) .. controls (0.88,2.4) and (0.54,2.4) .. (0.24,1.88) .. controls  (0.24,1.8) and (0.9,1.8) .. cycle;
\draw [thick] (-1.08,2) arc (180:0:0.54 and 0.15) arc (180:360:0.54 and 0.15) .. controls (0.88,2.4) and (0.54,2.4) .. (0.24,1.88);
\draw [thick] (0,1.5) .. controls (-0.4,1.3) and (-1.08,1.7) .. (-1.08,2);
\filldraw (0,1.5) circle (2pt);
%%%%%%%%%%%%%%%%%%%%%
\shade[ball color=black!30!white] (-4.3,1) ellipse (1.3 and 1.8);
\filldraw [white] (-5.7,2) -- (-3.1,2) -- (-3.1,2.8) -- (-5.7,2.8) -- cycle;
\draw [thick, top color=white,bottom color=pink] (-4.84,2) ellipse (0.54 and 0.15); 
\draw [thick, top color=white,bottom color=pink] (-3.76,2) ellipse (0.54 and 0.15); 
\filldraw [red] (4.3-2.4,1.2) -- (4.3-2.1,1.2) -- (4.3-2.1,1.35) -- (4.3-1.9,1) -- (4.3-2.1,0.65) -- (4.3-2.1, 0.8) -- (4.3-2.4,0.8) -- cycle;
%%%%%%%%%%%%%%%%%%%%%
\draw [thick, ->] (-4.94,2+0.15) -- (-4.74,2+0.15); 
\draw [thick, ->] (-4.74,2-0.15) -- (-4.94,2-0.15);
\draw [thick, ->] (-3.86,2-0.15) -- (-3.66,2-0.15); 
\draw [thick, ->] (-3.66,2+0.15) -- (-3.86,2+0.15); 
%%%%%%%%%%%%%%%%%%%%%
\shade[ball color=black!30!white] (4.3,1) ellipse (1.3 and 1.8);
\filldraw [white] (4.3-1.4,2) -- (4.3+1.4,2) -- (4.3+1.4,2.8) -- (4.3-1.4,2.8) -- cycle;
\filldraw [red] (4.3-2.4,1.2) -- (4.3-2.1,1.2) -- (4.3-2.1,1.35) -- (4.3-1.9,1) -- (4.3-2.1,0.65) -- (4.3-2.1, 0.8) -- (4.3-2.4,0.8) -- cycle;
\shade [top color=white,bottom color=pink] (4.3,0.5) .. controls (4.3-0.4,0.4) and (4.3-1.08,1.7) .. (4.3-1.08,2) arc (180:0:0.54 and 0.15) -- cycle; 
\filldraw [white] (4.3+0.54,2) ellipse (0.54 and 0.15); 
\shade [top color=white,bottom color=pink] (4.3+1.08,2) .. controls (4.3+0.88,2.4) and (4.3+0.74,2.4) .. (4.3+0.64,1.86) .. controls  (4.3+0.69,1.87) and (4.3+0.9,1.8) .. cycle;
\draw [thick] (4.3-1.08,2) arc (180:0:0.54 and 0.15) arc (180:360:0.54 and 0.15) .. controls (4.3+0.88,2.4) and (4.3+0.74,2.4) .. (4.3+0.64,1.86);
\draw [thick] (4.3,0.5) .. controls (4.3-0.4,0.4) and (4.3-1.08,1.7) .. (4.3-1.08,2);
\filldraw (4.3,0.5) circle (2pt);
\draw [thick, ->] (4.3-4.94,2+0.15) -- (4.3-4.74,2+0.15); 
\draw [thick, ->] (4.3-3.86,2-0.15) -- (4.3-3.66,2-0.15); 
%%%%%%%%%%%%%%%%%%%%%
\filldraw [red,rotate around={-90:(6.8-2.4,-2.3+1)}] (6.8-2.4,-2.3+1.2) -- (6.8-2.1,-2.3+1.2) -- (6.8-2.1,-2.3+1.35) -- (6.8-1.9,-2.3+1) -- (6.8-2.1,-2.3+0.65) -- (6.8-2.1, -2.3+0.8) -- (6.8-2.4,-2.3+0.8) -- cycle;
%%%%%%%%%%%%%%%%%%%%%
\shade[ball color=black!30!white] (4.3,-5+1) ellipse (1.3 and 1.8);
\filldraw [white] (4.3-1.4,-5+2) -- (4.3+1.4,-5+2) -- (4.3+1.4,-5+2.8) -- (4.3-1.4,-5+2.8) -- cycle;
\filldraw [red,rotate around={180:(2.15,-4)}] (4.3-2.4,-5+1.2) -- (4.3-2.1,-5+1.2) -- (4.3-2.1,-5+1.35) -- (4.3-1.9,-5+1) -- (4.3-2.1,-5+0.65) -- (4.3-2.1,-5+ 0.8) -- (4.3-2.4,-5+0.8) -- cycle;
\shade [top color=white!89!pink,bottom color=pink!80!black] (4.3,-6.25+0.5) .. controls (4.3-0.4,-6.25+0.4) and (4.3-1.08,-5+1.7) .. (4.3-1.08,-5+2) arc (180:0:0.54 and 0.15) -- cycle; 
\filldraw [white] (4.3+0.54,-5+2) ellipse (0.54 and 0.15); 
\shade [top color=white,bottom color=pink] (4.3+1.08,-5+2) .. controls (4.3+0.98,-5+2.4) and (4.3+0.94,-5+2.4) .. (4.3+0.94,-5+1.89) .. controls  (4.3+0.95,-5+1.89) and (4.3+1,-5+1.89) .. cycle;
\draw [thick] (4.3-1.08,-5+2) arc (180:0:0.54 and 0.15) arc (180:360:0.54 and 0.15) .. controls (4.3+0.98,-5+2.4) and (4.3+0.94,-5+2.4) .. (4.3+0.94,-5+1.89);
\draw [thick] (4.3,-6.25+0.5) .. controls (4.3-0.4,-6.25+0.4) and (4.3-1.08,-5+1.7) .. (4.3-1.08,-5+2);
\filldraw (4.3,-6.25+0.5) circle (2pt);
\draw [thick, ->] (8.6-4.94,2+0.15) -- (8.6-4.74,2+0.15); 
\draw [thick, ->] (8.6-3.86,2-0.15) -- (8.6-3.66,2-0.15); 
\draw [thick, ->] (8.6-4.94,-3+0.15) -- (8.6-4.74,-3+0.15); 
\draw [thick, ->] (8.6-3.86,-3-0.15) -- (8.6-3.66,-3-0.15); 
%%%%%%%%%%%%%%%%%%%%%
\filldraw [pink!60!white] (0,-2.2) arc (90:270:1.8) -- cycle;
\filldraw [gray!90!white] (0,-2.2) arc (90:-90:1.8) -- cycle;
\draw[thick] (0,-5+1) circle (1.8);
\draw [thick, ->] (-1.8,-4.1) -- (-1.8,-3.9); 
\draw [thick, ->] (1.8,-3.9) -- (1.8,-4.1); 
\end{tikzpicture}
\caption{In the first line, after the homotopic product of two equally oriented boundaries, the result is reduced to a disc. The product doesn't preserve the homotopy; in the second and third lines, after the homotopic product of two oppositely oriented boundaries, the result is reduced to a disc. Neither the product nor the deformation (obtained through a self-crossing with cutting and sewing of the surface) preserves the homotopy.}
\label{fig:opporiented}
\end{center}
\end{figure}
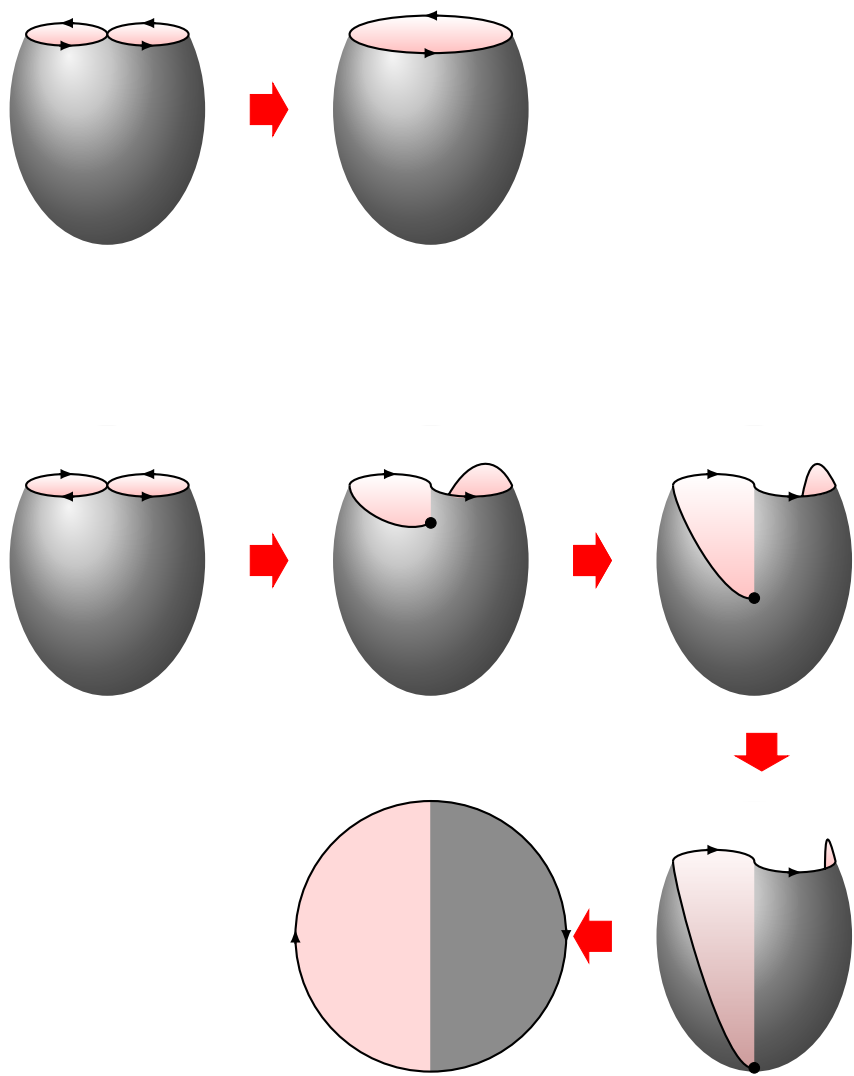

It is easy to see that the statement is valid with some generators reversing their orientation in the expression of the differential (see Fig.\eqref{fig:opporiented} for intuition on this point). So, obstructions become contractable to a point when an additional dimension is added.\smallbreak

The same argument is also valid for a non-trivial maximum bouquet. The operation in the relative homotopy group is used to compute the differential (one element in $\pi_n(G)$) that can always be contracted to the bouquet due to the additional dimension and the construction of relative homotopy (see Appendix~A). \qed\smallbreak

\begin{rem} Our definition of obstruction $\pi_n$-valued n-forms differs from the standard definition of obstruction co-cycle in algebraic topology \cite{Whitehead1978, Hatcher}, which is $\pi_{n-1}$-valued n-cocycle. 
\end{rem}

Similar considerations imply that:

%%. \begin{prop}The obstruction for a k-cycle in $M$ to be lifted to a k-cycle in $s(M)$ is determined by elements of $H^1(\pi_1(G))\times H^2(\pi_2(G))\times...\times H^{k-1}(\pi_{k-1}(G))$.\end{prop} NON MI RICORDO PERCHE'.... QUESTO CI SERVE???

\begin{defi} In view of Theorem~\eqref{triviality} we call the cohomology classes on $M$ generated by the obstruction forms on $M$ the characteristic classes of the semidiscrete principal bundle $P$ oven $M$.\end{defi}

\begin{defi} \label{equivalentPQ} Semidiscrete principal $G$-bundles over $M$ are equivalent if the total spaces are isomorphic and their sections can be globally isomorphically mapped.\end{defi}

Equivalence classes of semidiscrete principal bundles on $M$ are determined by the assigned structural data. 

\begin{teor} The classifying space of semidiescrete principal $G$-bundles on a discrete manifold $M$ is the set of parachutes defined by the intersections of faced simplices in the base space. \label{ClassSpace}\end{teor} 

\nit{Proof} This follows from the construction. The possible assignments of structural data are parametrised by the maximum bouquets and the parachutes over common faces of faced simplices in $M$. \qed\medbreak

This construction is a generalisation to the discrete case of the classification of principal bundles over a paracompact manifold $X$. In that case, the classification is determined by the homotopy classes $[X,BG]$ where $BG$ is the classifying space. For example, for $O(n)$ bundles $BG=G_n(\mathbb R^\infty)$ is the infinite Grassmannian \cite{DNF-volII}.  \smallbreak

A fibre bundle morphism is a map that commutes with the bundle projections and preserves the fibres.

\begin{defi}\label{gaugedefi}Given a semidiscrete principal $G$-bundle $P$, we call a local bundle automorphism or a gauge transformation a map $\Phi: P\longrightarrow P$ which induces the identity on each faced simplex in the base space and has the following properties:$$\Phi(p)=\phi(p)p$$\noindent where $p$ is a point in $P$ and $\phi: P\longrightarrow G$ is an equivariant map defined by $\Phi$, i.e. $\phi(gp)=g\phi(p)g^{-1}.$ \end{defi}

We emphasise the fact that given a faced simplex $\sigma\in M$, $\phi(p)$ depends on $\pi^{-1}\sigma$, so care must be taken when considering common faces of faced simplices. 

\begin{defi}\label{gaugefix} The choice of a section $s$ of a semidiscrete principal bundle $P$ is called gauge fixing.\end{defi}

%%%%%%%%%%%%%%%%%%%%%%%%%%%%%%%%%%%
\section{Ways to assign structural data to a semidiscrete principal bundle}\label{computingClasses}

In this section, we discuss various ways to assign the topological structural data in a semidiscrete principal bundle. These examples have no ambition to be exhaustive, and aim to illustrate the fact that the theory is non-trivial. \medbreak

\textbf{``Pure modelling'' approach:}  The parachutes defined by common faces of faced simplexes in $M$ are ``populated'' with topological obstructions with some modelling criterion, for example, assigned randomly or placed in relevant positions. The behaviour of the system is then analysed by applying different configurations of the topological obstructions, i.e. different realisations of the characteristic classes. \smallbreak

%The topological obstructions are consistently assigned i.e. when for $\sigma\in M$ a section $s(\sigma)$ contains multiple simplices, the topological term in the transition maps should satisfy the cocycle rule, which means that topological obstructions/generators should be assigned to a sufficient number of couples of simplexes in $s(\sigma)$ and the rest extended by the cocycle rule.    

The topological correction applied to the $\epsilon(G)$ product depends on a map $\alpha_n:\pi_n(G)\longrightarrow \epsilon_n(G)$. The only requirement this map has to satisfy is $\alpha(Id)=Id$. At this stage, we are not going to specify this map further. This is one of the model-like aspects of our construction, which is left arbitrary to some extent.\bigbreak 

\textbf{Natural assignment:}  Consider a section of a semidiscrete principal bundle. A (generic) section maps the vertices of each faced simplex to the structure group $G$. A classical compact Lie group $G$ is a smooth Riemannian manifold with respect to the invariant metric associated with the Killing form. Once a natural invariant metric and a volume form are selected, simplices geometrically embedded in $G$ can be intuitively defined. Two points (0-simplices ) in $G$ define a 1-simplex by the shortest geodesic segment that connects the points. Three points in a generic position define a triangle of geodesics, and a 2-simplex can be defined as a portion of a surface with minimal area with a boundary determined by geodesics. By induction, given a boundary of (n-1)-faces, for $n\le dim(G)$, an n-simplex is the locus of points that minimises the volume in the convex hull of the boundary.\smallbreak %{(notare che qui ho aggiunto compact, altrimenti $G$ non \`e Riemanniano ma solo pseudo Riemanniano e le affermazioni successive cadono)}

An important caveat is that points in a generic position in $G$ might not define a geometric simplex. In fact, the ($n-1$)-faces could ``embrace'' a topological obstruction. As an example, three points on a geodesic circle on a 2-torus satisfy the definition of a geodesic triangle but do not define a 2-simplex on the torus. If the points in $G$ are selected ``closely enough'', they define a geometric realisation of a simplex in $G$. \smallbreak

\begin{rem}In this context, we can impose the local condition that, compatibly with the dimensions, regular sections of a semidiscrete principal bundle map each faced simplex in $P$ to a geometric simplex in $G$. \label{regular} \end{rem}

 Consider a geometric simplex $\sigma$ in a Lie Group defined by the volume-minimising procedure with boundary conditions defined above. We can transform $\sigma$ by acting on each of its vertices by some 1-parameter continuous subgroup in $G$ and by redefining the simplex with continuity at each point by the same volume-minimising procedure. Recall that geodesics in compact Lie groups correspond to exponential maps \cite{Bump}, Thm. 16.2. \smallbreak

 With the notations adopted in the paper, consider a section $s$ of a principal bundle $P$, two distinct n-simplices  $\sigma_1$ and $\sigma_2$ that belong to $s(\sigma)$, where $\sigma$ is a common n-face of two faced simplices in $M$. We can restrict the $\epsilon(G)$-action in $P$ to the continuous transformation that identifies each pair of vertices in $\sigma_1$ and $\sigma_2$ projected onto the same vertex in $M$ by the shortest geodesic segment between the vertices. The vertices of the two simplices can always be identified by this process, but this is not guaranteed for the faces of higher dimensions. Suppose that the boundaries of a k-face of $\sigma_1$ and $\sigma_2$ are identified by the above process. The k-faces in $\sigma_1$ and $\sigma_2$ define a topological k-sphere in $G$. If faces cannot be identified, then a non-trivial homotopy class in $\pi_k(G)$ (its representative) is detected by this procedure. This construction associates an element in $\pi_k(G)$ to a k-face in $M$. Given that $\sigma_1$ or $\sigma_2$ can be contracted to a point, this procedure naturally defines the maximum bouquet of spheres associated with $\sigma$. \smallbreak

In these circumstances, we will define a way to assign a topological correction in $G$ to each pair of k-faces $\sigma^*_1\subseteq\sigma_1$ and $\sigma^*_2\subseteq\sigma_2$. We will start with simple and intuitive examples to introduce the general case:\bigbreak

1. $\sigma^*_1$ and $\sigma^*_2$ are identifiable.\medbreak

2. $\sigma^*_1$ and $\sigma^*_2$ are not identifiable, but their boundaries are (i.e., they generate elements of the maximum bouquet $\mathcal{B}$ associated with $\sigma$).\medbreak

3. $\sigma^*_1$ and $\sigma^*_2$ contain unidentifiable faces of lower dimension as in case 2 (one or more elements of the maximum bouquet are associated with the projection of $\sigma^*_1$).\medbreak

In case 1, the topological correction assigned to the pair of faces is $Id\in G$. \smallbreak

In case 2, one can arbitrarily select a common vertex in $\sigma^*_1$ and $\sigma^*_2$. The two k-faces contain a unique common (k-1)-face which does not contain the selected vertex. In the simplex $\sigma^*_1$, there is a unique geodesic segment $\gamma_1\subset G$ that connects the selected vertex to the barycentre of the opposite (k-1)-face. Analogously $\gamma_2$ in $\sigma^*_2$. The selected vertex can be continuously translated along the geodesic in $\gamma_1$ and then along $\gamma_2$ and transform $\sigma_1$ (note that the faced simplex is consistently transformed) by following the volume-minimising procedure. The result of this operation is that $\sigma^*_1$ and $\sigma^*_2$ are identified and the corresponding topological obstruction in $\pi_k(G)$ is ``resolved''. This construction provides a map $\pi_k(G)\longrightarrow G$, which depends on the choice of a vertex and associates to the detected homotopy class the element $\exp\xi_1\theta_1*\exp\xi_2\theta_2\in G$.\smallbreak 

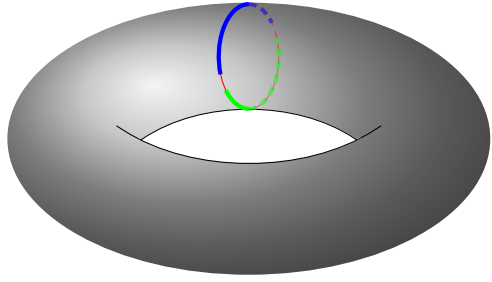
\begin{figure}[!htbp]
\begin{center}
\begin{tikzpicture}[>=latex,decoration={zigzag,amplitude=.5pt,segment length=2pt}]
\begin{scope}[scale=0.8]
\shade [ball color=black!30!white] (0,1.05*1.25*2) ellipse (1.605*1.25*2 and .905*1.25*2);
\filldraw[white,rounded corners=24*2pt] (-.9*2,1.3*2)--(0,1.9*2)--(.9*2,1.3*2);
\filldraw[white,rounded corners=24*2pt] (-.9*2,1.3045*2)--(0,0.82*2)--(.9*2,1.3045*2);
\path[rounded corners=24*2pt] (-.9*2,1.3*2)--(0,1.9*2)--(.9*2,1.3*2) (-.9*2,1.3*2)--(0,.92*2)--(.9*2,1.3*2);
\draw[rounded corners=28*2pt] (-1.1*2,1.42*2)--(0,.72*2)--(1.1*2,1.42*2);
\draw[rounded corners=24*2pt] (-.9*2,1.3*2)--(0,1.9*2)--(.9*2,1.3*2);
\end{scope}
%\draw[opacity=0.5,thick,red,densely dashed] (0,0.152*2) arc (270:90:.2*2 and .365*2);
%\draw [red, thick] (0,0.152*2) arc (-90:90:.2*2 and .365*2);
\draw [red] (0,1.945*2) arc (90:270:.2*2 and .348*2);
\draw [opacity=0.5,red,densely dashed] (0,1.945*2) arc (90:-90:.2*2 and .348*2);
\draw [blue,ultra thick] (0,1.945*2) arc (90:200:.2*2 and .348*2);
\draw [green,ultra thick] (0,1.945*2-.348*4) arc (270:220:.2*2 and .348*2);
\draw [opacity=0.5,ultra thick,blue,densely dashed] (0,1.945*2) arc (90:40:.2*2 and .348*2);
\draw [opacity=0.5,ultra thick,green,densely dashed] (0,1.945*2-.348*4) arc (-90:20:.2*2 and .348*2);
\end{tikzpicture}
\caption{The blue simplex and the green simplex on the torus have two common vertices. They cannot be identified by moving their vertices along the shortest geodesic segments between them. The procedure of moving vertices along geodesics allows us to identify the two 1-simplices and avoid the $S^1$ topological obstruction. }
\label{fig:torus}
\end{center}
\end{figure}

\noindent\emph{Example:} Consider two segments of a geodesic circle (1-simplices) on a torus as displayed in Fig. \eqref{fig:torus}. If the vertices of the segments are identified along the shortest geodesic path, the two simplices form an $S^1$ that ``embraces a homotopy generator''. Select one of the vertices (now common to both simplices) and apply the process to move it along a geodesic segment until it coincides with the other common vertex (barycentre of the opposite face). One of the simplices collapses to a point. Then we move the same vertex along the second geodesic segment until it reaches the other vertex. The two 1-simplices are identified through this process and the topological obstruction is avoided.\smallbreak   

\begin{figure}[!htbp]
\begin{center}
\begin{tikzpicture}[>=latex,decoration={zigzag,amplitude=.5pt,segment length=2pt},scale=1.3]
%%%%%%%%%%%%%%%%%%%%%%%%%%%%%%%%%%%%%%%%%%%%%%%%%%%%%%%%%%
%\draw[thick,dashed] (-1.5,-6) -- (1.8,-4.9);
\draw [ultra thick,-] (-1.5,-6) .. controls (-0.5,0.8-6) and (-0.1,1.1-6) .. (1.0666,-5.7+2.598/3);
\shade [top color=gray!20!black,bottom color=black] (-1.5,-6) -- (1.8,-4.9) -- (1.2,-6.2) -- cycle;
\shade [top color=white,bottom color=gray!40!black,opacity=0.8] (-1.5,-6) -- (1.8,-4.9) -- (1.2,-6.2) -- cycle;
\shade[ball color=cyan,opacity=0.8] (-1.5,-6) -- (1.2,-6.2) -- (0,2.598-6) -- cycle;
\draw [thick] (-1.5,-6) -- (1.2,-6.2) -- (0,2.598-6) -- cycle;
\draw [ultra thick] (-1.5,-6) -- (0,2.598-6);
\shade [top color=white,bottom color=red!40!black,opacity=0.8] (1.2,-6.2) -- (0,2.598-6) -- (1.8,-4.9) -- cycle;
\draw [thick] (1.2,-6.2) -- (0,2.598-6) -- (1.8,-4.9) -- cycle;
\filldraw (1.0666,-5.7+2.598/3) circle (2pt);
%\draw[thick,dashed] (5+1.5,-6) -- (5-1.8,-4.9);
\shade [top color=gray!20!black,bottom color=black] (5+1.5,-6) -- (5-1.8,-4.9) -- (5-1.2,-6.2) -- cycle;
\draw [ultra thick,-] (5+1.5,-6) .. controls (5.5,0.8-6) and (5.1,1.1-6) .. (5-1.0666,-5.7+2.598/3);
\shade[ball color=yellow,opacity=0.8] (5+1.5,-6) -- (5-1.2,-6.2) -- (5-0,2.598-6) -- cycle;
\draw [thick] (5+1.5,-6) -- (5-1.2,-6.2) -- (5-0,2.598-6) -- cycle;
\draw [ultra thick] (6.5,-6) -- (5-0,2.598-6);
\shade [top color=white,bottom color=red!40!black,opacity=0.8] (5-1.2,-6.2) -- (5-0,2.598-6) -- (5-1.8,-4.9) -- cycle;
\draw [thick] (5-1.2,-6.2) -- (5-0,2.598-6) -- (5-1.8,-4.9) -- cycle;
\filldraw (5-1.0666,-5.7+2.598/3) circle (2pt);
%%%%%%%%%%%%%%%%%%%%%%%
\node[opacity=0.3] at (0,1-5.8) {\Large$\pmb {\gamma_1}$};
\node[opacity=0.3] at (4.8,1-5.8) {\Large$\pmb {\gamma_2}$};
%%%%%%%%%%%%%%%%%%%%%%%%%%%%%%
%\shade [ball color=yellow] (10,-1-6) circle (4pt);
\end{tikzpicture}
\caption{The red faces are identifiable. The dark 2-faces on the bottom and the thick 1-faces are not.}
\end{center}
\label{tetraedri e geodetiche}
\end{figure}
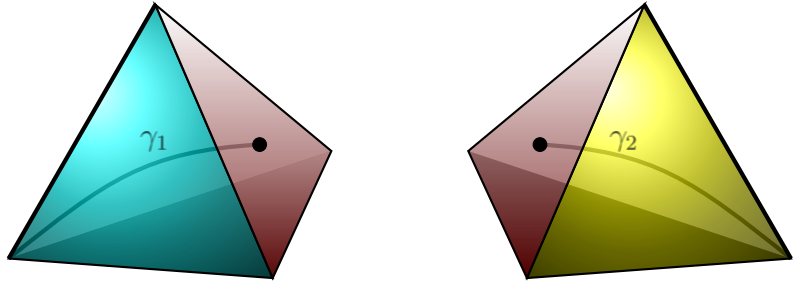

In case 3, suppose that k-faces $\sigma^*_1$ and $\sigma^*_2$ contain at least one identifiable common face of dimension k-1. All topological obstructions detected by faces of lower dimensions contained in $\sigma^*_1$ and $\sigma^*_2$ are ``resolved'' by the procedure described in case 2 by the geodesics defined by the barycentre of the identifiable (k-1)-faces and the opposite vertex (see Fig.\eqref{tetraedri e geodetiche}). The same process defines a topological correction in $G$. \smallbreak  

%%%%%%%%%%%%%%%%%%%%%%%%%%%%%%%%%%%%%%%%%%%%%%%%%%%%%%%%%%%%%%%%%%%%

In general, in case 3 $\sigma^*_1$ and $\sigma^*_2$ can contain identifiable (k-m)-faces. In this case, the resolving strategy is to collapse $\sigma^*_1$ to a identifiable face by a series of ``geodesic-to-the-barycentre'' transformations and then rebuild $\sigma^*_1$ on $\sigma^*_2$ by following the inverse order of the corresponding geodesics in $\sigma^*_2$ (the vertices of the simplices are always identifiable, so the correspondence between geodesics is unambiguous). In more detail, we start with $\sigma^*_1$ and arbitrarily select a vertex, and collapse it to the opposite face of $\sigma_{k-1} \subset\sigma^*_1$ along the geodesic $\gamma_1\subset G$ from the vertex to the barycentre of the face. Then we proceed inductively by selecting a vertex in $\sigma_{k-1}$ and so on, terminating with the geodesic $\gamma_m$ along which $\sigma^*_1$ eventually collapses to an identifiable face. We now have two sequences of geodesics defined by $\sigma^*_1$ and $\sigma^*_2$, respectively,

\begin{align*}
    \Gamma^\alpha=\{\gamma_j^\alpha,\ldots,\gamma_m^\alpha\}, \qquad \alpha=1,2
\end{align*}
through which we can extend our resolving strategy, i.e. $\sigma_1^*$ can be deconstructed along $\Gamma^1$ and reconstructed as identifiable to $\sigma^*_2$ via $\Gamma^2$.\smallbreak

Note that there are multiple choices for the sequences of geodesics, even when the identifiable face is unique. Once a sequence is selected, this procedure determines via the (exponential map) the correction of the transition map between the projected $\pi(\sigma^*_1)$ and $\pi(\sigma^*_2)$ that resolves the obstruction. It is easy to see that different choices are equivalent, up to adjoint actions in $G$. \smallbreak

We preferred to introduce the resolution strategy starting from simpler and intuitive cases, but all the situations described above can be consolidated as special cases of the last (k-m) case. We call this resolving strategy a \emph{cascade of geodesics}, and we have proved that:

\begin{teor}%[Top-down Smutandation Theorem] 
  Consider a semidiscrete principal bundle $P$ over $M$. Given a section of $P$, the geometrically embedded simplices defined by the boundary and volume minimising process and geodesic translation of the vertices define maximum bouquets of spheres associated with generators of $\pi_\bullet (G)$ up to $dim(G)$. The cascade of geodesics resolving procedure consistently defines the topological corrections of the transition maps of $P$.\label{smutandation}\end{teor}

In conclusion, the ``geodesics and barycenters'' procedure defined above uses natural elements of the geometry of the structural group as ingredients to define the structural data of a semidiscrete principal bundle. We have defined the topological corrections without specifically identifying the topological obstructions as elements of the relative homotopy groups.  \smallbreak 

The construction is highly intuitive. If the dimension of the simplex is smaller than the dimension of $G$,  we have geometrically embedded simplices and a well-defined identification problem between repeated (common) faces. The case in which the dimension of the simplices is greater than the dimension of $G$ translates into mapping a sphere of a higher dimension into $G$ with some known points defined by a section of the bundle, ultimately defined by the vertices. This will be the object of future analysis and development of our theory.\smallbreak  

\bigbreak

\textbf{A data-driven assignment:} As a conclusion of this section, we sketch an idea for a possible data-driven generation of the topological structural data of a semidiscrete principal bundle. This idea can be further developed and formalised.\smallbreak

Suppose that a set of sections of a principal bundle $P$ on $M$ can be sampled from some data. This means that the vertices of each faced simplex in $M$ are mapped in $G$ and the simplices in the sections satisfy the local regularity condition (\eqref{regular}), and therefore can be defined (for example), by the procedure introduced above.\smallbreak

The structural group $G$ is equipped with the standard $G$-invariant metric associated with the Killing form on $G$. For $dim(\sigma)\leq dim(G)$, the sampled sections $\pi^{-1}(\sigma)$ define a simplicial complex $\Sigma$ of disjointed simplices embedded in $G$. These disjoined simplices can be mapped onto each other by the $\epsilon(G)$-action.\smallbreak

More simplices can be generated and added to $\Sigma$ applying standard generation methods used in topological data analysis, such as \v{C}ech or Vietoris–Rips algorithms \cite{VietorisRips}. Geodesic balls can be centred on the vertices of each simplex in $\Sigma$ and their radii increased. Given that geodesics in $G$ are also the orbits of the exponential map, this process is in a way ``dual'' to moving the simplices with the $\epsilon(G)$-action exploited above.\smallbreak 

The analysis of the persistent homology elements in $H_n(\Sigma)$ detected/generated by this process can be exploited to assign structural data to $\sigma$ or its lower-dimensional faces. We expect that persistent elements of $H_n(\Sigma)$ approximate or detect elements of $H_n(G)$ and indicate obstructions to the identification of simplices while translating their vertices along geodesics. Finally, the Hurewicz homomorphism (see \cite{Hatcher}) can be exploited to assign elements in $\pi_n(G)$ to the structural data of the bundle. \smallbreak

This intuitive construction can also be extended and explored in the case $dim(\sigma)>dim(G)$ if we consider an isometric embedding of $G$ in a Euclidean space $\R^m$ for $m$ sufficiently high, where higher-dimensional simplices can be defined and the persistent homology analysed. The Nash $C^k$ Theorem guarantees the existence of isometric embeddings in higher dimensions \cite{Nash}. There, it is proved that if $M$ is a compact manifold of dimension $n$ and with a $C^k$ Riemannian metric $g$, with $k\geq 3$ or smooth, then there exists a $C^k$ (with $k=3,4,\ldots,\infty$) embedding of $M$ in the Euclidean space $\mathbb R^{\frac {3n^2+11n}2}$, which is isometric for $g$. These results have been further generalised to the analytic compact case in \cite{GreeneJacobowitz}.

%%%%%%%%%%%%%%%%%%%%%%%%%%%%%%%%%%%%%%%%%%%%%%%%%%%%%%%%%%%%%%%%%%%%%%%%%%%%%%%%%%%%%%%%%%%%%%%%%%%%%%%%%%%%%%%

\section{Associated bundles}

The definitions of discrete Lie integral structure allow us to generalise the classical concept of an associated bundle. Differential forms with special characteristics defined on the total space of a principal bundle locally give rise to analogous forms on the base space. We construct this relevant class of differential forms on $P$ and define their behaviour under gauge transformations. Recall that on a principal $G$-bundle $P$, the action $\epsilon(G)$ on simplices in the total space of $P$ is well defined and geometrically interpretable.\smallbreak

Consider a discrete $V$-valued ($G$-valued, respectively) differential k-forms $v$ and $\omega$ on the total space of a principal bundle $P$ such that locally, for each vertex of any k-simplex $XYZ\ldots$ and $g\in G$, \begin{equation}\begin{array}{rcl} v[(gX)YZ\ldots]&=& gv(XYZ\ldots)\\ 
\omega((gX)YZ\ldots)&=&g\omega((XYZ\ldots)g^{-1} \end{array}\label{associated1}\end{equation}

This vertex-wise behaviour is in line with the standard definition of equivariance under the $G$-action on the vertices in the simplex. In addition, we allow elements that act on different vertices to commute under $\epsilon(G)$-transformations. In other words, for a given order of the vertices of a simplex, the composition of $\epsilon(G)$ transformations allows us to group together and multiply in a given order the elements of $G$ which act on the same vertex:  \begin{equation} \begin {array}{rcl} v[(g_2g_1X)(h_2h_1Y)Z\ldots]&=&g_2g_1h_2h_1v(XYZ\ldots)\\  
\omega((g_2g_1X)(h_2h_1Y)Z\ldots)&=&g_2g_1h_2h_1\omega((XYZ\ldots)h_1^{-1}h_2^{-1}g_1^{-1}g_2^{-1}.\end{array}\label{associated2}\end{equation}

\iffalse If a k-form on $P$ is equivariant, then $$\begin{array}{rcl} v[(g_1X)(g_2Y)(g_3Z)\ldots]&=&\epsilon(g_i)\,v(XYZ\ldots)\\  \omega((g_1X)(g_2Y)(g_3Z)\ldots)&=&\epsilon(g_i)\omega((XYZ\ldots)(\epsilon(g_i))^{-1} \end{array}$$\noindent where $\epsilon(g_i)\in G$ is any ordering of the $g_i$-s. 
In other words, given $\epsilon_1(g_i)$, $\epsilon_2(g_i)$ and $\epsilon_3(g_i)$ permutations of elements acting on vertices of $\sigma\in P$, we etsblish the equivalence relation:\begin{equation}\begin{array}{rcl} \epsilon_1(g_i)\,v(XYZ\ldots) &\cong&  \epsilon_2(g_i)\,v(XYZ\ldots)\\  \epsilon_1(g_i)\omega((XYZ\ldots )(\epsilon_1(g_i))^{-1} &\cong& \epsilon_2(g_i)\omega((XYZ\ldots)(\epsilon_2(g_i))^{-1}.\end{array}\label{simpeq}\end{equation}\noindent The commutativity of elements of $G$ acting on fibres over distinct vertices implies that: $$\begin{array}{rcl} \epsilon_1(h_i)\epsilon_2(g_i)\,v(XYZ\ldots) &\cong&\epsilon_3(h_ig_i)\,v(XYZ\ldots)\\  \epsilon_1(h_i)\epsilon_2(g_i)\omega((XYZ\ldots)(\epsilon_2(g_i))^{-1}(\epsilon_1(h_i))^{-1}&\cong& \epsilon_3(h_ig_i)\omega((XYZ\ldots)(\epsilon_3(h_ig_i))^{-1}.  \end{array}$$\noindent This convention gives rise to a well-defined vertex-wise $G$-action induced by the one on fibres:$$\begin{array}{rcl} v[(h_1g_1X)(h_2g_2Y)(h_3g_3Z)\ldots]&=&\epsilon(h_ig_i)\,v(XYZ\ldots)   \\  \omega((h_1g_1X)(h_2g_2Y)(h_3g_3Z)\ldots)&=&\epsilon(h_ig_i)\omega((XYZ\ldots)(\epsilon(h_ig_i))^{-1}. \end{array}\label{equieq}$$
\fi

This convention consistently defines a local $\epsilon(G)$-action on the space of differential forms modulo permutations of the vertices. The definition of the transition maps on $P$ given below, (\eqref{transition}), naturally extends this local action to an action (including the topological correction $\chi$) on the total space of the bundle. We denote this action by $\epsilon^\ast(G)$ and introduce the following:

\begin{defi} \label{defass}Given a principal $G$ bundle $P$, we denote by $B$ a $G$-module or $G$ itself and by $P_k$ the set of k-simplices  in $P$. The space $$E:=(P_k\times B)/\epsilon^\ast(G)$$\noindent  is called a bundle of k-forms associated with $P$.\end{defi}

\begin{defi} We call the equivalence classes in the above quotient equivariant differential forms on $P$.\end{defi}

The structure of these equivalence classes is obvious. In fact, since the $\epsilon^\ast(G)$ action allows elements acting on different vertices to commute, an equivariant k-form has $k!$ realisations in correspondence to the permutations of the vertices of each simplex. 

\begin{defi}An equivariant k-form $\eta$ which belongs to an equivalence class in $P_k\times B$ is called a section of the associated bundle of k-forms. Given a section $s$ of the principal bundle, the local pullback $\eta^*$ to $M$ is called the local expression of the section.\end{defi}\smallbreak

In other words, the equivariant sections of a bundle $E$ associated with $P$ determine a fibre bundle over the set of k-simplices in $M$ with fibre $B$ and a $G$-representation on each fibre, which comes from the underlying principal bundle. The fibre bundle projection $\pi$ maps $B$ onto a k-simplex in $M$.\smallbreak

\begin{rem}\label{differential_transitions} The differential is a local operation on forms on $P$. The topological correction $\eta$ in Definition~\eqref{localdef} is applied on the left on the realisation of the differential of a form. A topological correction is applied to all realisations ``after'' differentiation on a given simplex. \end{rem}\smallbreak

This means that the values of a k-form $\omega$ can coincide on identifiable k-faces of faced simplices in $P$, but after differentiation, a topological (k+1)-obstruction will introduce a topological correction on $d\omega$ while transitioning between unidentifiable (k+1)-faces. \medbreak

A gauge transformation is determined by a section $\Phi$ of an associated bundle of 0-forms $P\times_G G$. 

\begin{prop}Given a section of a principal bundle and a local expression $v^*$ and $\omega^*$ of a section of an associated bundle of k-forms, the local expressions vary under a gauge transformations $\Phi(X)$ according to:\begin{equation}\label{covariant}\begin{array}{l} \Phi(v^*)=\epsilon(\Phi)\,v^* \\ \Phi(\omega^*)=\epsilon(\Phi)\,\omega^*\,\epsilon(\Phi)^{-1}. \end{array}\end{equation}\label{gaugeK}\end{prop}The pull-back of an equivariant k-form and the local expressions of the sections of associated bundles on intersections of faced simplices depend on the choice of the faced simplex. Distinct local pull-backs are transformed into each other through a transition map \eqref{transition}, which includes the topological correction $\chi$.

\begin{prop}Given a section $s$ of a principal bundle, the local expressions $v^*_i$ and $\omega^*_i$ obtained by restriction of sections of associated bundles to the i-th common k-face of faced simplices in $s$ transform according to:\begin{equation}\label{trans}\begin{array}{l} v_j^*=\epsilon(\xi^*_{ij})\,v^*_i \\ \omega_j^*=\epsilon(\xi^*_{ij})\,\omega_i^*\,\epsilon(\xi^*_{ji})^{-1}. \end{array}\end{equation}\label{transitionK}\end{prop}In summary, the intrinsic geometric meaning of the above construction is based on the standard $G$-action on the fibres of a principal bundle. Any action functional, which includes sections of associated bundles of k-forms, should be invariant under gauge transformations $Gv$ and $G\omega G^{-1}$. \smallbreak

In the case of 0-forms, the notion of covariance and the construction of associated bundles are completely analogous to the standard geometric case. For example, the local expression $v^*(X)=v(s(X))$  of a $V$-valued section of an associated 0-bundle varies under gauge transformation and transition maps by:\begin{equation}\label{gaugeass}\begin{array}{l} [\Phi(v)](X)=\Phi(X)v(X) \\v^*_j(X)=\zeta^*_{ij}(X)v^*_i(X). \end{array}\end{equation}

The operations that characterise Lie integral structures are compatible with the above ``vertex-wise'' definitions of equivariant forms. Let $\omega$ and $\eta$ denote $G$-valued equivariant k-forms and $v$ an equivariant $V$-valued k-form on $P$. Then the $V$-valued differential k-form $\omega\eta v$ is equivariant;$$\begin{array}{rr}\omega((gX)YZ\ldots)\eta((gX)YZ\ldots)v((gX)YZ\ldots)=& g\omega(XYZ\ldots)g^{-1}g \eta(XYZ\ldots)g^{-1}gv(XYZ\ldots)=  \\
&=g(\omega(XYZ\ldots)\eta(XYZ\ldots)v(XYZ\ldots)).\end{array}$$ On the contrary, we will show below that the differential of an equivariant form in general is not equivariant. With the intent to highlight the reason for this, we introduce the notion of horizontal differential form.
\begin{defi}A differential k-form on the total space of a semidiscrete principal bundle is called horizontal if it is trivial on each type I and each generic simplex in $P$. An equivariant horizontal k-form is called tensorial.\end{defi}
Consider two equivariant horizontal k-forms $v$ and $\omega$ on $P$ and a type II k-simplex $XYZ\ldots $. Then we take $A=gX$ and compute \begin{equation}\begin{array}{l}dv(AXYZ\ldots)=v(XYZ\ldots)-v((gX)YZ\ldots)+0\ldots=v(XYZ\ldots)-gv(XYZ\ldots)\\ d\omega(AXYZ\ldots)=Id\ldots\omega((gX)YZ\ldots)^{-1}\omega(XYZ\ldots)=g\omega(XYZ\ldots)^{-1}g^{-1}\omega(XYZ\ldots).\end{array}\label{hor1}\end{equation}
\noindent In analogy to Equation~\eqref{defi0} these expressions can be interpreted as the finite variation of $v$ and $\omega$ in the (vertical) direction of the simplex $X(gX)$.\smallbreak

Consider equivariant k-forms $v$ and $\omega$ and realisations of their differentials on the (k+1)-simplex $XYZ\ldots$.  $$\begin{array}{l} dv(XYZ\ldots)=v(YZ\ldots)-v(XZ\ldots)+v(XY\ldots)+\ldots\\ d\omega(XYZ\ldots)=\ldots\omega(XY\ldots)\,\omega(XZ\ldots)^{-1}\,\omega(YZ\ldots)\end{array}$$\noindent then we evaluate these expressions on the simplex $(gX)YZ\ldots$\begin{equation}\begin{array}{l} v(YZ\ldots)-v((gX)Z\ldots)+v((gX)Y\ldots)+\ldots=v(YZ\ldots)-gv(XZ\ldots)+gv(XY\ldots)+\ldots\\ \ldots\omega(gXY\ldots)\,\omega(gXZ\ldots)^{-1}\,\omega(YZ\ldots)=g\ldots\omega(XY\ldots)g^{-1}g\omega(XZ\ldots)^{-1}g^{-1}\,\omega(YZ\ldots)\end{array}\label{equidiff1}\end{equation}\noindent On the other hand, if $dv$ and $d\omega$ were equivariant, this would mean that\begin{equation}\begin{array}{l} dv((gX)YZ\ldots)=gv(YZ\ldots)-gv(XZ\ldots)+gv(XY\ldots)+\ldots\\ d\omega((gX)YZ\ldots)=g\ldots\omega(XY\ldots)\,\omega(XZ\ldots)^{-1}\,\omega(YZ\ldots)g^{-1}.\end{array}\label{equidiff2}\end{equation}\noindent Obviously \eqref{equidiff1} is not equal to \eqref{equidiff2} and for this reason we introduce the following:
\begin{defi}For any equivariant k-forms $v$ and $\omega$ and a (k+1)-simplex $XYZ\ldots$ we call $$\begin{array}{l}  d^*v(XYZ\ldots) = dv(XYZ\ldots) - v(Y Z\ldots)+gv(Y Z\ldots) \\ d^*\omega(XYZ\ldots)=d\omega(XYZ\ldots)\,\omega(Y Z\ldots)^{-1}\,g\omega(Y Z\ldots)g^{-1},\end{array}$$\noindent the equivariant differential of $v$ and $\omega$. \end{defi}

\begin{rem}
In view of Equation~\eqref {hor1}, one can interpret these expressions as if the equivariant differential is obtained from the standard differential corrected by the finite variation of a k-form in the vertical direction.
\end{rem}

\section{Connections on principal bundles}

In this section, we introduce some relevant elements of our construction, such as discrete parallel transport and holonomy based on the proper use of integral structures on discrete differentiable manifolds. In this context, some properties of integral structures (and not equivalence classes) will be exploited in their own right. The definitions introduced in this section are mostly local, that is, they refer to a determined faced simplex.\smallbreak 

Classically, the concept of a connection on a principal bundle enables the definition of a covariant derivative. This is done by introducing a ``correction'' in the derivation in the direction tangent to the fibres of the bundle (called vertical). Strictly speaking, the structural group of the semidiscrete principal bundle introduced above is $(\chi\epsilon(G),\pi_\bullet(G))$ ($G$ over 0-simplices). In our construction, we consider the differential (and also the covariant derivative) as local operations. This means that they act ``before'' the topological obstructions affect the transition maps of the bundle. Therefore, the construction of the connection introduced below implicitly considers that the topological term of the structural group is trivial. For simplicity, we omit the topological term in the calculations in this section.\smallbreak    

\begin{defi} \label{defi1} A connection on the principal bundle $P$ is defined by a $G$-valued 1-form $\varphi$ on $P$, which locally satisfies the following properties:\smallbreak

1. Let $XY$ be a 1-simplex of type I (there exists a unique $g\in G$ such that $Y=gX$), then $\varphi(XY)=g$.

2. Given  1-simplices  $XY_1$ and $XY_2$ both of type II and such that $Y_2=gY_1$, then $\varphi(XY_2)=g\varphi(XY_1)$. \end{defi}

The vertical simplices are mapped to the element of $G$ that ``generates'' them and,

\begin{defi} A 1-simplex $XY$ such that $\varphi(XY)=\varphi(YX)=Id$ is called horizontal.
\label{defi2}\end{defi}

\begin{defi} Given a connection on a principal bundle, a 1-simplex $XY$ in the base space and a point $X_1\in\pi^{-1}(X)$, the horizontal lift of $XY$ is the horizontal 1-simplex attached to $X_1$.
\end{defi}

The following fact is an important implication of Definition~\eqref{defi1}:

\begin{prop} The set of connections on a semidiscrete principal bundle $P$ is modelled over the set of local maps from the set of 1-simplices  in $M$ to $G$.
\end{prop} \nit{Proof} Consider a connection on $P$ and apply a constant element of $G$ on its value on each type II simplex. This operation defines another connection on $P$.\qed\smallbreak

We will call \emph{covariance} the following property of a connection on a principal bundle:

\begin{prop}  If $XY$ and $X_1Y_1$ are horizontal lifts of a 1-simplex in $M$ such that $X_1=gX$ then $Y_1=gY$.\label{cov} \end{prop}

\nit{Proof} Consider the 2-simplex $XX_1Y_1$. Item 2 in
Definition~\eqref{defi1} implies that $\varphi(XY_1)=g$. Now
consider the 2-simplex $XYY_1$, the same consideration implies that
$\varphi(YY_1)=g$, which in view of item 1 in Definition~\eqref{defi1}
leads to the claim. \qed\smallbreak

This is a finite version of the standard infinitesimal construction in which connection 1-forms take values in the Lie algebra of the structure group. It associates each vector $v$ tangent to $P$ with the element in the Lie algebra that generates the fundamental vector field corresponding to the vertical component of $v$. The horizontal space is the kernel of the 1-form.  \smallbreak

The definition of the horizontal lift of a discrete curve is obvious.

\begin{defi} The horizontal lift of a curve in $M$ going from $X$ to $Y$ defines a map between the fibres $\pi^{-1}(X)$ and $\pi^{-1}(Y)$ called parallel transport.
\end{defi}

The parallel transport determines the following path-dependent mapping.

\begin{defi} The maps induced on a fibre $\pi^{-1}(X)$ by the parallel transport along closed curves with a base point in $X$ generate a discrete subset of $G$ of transformations of the fibre. We call it the holonomy set in $X$ and denote it by $\mathrm{Hol}_X$.
\end{defi}

We can introduce a standard group structure on the set of discrete curves in $M$ in which the product is determined by the composition, the identity element is the trivial curve (a point), and the inverse of a curve is the same curve taken with the opposite direction. There is a natural group homomorphism between the set of closed curves based in $X$ and the set $\mathrm{Hol}_X$. For this reason (with a slight abuse of language), we also call the holonomy set \emph{holonomy group}. \smallbreak

In Section~2, we interpreted the discrete derivative of a 0-form as the 1-form, which associates to each oriented 1-simplex the difference of the values of $f$ on the boundary (recall Equation~\eqref{defiV0} and Equation~\eqref{defi0}). The essence of the classical definition of a covariant derivative is the substitution of the infinitesimal variation of a tensor field in the direction determined by a vector tangent to $M$ (since it is a priori impossible to compare quantities in different points in $M$) with the one determined by its horizontal lift. In differential geometry, a connection on a principal bundle induces a connection on each associated bundle. This intuitive geometric meaning of the covariant derivative can be directly implemented in this case.

\begin{defi} Given a type II k-simplex $XYZ\ldots$, the point-based horizontal projection $p|_X(XYZ\ldots)$ in $X$ is the unique type II k-simplex with vertex $X$ in which the edges attached to $X$ are horizontal.
\end{defi}
The point-based value of the covariant derivative of a differential k-form on a type II k-simplex is the k-form evaluated on the horizontal projection of the k-simplex. This means that the differential of the form is evaluated on specific elements in the $\epsilon(G)$-orbit of the k-simplex. In symbols:

\begin{defi} Given a connection $\varphi$ on a principal bundle and a differential k-form $\xi$ on its total space, the covariant derivative associated to $\varphi$ is  $\nabla_\varphi \xi|_X:=d\xi(p|_X(XYZ\ldots)).$
\label{CovDer}\end{defi}

\begin{prop}The covariant derivative of a k-form on $P$ is trivial on generic simplices  in $P$.\end{prop}

\nit{Proof} Consider a generic (k+1)-simplex $\sigma$ with at least one vertical 1-face $YZ$ and take a vertex $X$. Since $p(XY)=p(XZ)$ a generic k-simplex collapses under horizontal projection to a lower-dimensional type II simplex. The same occurs with the horizontal projection of $\sigma$ if based on $Y$ or $Z$. So, by convention, the differential of a k-form evaluated on $\sigma$ equals the identity on $p(\sigma)$.\qed\smallbreak

\begin{prop}Consider equivariant $V$-valued and $G$-valued differential k-forms $v$ and $\omega$, then $$\begin{array}{ll} \nabla_\varphi v|_X(X(gY)Z\ldots)=\nabla_\varphi v|_X(XYZ\ldots),& \nabla_\varphi v|_X((gX)YZ\ldots)= g^k\nabla_\varphi v|_X(XYZ\ldots), \\ \nabla_\varphi \omega|_X(X(gY)Z\ldots)=\nabla_\varphi \omega|_X(XYZ\ldots),& \nabla_\varphi\omega|_X((gX)YZ\ldots)= g^k\nabla_\varphi\omega|_X(XYZ\ldots)(g^k)^{-1}.\end{array}$$\noindent where $XYZ\ldots$ is a type II (k+1)-simplex.\label{covgen}\end{prop}

\nit{Proof} The simplex $X(gY)(hZ)\ldots$ and $XYZ\ldots$ have the same horizontal projection based on $X$. Evaluating a covariant derivative in $(gX)YZ$ based on $gX$ presumes recognising the horizontal projection of that simplex in $gX$. Now denote by $X_1Y_1Z_1\ldots:=p|_X(XYZ\ldots)$ and by  $X_2Y_2Z_2\ldots:=p|_{gX}(XYZ\ldots)$. The covariance property (Proposition~\eqref{cov}) implies that $X_2Y_2Z_2\ldots=(gX_1)(gY_1)(gZ_1)\ldots$. A straightforward computation of the differential of an equivariant differential form on $X_2Y_2Z_2\ldots$ concludes the proof.\qed\smallbreak

We will call differential forms with this property \emph{covariant} forms.

\begin{defi} The curvature form $R$ of a principal bundle is the 2-form $R=\nabla_\varphi\varphi.$
\end{defi}

\begin{corol} The curvature form is a horizontal covariant form.\label{curvcov}\end{corol}

The fact that the following Theorem arises naturally from our construction is a key result of this paper.

\begin{teor}(a l\'a Ambrose-Singer) The curvature is a $\mathrm{Hol}$-valued 2-form, $R_X(XYZ)=\varphi(ZX)\varphi(YZ)\varphi(XY).$\label{hol}\end{teor}\nit{Proof} Consider a non-degenerate type II 2-symplex $XYZ$ in the total space of the bundle. Let us compute the point-based value $R|_X$. Following the definition, we take the horizontal projection $h|_X(XYZ)=XY_1Z_1$, where $h|_X(XY)=XY_1$, $h|_X(XZ)=XZ_1$. So $$R|_X(XYZ)=d\varphi|_X(XY_1Z_1)=\varphi(XY_1)\varphi^{-1}(Z_1X)\varphi(Y_1Z_1)=\varphi(Y_1Z_1).$$ \noindent Consider the horizontal projections $h|_{Y_1}(Y_1Z_1):=Y_1Z_2$ and $h|_Y(YZ):=YZ_3$. Definition~\eqref{defi1} 
implies that $\varphi(Y_1Z_1)=\varphi(Z_2Z_1)$, where $Z_2Z_1$ represents the obstruction for the horizontal projection of the boundary of $XYZ$ to be horizontal. 
The same fact implies that $\varphi(Y_1Y)=\varphi(XY)$, $\varphi(Z_1Z)=\varphi(XZ)$ and $\varphi(Z_3Z)=\varphi(YZ)$. Furthermore, by construction 
$$Z_1=\varphi(ZZ_1)\varphi(Z_3Z)\varphi(Z_2Z_3)Z_2.$$ \noindent So we have proved that:  $$\varphi(Z_2Z_1)=\varphi(ZX)\varphi(YZ)\varphi(XY).$$ \noindent This expression is compatible with the fact that $R$ is a horizontal form. \qed\smallbreak

\begin{corol}The point-based expressions of the curvature are $G$-adjoint. \end{corol} 

\nit{Proof}  $R|_X(XYZ)=  \varphi(ZX)\varphi(YZ)\varphi(YZ)\varphi(XY)[\varphi(ZX)\varphi(ZX)^{-1}]=\varphi(ZX)R|_Z(ZXY)\varphi(ZX)^{-1}.$\qed\smallbreak

\begin{rem} The holonomy phenomenon involves a direct use of the integral structure. Curvature is the local expression of holonomy. More precisely, the different realisations of the curvature 2-form on a 2-simplex correspond to the holonomy transformations induced on the vertices by the loops determined by the boundary of each 2-simplex. The values of the different realisations of $R$ on each 2-simplex differ by conjugation in $G$. This phenomenon is analogous to what occurs in differential geometry, where the holonomy group is defined up to conjugation depending on the base point of the closed curves.\end{rem}

In general, for any k-form we have:

\begin{prop}
\label{change-vertex} Consider equivariant $V$-valued and $G$-valued differential k-forms $v$ and $\omega$. Then, 
$$\begin{array}{l} \nabla_\varphi v|_{X_2}(X_1X_2X_3\ldots)= A \nabla_\varphi v|_{X_1}(X_1X_2X_3\ldots), \\  \nabla_\varphi\omega|_{X_2}(X_1X_2X_3\ldots) =A \nabla_\varphi\omega|_{X_1}(X_1X_2X_3\ldots) A^{-1}\end{array}$$

\noindent where $A=\varphi(X_1X_2)\prod_{i=3}^n\varphi(X_1X_i)R|_{X_i}(X_iX_1X_2)$.
\end{prop}

\nit{Proof} The transformation between the horizontal projections performed in the vertices $X_1$ and $X_2$ of a k-simplex in $P$ is realised by the $\epsilon(G)$-action. If $X^*_1=\varphi(X_1X_2)X_1$, then $X^*_1X_2$ is horizontal. Similarly, we obtain a horizontal 1-simplex $X_1^*, X^*_i$. The simplex $X_2,X^*_i$ is not in general horizontal, but the curvature form computed on the 2-simplex $X_iX_1X_2$ provides the necessary ``correction'' on $X^*_i$ to obtain a horizontal 2 simplex attached to the vertex $X_2$. \qed  \smallbreak

Given that the trace is Ad-invariant, we can consistently define a scalar curvature in the following way: 

\begin{defi}The scalar curvature is $ \mathcal{R}(XYZ)=Tr R(XYZ)$. \end{defi}

If we compare $R$ to the point-based value of the differential $d\varphi_X(XYZ)=\varphi(XY)\varphi(ZX)^{-1}\varphi(YZ)$, we see that the curvature form and the differential of the connection form do not belong to the same conjugation class.\smallbreak

The standard skew-symmetry of the curvature tensor translates into the following:

\begin{corol}$R|_A(ABC)=R|_A(ACB)^{-1}$\end{corol}

Another important property of the curvature form on a principal bundle is expressed by the second Bianchi identity (the vanishing of its covariant derivative). In our case, we can prove an analogous result.

\begin{teor} (a l\'a second Bianchi identity) The curvature form is covariantly closed.\label{bianchi}
\end{teor}
\nit{Proof} Consider a type II simplex $ABCD$ in $P$ (the degenerate case is trivial). We take as before the horizontal projections $AB_1$, $AC_1$, and $BD_1$ of the edges attached to $A$. These horizontal projections single out a type II 2-simplex $AB_1C_1D_1=h(ABCD)$ (recall the construction of the induced triangulation). By definition $\nabla R(ABCD)=dR(AB_1C_1D_1)$. We compute one of the point-based values $dR_A$:\begin{equation}\label{R1}\nabla =dR|_A(AB_1C_1D_1)= R(AB_1C_1)^{-1}R(AB_1D_1)R(AC_1D_1)^{-1}R(B_1C_1D_1).\end{equation} Following the construction described in the proof of Theorem~\eqref{hol} we see that:\begin{equation}\label{R2} R|_A(AB_1C_1)=\varphi(B_1C_1),\,\,\,\,\,R|_A(AB_1D_1)=\varphi(B_1D_1),\,\,\,\,\, R|_A(AC_1D_1)=\varphi(C_1D_1).\end{equation} Observe that in the generic case $B_1C_1D_1$ is not a horizontal simplex, so
\begin{equation}R(B_1C_1D_1)=\varphi(D_1B_1)\varphi(C_1D_1)\varphi(B_1C_1).\label{R3}\end{equation}
So, substituting \eqref{R2} and \eqref{R3} in \eqref{R1} we prove the claim.\qed\medbreak

A powerful framework for the computation of relevant characteristic classes is provided by the Chern-Weil theory. Characteristic classes of principal bundles appear as elements in the De Rham cohomology of the base manifold defined by differential forms with values in the Lie algebra associated with the structural group. The representative forms are constructed starting from a curvature form on the bundle and certain $G$-invariant polynomials. The first Chern class is generated by the curvature (a tensor form which is covariantly closed). Similar phenomena occur in the context of our theory.\smallbreak  

\begin{rem}
If the structural data of the bundle contain obstructions in $H^2(\pi_2(G))$, then there are common 2-faces of faced simplices that cannot be identified in $P$, while their boundaries can be identified. By combining Remark \eqref{differential_transitions} and Theorem~\eqref{bianchi} (the curvature form is covariantly closed), we can define the first Chern class for semidiscrete principal bundles as the characteristic class generated by the curvature tensor on $M$. Similarly to the classical case, the first Chern class is trivial if the connection is flat and the structural data of $P$ do not contain topological obstructions of dimension 2.\smallbreak
\end{rem}

The fact that an analogue of the first Chern class arises naturally, has the expected properties, and provides a link between local obstructions (curvature) and topological obstructions (bundle structure data) shows the consistency of our theory of semidiscrete principal bundles. \smallbreak

We don't see a straightforward way to define characteristic classes of higher order (detecting unidentifiable simplices of higher dimension) in terms of invariant polynomials of the curvature tensor. Instead, this approach can be generalised in terms of k-connections and (k+1)-curvatures, which are formalised in the context of the theory of gerbes. In fact, Definition~\eqref{defi1}, Definition~\eqref{defi2} and the following can be directly generalised to define k-connections as $\epsilon(G)$-valued k-forms, and horizontal k-simplices. Similarly, Definition~\eqref{CovDer} can be directly generalised by letting the covariant derivative run over the horizontal k-simplices in the $\epsilon(G)$-orbit of k-simplices in $P$. Nevertheless, the development of a complete theory in these terms is out of the scope of this paper and will be the focus of future work. \smallbreak

We conclude the section by introducing a concept of torsion that naturally arises in this context. In Section~4, we observed that when the structural data of a principal bundle (more specifically, the topological corrections in consecutive dimensions) can be assigned in a way that the transition maps $\eta_{ij}$ do not commute with the differential on the total space of the bundle. This deviation can be measured as follows.\smallbreak

We consider an n-form on $M$, which assigns the identity element $Id\in G$ on each n-simplex. This form can be lifted to an equivariant n-form on $P$, which we call the canonical n-form $\theta$.\smallbreak    

Consider an n-simplex $\sigma\in M$, a section $s$, $\sigma_i$ and $\sigma_j$ in $s(\sigma)$. For a transition map $\eta_{ij}$ and a connection $\varphi$, we can compute $(\eta_{ij}\nabla_\varphi \theta(\sigma_i))^{-1} \nabla_\varphi \eta_{ij} \theta(\sigma_i)$. This expression is equivariant, and if it differs from the identity, it detects those situations in which different topological corrections are applied to n-forms and (n+1)-forms.\smallbreak

Classically, the torsion form is defined as the covariant derivative of the canonical 1-form $\theta$ and measures the failure of $\theta$ to induce an absolute parallelism on $M$ (see, for example, \cite{Salamon}). In other words, the (intrinsic with respect to the choice of connection ) torsion of a geometric structure defines an obstruction to its integrability. Similarly, the expression introduced above measures how the covariant derivative of a canonical form is affected by moving to adjacent open sets on the base manifold, and how stable/similar to itself the bundle structure remains between open sets of $M$. \smallbreak

An expression with similar properties is needed in the case in which $\sigma$ belongs to the intersection of multiple-faced simplices. The definitions and objects introduced in the previous sections (transition maps, obstruction forms, etc.) allow us to generalise the above expression:

\begin{defi}
On a semidiscrete principal bundle $P$ we call n-torsion the form $$\tau_n(\sigma)  = (\beta_{n+1}(\chi_{n+1})\nabla_\varphi \theta_n(\sigma))^{-1}\,\,\nabla_\varphi\beta_n(\chi_n)\theta_n(\sigma) $$\noindent where $\beta_n$ is the injective group map defined in \eqref{transitions} and $\chi_n$ are the obstruction forms (Definition~\eqref{ObstructionForm}) associated with the transition maps of the bundle.  \end{defi}

By construction, this form is equivariant on $P$ and can be projected to a form on $M$. We call a connection on a principal bundle for which $\tau_n=Id$ n-torsion-free. Bundles might or might not admit torsion-free connections.   

%%%%%%%%%%%%%%%%%%%%%%%%%%%%%%%%%%%%%%%%%%%%%%%%%%%%%%%%%%%%%%%%%%%%%%%%%%%%%%%%%%%%%%%%%%%%%%%%%%%%%
\section{Local expressions}

Once a gauge-fixing is selected (recall Definition~\eqref{gaugefix}), our theory can be expressed in terms of $G$-valued forms and tensors on the base manifold $M$. We denote the local expression of an equivariant form $\omega$ by $\omega^\ast$.\smallbreak

\begin{prop}\label{CovDerLoc} Consider a section $s$ and a local expression $\varphi^\ast$ of a connection form $\varphi$ on $P$. The local expressions of the point-based values of the covariant derivative of a $G$-valued differential k-form $\omega$ and a $V$-valued k-form $v$ are:$$\begin{array}{l} \nabla_\varphi v^*|_{X_1}(X_1X_2X_3\ldots)=  \prod_{i=1}^{n-1}\varphi^\ast(X_1X_i)^{-1} dv^\ast(XYZ\ldots), \\  \nabla_\varphi\omega^*|_{X_1}(X_1X_2X_3\ldots)=\prod_{i=1}^{n-1}\varphi^\ast(X_1X_i)^{-1}d\omega^\ast(X_1X_2X_3\dots) \prod_{i=1}^{n-1}\varphi^\ast(X_1X_i),  \end{array}$$

\noindent where the order in the product follows the rules \eqref{associated1} and \eqref{associated2} on sections of associated bundles.

\end{prop}

\nit{Proof} The horizontal projection within the covariant derivative is realised by the $\epsilon(G)$-action on the vertices determined by the values that $\varphi$ assume on the 1-simplices that contain the reference vertex.\qed \smallbreak

The expressions that determine the local form of the covariant derivative play an analogous role to the Christoffel symbols. 

\begin{prop} Consider a section $s$ and a local expression $\varphi^\ast$ of a connection form $\varphi$. On each 2-simplex $XYZ$ in the base manifold, the point-based values of the local expression $R^\ast$ are: $$R^\ast(XYZ)|_X=\varphi^\ast(ZX)\varphi^\ast(YZ)\varphi^\ast(XY).$$ \end{prop}

\nit{Proof} A section is a local simplicial isomorphism, Theorem~\eqref{hol} implies the claim.\qed\smallbreak

The variations of the local expressions of connection and curvature forms under a gauge transformation are determined by:

\begin{teor}\label{gaugetr}Given a section $s$ and a gauge transformation $\Phi$, the point-based local expressions of the connection and the curvature with respect to the section $\Phi s$ vary by the following rule:

$$\begin{array}{rcl} \varphi_{\Phi s}^\ast(XY)=\Phi(Y)\,\,\varphi_s^\ast(XY)\,\,\, d\Phi(XY)\,\,\Phi(Y)^{-1}\\ R_{\Phi s}^\ast|_X=  \Phi(X)\,\, R_s^\ast|_X\,\,\Phi(X)^{-1}.\end{array}$$
\end{teor}

\nit{Proof} Suppose $\Phi(X)=g_1$, $\Phi(Y)=g_2$, $X_1, X_2\in \pi^{-1}X$ and $Y_1, Y_2\in \pi^{-1}Y$ such that $X_2=g_1X_1$ and $Y_2=g_2Y_1$. Then item 2 of Definition~\eqref{defi1} implies that \begin{equation}\varphi(X_2Y_2)=g_2\varphi(X_1Y_1)g_1^{-1}.\label{veragauge}\end{equation}\noindent This fact, combined with the convention introduced by Definition~\eqref{defi0} proves the first statement. The second claim follows from Corollary~\eqref{curvcov} and Proposition~\eqref{covgen}, but can also be proved by a simple straightforward computation.\qed\smallbreak

Similarly, for a given set of structural data for $P$, transition maps between faced simplices transform the connection and the curvature as follows:\begin{equation}\begin{array}{rcl} \varphi^*_j(XY)=\beta_1(\mathcal{S}_1)\zeta_{ij}(Y)\,\varphi^\ast_i\,\zeta_{ij}(X)^{-1}(\beta_1(\mathcal{S}_1))^{-1}\\ R^*_j(XYZ)|_X=\beta_2(\mathcal{S}_2)\zeta_{ij}(X)\, R^*_i(XYZ)|_X\,\zeta_{ij}(X)^{-1} (\beta_2(\mathcal{S}_2))^{-1} \end{array}\label{transition}\end{equation}

\noindent where, in line with Definition~\eqref{transitions}, $\beta_1(\mathcal{S}_1)$ and $\beta_2(\mathcal{S}_2)$ are the topological corrections for $\alpha_1\in \pi_1(G)$ and $\alpha_2\in\pi_2(G)$, which is then extended to the $\epsilon(G)$-action.

\medbreak

\begin{defi}A connection that satisfies the condition $R\equiv Id$ is called flat. \end{defi}

\begin{prop}Every semidiscrete principal $G$-bundle $P$ admits flat connections. Locally flat connections on $P$ are in one-to-one correspondence with its sections modulo constant gauge transformations.\label{flatsec}\end{prop}
\nit{Proof} A section $s$ maps each faced n-simplex $\sigma$ in $M$ into a type II n-simplex in $P$. There are no obstructions to assign the identity in $G$ to each 1-dimensional face of $s(\sigma)$ and extend this assignment covariantly to a form on $P$. From Equation~\eqref{veragauge} follows that the value of this flat connection on the 1-simplices  of a distinct section $s_1$ is not the identity, except if $s_1$ is obtained from $s$ via a constant gauge transformation.\qed\smallbreak

In particular, Theorem~\eqref{gaugetr} implies that flatness is a gauge-invariant property of a connection. Let us call \emph{trivial} a local expression $\varphi^\ast$ of a connection if it is everywhere equal to $Id\in G$. Elementary examples can be constructed to illustrate the fact that a flat connection can have a non-trivial local expression. A non-trivial local expression of a connection gives rise to a non-trivial parallel transport. This fact might be problematic as, apparently it would mean that curvature does not provide significant constraints on parallel transport. Nevertheless, our model remains consistent, in fact: 

\begin{corol} Every flat connection admits a trivial local expression with an opportune gauge-fixing.\label{trivsec}\end{corol}
\nit{Proof} The statement is an immediate consequence of Proposition~\eqref{flatsec}. The existence of a flat connection is equivalent to the existence of type II 2-simplices  with entirely horizontal boundaries. A trivial local expression comes from a section that contains exactly those 2-simplices. \qed\smallbreak

An interesting corollary of Proposition~\eqref{flatsec} is the classical fact that connections on semidiscrete principal bundles are entirely determined by holonomies around noncontractible cycles on the base space $M$. This result provides a global characterisation of flat connections.
    
\begin{corol}
Flat connections are in one-to-one correspondence with equivalence classes of homomorphisms from the fundamental group of $M$ to the gauge group $G$ up to conjugation.\end{corol}

\nit{Proof} A noncontractible cycle in $M$ can only be realised by 1-simplices that belong to different faced simplices. Suppose that we transform a flat connection into a locally trivial connection by the natural gauge fixing as per Proposition~\eqref{flatsec}. In general, it is not possible to lift a noncontractible cycle in $M$ to a closed horizontal curve using locally constant gauge transformations. The holonomy around the cycle will be determined by the transition maps between the faced simplices. As before, following the 1-cycle from a different starting point changes the resulting element in $G$ via conjugation.\qed\smallbreak

\medbreak 

Let us consider in more detail the associated bundles of vector-valued 0-forms. The covariant derivation is a substitution of an infinitesimal variation by the variation in the horizontal direction. So, given item 2 of Definition~\eqref{defi1} we have:

\begin{corol} Given a connection and a section of a principal bundle, the local expression of the covariant derivative of a section $v$ of a vector bundle on $M$ is: $$\nabla_\varphi v(XY)=\varphi^\ast(XY)^{-1}dv(XY)=\varphi^\ast(XY)^{-1}v(Y)-v(X).$$
\end{corol}
 
\nit{Proof} It follows directly from Proposition \eqref{CovDerLoc}. Observe that the value of $\varphi^\ast(XY)^{-1}\in G$ acts on one of the vertices to produce the horizontal projection.\qed 

Equation~\eqref{leib} guarantees a fundamental property of a connection on a vector bundle. Given a section $v$ and a scalar function $h$:\begin{equation}\label{leib2}\nabla(hv)=dh\,\,v+h\,\,\nabla v.\end{equation} This fact enables the definition of parallel transport of a tensor.

\begin{defi} \label{pardef} A tensor $v(X)$ on $M$ is parallely transported along a discrete curve $\gamma\subset M$ if $$\nabla_\varphi(\gamma)v=0.$$
\end{defi}

\begin{rem} We emphasise the fact that according to  Definition~\eqref{pardef}, the parallel transport provides a way to \textbf{identify} elements, which belong to fibres of a vector bundle over different points in $M$. In fact $\nabla v(XY)=0$ means that $v(Y)=\varphi^\ast v(X)$.
\end{rem}

With these definitions, the local expression of the parallel transport of $v(X)$ on $M$ along a 1-simplex (XY) is just $\varphi^\ast(XY)v(X)$.  The local expression on the base space of the \emph{parallel transport operator} along a continuous discrete curve $\gamma\subset M$ is just the composition of this action on 1-simplices :

\begin{equation}\label{par}\mathscr{P}_\gamma=\prod_\gamma\varphi^\ast.\end{equation}

For (associated) tensor bundles, holonomy is defined as a set of transformations induced on a fibre by the parallel transport along closed paths in the base manifold. The curvature of the associated bundle is a $V$-valued 2-form with point-based values defined by the expression:
$$R(XYZ)v(X)=\mathscr{P}_{\partial(XYZ)}v(X).$$
\noindent A statement analogous to the Ambrose-Singer Theorem becomes obvious.

\begin{rem}The construction described in this section does not require any particular regularity condition on the simplicial complex $M$. In fact, the discrete base manifold does not need to be a triangulation of a differentiable manifold; in particular, it could not have constant dimension.\end{rem}

These last remarks complete the construction that supports a rigorous formulation of a gauge theory on a semidiscrete principal bundle. This construction can be easily generalised to include metric or other structures on the base manifold. For example, any vector-valued k-form can be multiplied by a scalar function related to the volume of the simplices in the support. In such a case, the first derivative has to be redefined by dividing the variation of the field by the ``length'' of the 1-simplex. Equation~\eqref{leib} guarantees that a connection on the principal bundle induces a connection in a proper sense on any associated bundle, so the covariant derivative of sections of a vector bundle multiplied by scalar functions follows the standard rule \eqref{leib2}. The parallel transport of a vector can be easily modified by a scalar factor (on the right in Equation~\eqref{par}) on each discrete step.$$\mathscr{P}_\gamma=\prod_\gamma\varphi^\ast h.$$\noindent This modification gives ample possibilities to relate the deterministic and the random subsets of variables in our state-space description.

\section{Complexity}\label{glob}

Our main ansatz is that emergent phenomena in complex systems are related to the impossibility of performing any globally valid synchronisation (or consistent interpretation) of the internal states of the agents. In our theory, any relation between two agents in a complex system depends on the way the agents mutually interpret their internal states. Internal states can be consistently related only by means of a parallel transport on the manifold.\smallbreak

The curvature form and the holonomy phenomenon are direct analogues of geometric frustration on the one hand, and manifestations or consequences of non-trivial parallel transport/connection, i.e. a collective/effective gauge field on the other. A connection (a gauge field) on a semidiscrete principal bundle can be defined by some dynamical procedure, for example, an extremisation of a functional (toy examples are proposed in the following sections). Analogously, the topological obstructions to the triviality of a bundle can determine/cause ``loops" in various dimensions in the total space that cannot be closed. This is a direct analogue to the concept of topological frustration.\smallbreak  

In order to highlight the fundamental analogies with geometric and topological frustration, while considering the more general aspects of our construction, in the absence of a better or more widely adopted term, we refer to these mechanisms as \emph{exasperation}.  

\begin{defi} Dynamic exasperation is determined by non-trivial parallel transport on $M$.\end{defi}

A trivialisation of a semidiscrete bundle corresponds to a local synchronisation of states of the agents represented by the vertices of a faced simplex. This can be interpreted as a local multi-object interaction.

\begin{defi} Intrinsic exasperation is determined by a non-trivial underlying fibre bundle and the related topological obstructions. We call exasperation of order n the obstruction elements in $H_n(\pi_n(G))$.\end{defi}

Consider two agents $X_1$ and $X_2$, and their states $v(X_1)$ and $v(X_2)$. The agent $X_1$ can ``see'' the following superposition (with opportune normalisation) of the state $v_2$ transported parallelly along a family $\mathscr{B}$ of continuous discrete curves $\gamma$ going from $X_2$ to $X_1$. This is an analogue of the Wilson lines in classical gauge theories. The contribution to each line can be weighted, for example, by a probability density $f$ on the fibre. $$ \int_\mathscr{B}a\, f(\mathscr{P}_\gamma v_2).$$ \noindent We can also introduce a weight factor $a$, which makes the contribution of more direct paths (for example, inversely proportional to the power of the number of 1-simplices  contained in the curve) stronger. The choice of the set $\mathscr{B}$ of curves connecting the two vertices depends on the way in which one decides to treat a specific problem, on the computational power which can be employed, etc. In any case, there are two relevant possibilities to reduce the number of components in the set. One is based on introducing a cutoff on the number of 1-simplices  contained in the discrete curve.  The other possibility is to introduce a cutoff point on the value of the probability density evaluated $f$ in the state transported parallelly along a curve.\smallbreak

Some remarkable properties of parallel transport operators arise naturally in our theory.

\begin{teor}In the absence of topological obstructions introduced by the transition maps, the local expression of the parallel transport along the 1-simplex $AB$ is invariant under transition maps, i.e. independent of the choice of the local expression $\varphi_i(AB)$. \label{indip} \end{teor}

\nit{Proof} Without loss of generality, consider faced 2-simplices  in $M$. Take $ABC$ and $BDC$ with a common 1-simplex $BC$ and write the local expression of the parallel transport operator along the continuous discrete curve $(AB)(BC)(CD)$. Recall Equations~\eqref{gaugeass} and \eqref{transition} then compute $$\begin{array}{l}\varphi_2^*(CD)\varphi_1^*(BC)\varphi_1^*(AB)v(A)=\varphi^*(CD)\,\zeta_{12}(C)\,[\varphi_1^*(BC)\varphi^*(AB)v(A)] \\
                  \varphi^*_2(CD)\varphi_2^*(BC)\varphi_1^*(AB)v(A)=\varphi^*(CD)\,[\zeta_{12}(C)\varphi_1^*(BC)\zeta_{12}(B)^{-1}]\,\zeta_{12}(B)[\varphi^*(AB)v(A)].\end{array}$$The right-hand side terms of these equations are equal.\qed\smallbreak

\begin{corol}The parallel transport operator $\mathscr{P}_\gamma$ depends on $\gamma\subset M$ and is independent of the choice of local expression of the connection.\end{corol}

\begin{prop}Trivial connections on non-trivial principal bundles give rise to non-trivial parallel transport.\end{prop}

\nit{Proof} The transition functions of the bundle intervene every time that the lift of a path $\gamma\subset M$ passes from one faced simplex to another.\qed\smallbreak

\begin{prop} Given a gauge transformation $\Phi(X)$, we have $$\Phi[\mathscr{P}_\gamma v(X)]=\Phi(Y)\mathscr{P}_\gamma v(X),$$ \noindent for any continuous discrete curve $\gamma$, which joins the points $X$ and $Y$ in $M$. \label{parinv}\end{prop}

\nit{Proof} Combine Equation~\eqref{gaugeass}, and Equation~\eqref{veragauge} in Equation~\eqref{par}. \qed\smallbreak

\begin{corol}If two agents are connected by a unique curve, the effects of the parallel transport can be eliminated by a suitable gauge fixing.\end{corol}

In other words, the connection form along one-dimensional ``bridges'' can be considered trivial. If there is more than one path in the base space joining $X_1$ and $X_2$, then it is not possible to eliminate the contribution of a non-trivial parallel transport by means of a gauge transformation.\smallbreak 

Observe that including the contributions of different paths in $M$ to the way one agent ``sees'' the state and interacts with another agent is directly related to the topological complexity of the base space. In a gauge theory, the interactions are carried out by a gauge field, i.e. agents ``see'' parallel-transported states, so the topological complexity (TC) of the total space of the bundle becomes relevant. For this reason, in our fibre-bundle framework, topological frustration and topological complexity appear to be closely related.\smallbreak 

Recall that the topological complexity $TC(M)$ of a space $M$ is associated with the impossibility to define global sections of the space of paths $P(M)$ on $M$, which is a fibre bundle over the space of endpoints $M\times M$. The topological complexity is a homotopy invariant and is closely related to the Lusternik-Schliemann category. The Lusternik-Schliemann category is a sectional category which measures how many local trivialisations or sections of a fibre bundle are needed to describe the global behaviour. \smallbreak

An equivariant version of the Lusternik-Schliemann category has been introduced in \cite{Marzantowicz1989}. The topological complexity of fibrations has been analysed in several articles \cite{Dranishnikov2015, Grant2012, Naskar2020}. The relations between the topological complexities of the base space, the fibre and the total space of the fibrations have been investigated. For example, in \cite{Daundkar2024} it is shown that for a class of fibre bundles $F\longrightarrow P \longrightarrow M$ the following inequality holds: $$TC(P)\leq TC(M) +TC(F)-1.$$ Although this relation has not been proven in general for fibre bundles, we can expect that there are topological upper bounds on the TC of the total space determined by the base space and the fibres. Ultimately, these relations imply topological constraints on the parallel transport on a principal bundle.\smallbreak

Very interestingly, in more recent years, discrete versions of the theory behind topological complexity have been developed. In \cite{Ternero2015, Ternero2019}, a simplicial version of the Lusternik-Schnirelmann category has been developed, and in \cite{Ternero2018, Gonzalez2018}, simplicial formulations of topological complexity have been introduced. \smallbreak

The so-called contiguity property of maps (and contiguity classes of maps) between simplicial sets is exploited as an analogue of homotopy. The endpoint set is modelled in terms of the categorical product of simplicial sets.  The LS category and the TC are defined as the minimum number of simplicial complexes with specific contractability properties (Faber complexes) that can cover the relevant simplicial set.\smallbreak

Higher-dimensional analogues of discrete (simplicial) topological complexity have been introduced in \cite{Alabay2024}.\smallbreak

Finally, given that the concept of topological complexity itself is directly related to the non-triviality of a fibre bundle, we could leverage the simplicial TC theory developed so far and hypothesise that the topological complexity over a discrete manifold (simplicial complex) $M$ can be efficiently described by the obstructions for the triviality of fibre bundles over $M\times M$. The impossibility of extending contiguity families of simplicial maps could be potentially interpreted in terms of geometric exasperation. Exploring and formalising this unified framework will be the focus of future work.

%%%%%%%%%%%%%%%%%%%%%%%%%%%%%%%%%%%%%%%%%%%%%%%%%%%%%%%%%%%%%%%%%%%%%%%%%%
\section{A gauge theory}

In this section, we describe a gauge theory of a complex adaptive system based on Lie integral structures. Our constructions exploit operations as simplex-wise products of $G$-valued forms and the action of a $G$-form on a $V$-form, included in the Lie integral structure. Our intention is to devote a separate article to the detailed study, applications, and computer simulations based on the models proposed in this section.\smallbreak

In a gauge theory, the dynamics of the sections of an associated bundle is determined by the structure of the underlying principal bundle.\smallbreak

We start with a discrete (finite) set of agents characterised by a set of variables, part of which can be considered as deterministic and the remaining part as random variables. We consider the corresponding set of points in the space of deterministic variables and build a simplicial complex (a discrete differentiable manifold $M$) by means of the \v{C}ech or Vietoris–Rips algorithm. The agents in the system are represented by 0-simplices. The internal states of the agents are parametrised by an n-dimensional vector space $V$ equipped with a continuous probability density function $f$.\smallbreak

A gauge theory manifests itself through a set of \emph{local data}. A crucial role in the construction is played by the set of faced simplices, which we call the \emph{local structure} of the system, plus the set of topological obstructions that make the principal bundle non-trivial. The set of faced simplices reflects the multi-object interaction between agents. The following dynamical data are assigned to each faced simplex:\smallbreak

- a material field $v(X)$ which represents the internal states of the vertices;\smallbreak

- material fields represented by higher-order tensors (potentially);\smallbreak

- a gauge field $\varphi^*(XY)$;\smallbreak

- a set of gauge transformations defined by Equation~\eqref{gaugeass}, Equation~\eqref{veragauge} and Theorem~\eqref{gaugetr}.\smallbreak

The assignments of the fields on the faces in the intersection of faced simplices must be compatible with the transition rules Proposition~\eqref{transitionK}, Equation~\eqref{gaugeass}, Equation~\eqref{veragauge}. With these ingredients, sections of a semidiscrete principal $G$-bundle $P$ and an associated vector bundle with fibre $V$ can be constructed and transformed. The gauge field is a $G$-valued 1-form, which is a local expression of a connection on $P$. Configurations of the local fields, which cannot be transformed one into another by a gauge transformation, are associated with sections of non-equivalent principal bundles.\smallbreak

A specific gauge theory can be formulated by means of a real-valued functional or \emph{action} defined on the set of configurations of the fields, which must be invariant under gauge transformations.\smallbreak

The approach of defining an effective interaction field between large sets of objects has been largely exploited. In our construction, the interaction between agents is carried out by the gauge field in a path-dependent way. \smallbreak

Examples of Ising and continuous spin glasses can be formalised in our framework. Consider, as an example, the Ising model with coupling between the nearest neighbours in the lattice. It is clear that in the case of antiferromagnetic coupling between the spins, not all pairs can be antiparallel and there is not a unique minimum of the energy, but rather a highly degenerate ground state, manifesting frustration. Each n-dimensional lattice can be ''triangulated'' i.e. transformed into a simplicial complex. We consider a (cyclic) $\mathbb{Z}_2$-principal bundle on the simplicial complex. The connection is a $\mathbb(Z_2)$-valued 1-form with values given by the products of the spins attached to each 1-simplex, the curvature is a 2-form defined by holonomy around each triangle, etc. 

In the standard Ising models, there is no topological frustration; however, phase transitions can still arise \cite{Wegner}. In dimensions higher than two, some non-trivial topologies can enter the play for wall crossing regions (walls separating regions with all spins up from regions with all spins down). Instead, a genuine topological order can be obtained by introducing discrete gauge theories coupled to the Ising model, \cite{fradkin-book,Nayak}. The topological aspects can be formalised in terms of a non-trivial principal bundle, i.e. topological obstructions in the transition maps of the bundle.

Our next goal is to construct a static toy model in which the states of the material field are fixed and determine the corresponding state of the gauge field. The definition of a gauge-invariant action depends on the specific choice of a Lie group $G$ and $G$-module $V$. Consider a vector field $v(X)$ and the point-based values of the curvature tensor $R^*|_X$.

\begin{defi} The holonomy orbit $R^*|_X\,v(X)$ of a tensor is the set of elements in the fibre $V_X$ obtained by transporting $v(X)$ parallely along the boundaries of the 2-simplices,  which contain $X$.
\end{defi}

In gauge theories in theoretical physics, these are usually called Wilson loops. \smallbreak

We require that the configuration of the gauge field is determined by minimising the following functional:

$$\mathscr{G}(\varphi,v)=\int_M d_{Maha}(v(X),R*|_X\,v(X)),$$

\noindent where $d_{Maha}$ denotes the Mahalanobis distance \cite{Mahalanobis} in the space of internal states induced by the second moment of $f$. This functional is manifestly $O(N)$-invariant, so it gives rise to an $O(N)$ (static) gauge theory.\smallbreak

In the dynamic case, the states of both the fields involved in the theory (the material field and the gauge field) evolve over time. One possible approach is to consider the time $t$ as a discrete parameter. This means that the base space is reproduced in different time moments and the space-time is a cylinder, i.e. a product $M\times t$. The dynamical behaviour of each agent can be modelled as a stochastic process with a time trend induced by the interaction with the gauge field. Consider, for example, an interaction determined by maximising the probability-valued functional $$\mathscr{G}(\phi(t), v(t))=\int_M f(R|_Xv(X,t)).$$\noindent The evolution of the system can be described via an iterative computation without formulating explicitly field equations. At a given moment $t_0$, one observes the state of the material field, computes the most probable state of the gauge field, and imposes a time trend in the evolution, which maximises the complete functional. Then the system evolves to the next $t_1$ etc. \smallbreak

In classical (differentiable) gauge theories, the dynamical evolution of the system is transversal to the orbits of the action of the gauge group. Principal bundles are classified into equivalence classes up to diffeomorphisms, parametrised by a suitable classifying space \cite{Milnor56a, Milnor56b}. The inequivalence of principal bundles is determined by the topology of the base space and the topology of the fibres (recall the definition of characteristic classes as $\pi(F)$-valued elements of a cohomology on $M$). The principal bundle equivalence class impacts the possible gauge field configurations, and makes some configurations accessible and others inaccessible. In classical gauge theories, the dynamics is local, so the topological invariants of the specific gauge theory do not evolve. In topological field theories, the field propagators contain topological information. \smallbreak

A gauge theory on a semidiscrete principal bundle can be defined in the classical sense if the dynamics is local and the intrinsic exasperation of the bundle is fixed and does not vary dynamically. This is the case with the toy examples proposed above.\smallbreak 

Our formalism also offers the possibility to assign a dynamical role/behaviour to the intrinsic exasperation. Precisely as mentioned above, a variation of the structural data of a semidiscrete principal bundle will generate variations in the configurations if the gauge fields are inaccessible through the local dynamics of the gauge theory. Similar phenomena exist in semi-classical gauge theories. For example, consider the decay of an unstable particle, like the case of a neutron decaying into a proton plus an electron and an antineutrino. For the sake of simplicity, let us concentrate on the $U(1)$ electromagnetic structure, forgetting the weak- and strong-interaction quantum numbers. Since the neutron is uncharged, it is described by a worldline in spacetime, with a zero section of the normal bundle, defining vanishing flux through the homological 2-spheres linked to the worldline. But after the decay, the system is described by three different worldlines: one with a normal bundle section having positive flux (the proton), one with negative flux (the electron), and a neutral one (the antineutrino). The topology is changed from the complement of a single wordline to the complement of three distinct worldlines. This change of topology looks like a non-continuous process only in light of the standard topology of a continuum. \smallbreak  

This mechanism (assigning a dynamical role of intrinsic exasperation) can be used to model metastable states or configurations of the gauge fields, which are mutually inaccessible, or to model or induce phase transitions in a complex adaptive system (for example, a transition between a trivial and a non-trivial bundle on $M$). This can be realised by introducing an independent dynamical mechanism on the intrinsic exasperation, or by coupling the intrinsic and the dynamical exasperation of the theory. There are not many known examples of actions that allow gauge fields to induce topological transitions. For example, in string theory compactified on a Calabi-Yau manifold, for a long time, the possibility that smooth topological transitions (flops changing the cohomology of the internal manifold) are admissible by string geometry has been shown free of obstructions, without however producing explicit examples. Only relatively recently, it has been provided a concrete example of dynamical topology change in presence of fluxes in a compactification of M-Theory down to five dimensions \cite{Greene2001}. There, the five dimensional solution is a wall domain and the transition appears moving along the fourth spatial dimension. See also \cite{Mohaupt2002}, \cite{Jarv2003}. More recently, topological transitions are shown to be required by the Cobordism Conjecture \cite{Ruiz2024}. Up to our knowledge, no explicit Lagrangian description of such a dynamical evolution of topology in time has been realised yet.
The introduction of a formal context to model these phenomena is an interesting novelty in our theory.\smallbreak  

In the spirit of showing that the theory is non-empty, we suggest the following examples. The simplest way to modify the structural data of a principal bundle, especially if topological obstructions are assigned by some random mechanism (see Section~5), is to allow a similar random mechanism to modify the assignment over time. A slightly more advanced idea is to unlock a random mechanism by the configurations of the gauge fields. In Section~7 above we explained how the curvature form generates a discrete analogue of the first Chern class. As an example, suppose that high values of the scalar curvature, above an assigned threshold, or following some probability distribution law, could modify the structural data of the bundle in correspondence to specific 2-simplices in the base space. The topological corrections of the bundle can also be ranked according to the trace of the element $Tr\beta_2(G)$, and higher values of the scalar curvature could unlock a higher probability of transitions to topological obstructions of higher rank. Particularly interesting are induced transitions from a trivial to a non-trivial principal bundle. This example realises a situation where sections of non-equivalent bundles are mutually inaccessible by means of gauge transformations.\smallbreak

This example can be directly generalised in terms of n-connections, (n+1)-curvatures and (n+1)-characteristic classes, as mentioned in Section~7.

Finally, a construction of a space-time covariant gauge theory in this context must be based on a simplicial complex (base manifold), which is naturally foliated in subsets representing a triangulation of the agents at a given moment. Simplicial triangulation of space-time introduces non-trivial considerations on causality and is very important for the gauge theory approach to graph neural networks described in \cite{MihaylovPost} in which GNN (and a more general class of Neural Network architectures) are interpreted as propagators of a Gauge field theory. 

\section{A network model}

One possible application of our theory is a network model analogous to the standard network fitness model. Suppose that the probability of the existence of a link between two agents is proportional to a function of the internal states of the agents, that is a shape function.\smallbreak

We can assume that the shape function $K$ is globally defined, but the probability of the existence of a directional link between two agents $X$ and $Y$ depends on how the agent $X$ ``sees'' the internal state of the agent $Y$ as it is delivered by parallel transport on the manifold. We assume that the probability that there exists a directed link from $X_1$ to $X_2$ is governed by the following expression: $$P(X_1, X_2)=K(v_1, \int_\mathscr{B}a\, f\mathscr{P}_\gamma v_2).$$ \noindent A non-trivial gauge-fitness theory arises when the shape function is \emph{not} a gauge-invariant quantity.\smallbreak

In our theory, the probability of the existence of a link depends on the state of the material field and the state of the gauge field in an indistinguishable way. The state of the links can be interpreted as a superposition of collective states of the gauge field integrated over a set of paths. It is therefore impacted by both the dynamical and intrinsic exasperation of the system.\smallbreak

We can propose a simple example of this network gauge theory. Consider a set of agents characterised by an n-dimensional internal (vector) space endowed with the Normal distribution with mean value in the origin. Suppose that the ``observers'' associated with each agent agree on their evaluations of the first and the second moment of the distribution. In this case, the requirement of preserving the first (mean value) and the second (shape) moment reduces the possible transformations of the internal space to those which rotate or reflect the level sets of the distribution with respect to the origin without deforming them. So, the admissible transformations are given by the orthogonal group $O(n)$. The level sets of the n-dimensional Gaussian are (n-1)-dimensional ellipsoids in, so because of this symmetry, we can reduce the structure group in our bundle just to the special orthogonal subgroup $SO(n)$.\smallbreak

The shape function in several network models reflects a suitable notion (dis)similarity between the states of the agents. As a natural shape function in our example, we can suggest the angle between the internal state vectors of the agents.

%$$\mathbf{x}(u)=\int_\Omega \Theta (u,v)\psi_{(v\longrightarrow u)} \mathbf{x}(v) dv$$

%$$ \psi(v,u)= \sigma (A\mathbf{x}(v)+B\mathbf{x}(u)+C)$$

\section{Conclusions and future work}

In this article, we present and motivate a construction of principal bundles over a discrete set, which naturally emerges from local multi-object interactions. Within this framework, we introduce and define key mathematical structures, including local objects such as connections, curvature, and torsion, as well as topological invariants like characteristic classes, and examine their fundamental properties. The construction encompasses non-trivial principal bundles and identifies obstructions to triviality. \smallbreak   

Associated bundles are also defined within this framework. The resulting construction is geometrically coherent, and its fundamental properties, transformations, and interrelations among various elements are both intriguing and analogous to, or suitably generalise the corresponding structures found in the classical theory of geometric structures on differentiable manifolds. Several elements in our construction are vector-valued and commutative; when necessary, these are extended to more general, non-commutative, group-valued objects. \smallbreak 

We introduce the essential elements for a rigorous construction of gauge theories of complex adaptive systems. In line with the geometric frustration mechanism for emergent phenomena, the theory associates the dynamical properties of systems with the geometric obstructions to the triviality of the bundles. We expect this approach to modelling, interpreting, and quantifying the complexity of real and artificial systems to be successful.\smallbreak  
We also propose a series of examples of topological obstruction assignment methods, candidates for structural groups, possible (toy) action functionals, etc. The examples aim to show that the theory is non-empty, and they have no ambition of being exhaustive.\smallbreak 

Future directions for the development of this work may include:\smallbreak

- Investigate the properties of relative homotopy for classical Lie groups with respect to non-trivial bouquets of spheres. In particular, examining how homotopy generators in $\pi_n(G)$, associated with specific spheres in maximal bouquets, influence or ``propagate'' through the long exact sequence of relative and standard homotopy groups.\smallbreak

- Conduct extensive numerical simulations involving various gauge groups and specific action functionals relevant to both real and artificial systems of interest. This includes modelling dynamic behaviour, intrinsic complexity, and introducing models with manifest time evolution.\smallbreak

- Explore the dependencies and theoretical consequences of modifying the simplicial complex that defines the base space $M$. This involves analysing the persistent structural properties of the theory while associating different simplicial complexes with the set of agents, i.e. configurations of multi-agent interactions. Analyse the effects on the structural data of the principal bundles on $M$, on dynamic and intrinsic exasperation.\smallbreak

- Explore and formalise the topological complexity of discrete manifolds in terms of non-trivial semidiscrete fibre bundles and geometric exasperation by leveraging the existing research on discrete (simplicial) topological complexity. Express the non-contiguity of simplicial maps by means of geometric exasperation. Study the relations between topological complexity and topological frustration of various orders.

- Formalise and expand the implications of this framework within the domains of geometric deep learning and the gauge-theoretic approach to Graph Neural Networks, as outlined in \cite{MihaylovPost}.

\section*{Acknowledgements} The first author thanks Giovanni Petri and Francesco Vaccarino for introducing him to this topic.

\begin{appendix}

\section{Relative homotopy}

In this appendix, we briefly recall the definitions and some relevant properties of the so-called relative homotopy groups of pointed pairs of spaces. The relative homotopy consistently generalises the concept of homotopy groups; for more details, we refer the reader to \cite{Hatcher, Gray, Hu}. We adopt the standard and geometrically intuitive definition of homotopy groups based on the maps $H:(I^{n}, \partial I^{n})\longrightarrow (A,x_0) $ of n-cubes, boundaries, and base points (and their natural compositions). \smallbreak

Consider a pointed space $(A,x_0)$ and a subspace $B\subset A$ that contains the base point $x_0$. We denote by $i:(B,x_0)\longrightarrow (A,x_0)$ the inclusion of pointed spaces, and $(A,B,x_0)$ is usually called a pointed pair of spaces.\smallbreak

We denote by $J_n$ the subset of the boundary of the (n+1)-cube $\partial I^{n+1}$ from which the ``interior" of the top  n-face is removed: $$J_n:=I^n\times \{0\} \cup \partial I^n\times I.$$

The relative homotopy of $A$ with respect to the subspace $B$ is defined as a map of triplets of spaces:

\begin{equation} H:(I^{n+1}, \partial I^{n+1},J^n)\longrightarrow (A,B,x_0) \label{triplets}\end{equation}

It is known that this homotopy relation and the induced equivalence classes of maps behave well under composition (concatenation) and base point selection. In fact, for $n\ge 2$, relative homotopies form groups denoted by $\pi_n(A,B)$, which are abelian for $n\ge 3$. 

The relative homotopy groups are elements of a long exact sequence of homotopy groups (and well-defined group morphisms) induced by the inclusions $i$, and $j:(A,x_0)\longrightarrow(A,B)$: 

$$\ldots\pi_{n+1}(A,B)\stackrel{\delta}{\longrightarrow}\pi_n(B,x_0)\stackrel{i_\ast}{\longrightarrow}\pi_n(A,x_0)\stackrel{j_\ast}{\longrightarrow} \pi_n(A,B)\stackrel{\delta}{\longrightarrow}\pi_{n-1}(B,x_0)\ldots$$\medbreak

\noindent where $\delta$ is called the connecting homomorphism. A class in $\pi_n(A,B)$ can be represented by a map of triplets~\eqref{triplets}, and $\delta$ is defined by restricting this map to the top face $I^{n-1}\times \{1\}$ to define $h=H|:(I^{n-1},\partial I^{n-1})\longrightarrow (B,x_0)$.\smallbreak

%The exact sequence can be exploited to understand and determine $\pi_n(G,\mathcal{B})$. First, if the faces of lower dimensions of $\sigma$ are identifiable, then $\pi_n(\mathcal{B},x_0)=Id$ for each $n$, obviously leading to 

%$$ Id \stackrel{i_\ast}{\longrightarrow}\pi_n(G,x_0)\stackrel{j_\ast}{\longrightarrow} \pi_n(G,\mathcal{B})\stackrel{\delta}{\longrightarrow}\pi_{n-1}(\mathcal{B},x_0)...$$

%$$...\pi_{n+1}(A,\mathcal{B})\stackrel{\delta}{\longrightarrow}\pi_n(\mathcal{B},x_0)\stackrel{i_\ast}{\longrightarrow}\pi_n(A,x_0)\stackrel{j_\ast}{\longrightarrow} \pi_n(A,\mathcal{B})\stackrel{\delta}{\longrightarrow}\pi_{n-1}(\mathcal{B},x_0)...$$\medbreak

On the one hand, the relative homotopy groups measure the deviation of the homomorphism $i_\ast$ from being injective. A homotopy class belongs to the kernel of $i_\ast$ if for any representative $f:(I^n,\partial I^n)\longrightarrow (B,x_0)$ the induced map $i\circ f:(I^n,\partial I^n)\longrightarrow (A,x_0)$ is homotopically equivalent to the identity.\smallbreak

On the other hand, the group homomorphism $j_\ast$ is in general neither injective nor surjective.\smallbreak    

In the case relevant to our model, the ambient space $A$ is a classical Lie group $G$ and the subspace is a bouquet of spheres of various dimensions with a common point $X_0\in G$. By construction in our case, the relevant elements of $\mathcal{B}$ are actual generators of $\pi_n(G)$ not contractible to a point.\smallbreak 

\begin{rem} By applying the restriction in the assignment of the topological data described in Section~4, the operations in $\pi_n(G,\mathcal{B})$ can be replaced by operations in $\pi_n(G)$. A generator of $\pi_n(A)$ can be intuitively interpreted as a topological n-sphere geometrically embedded in the manifold $A$. The exactness of the long sequence in $\pi_n(A,x_0)$ means that the generators of $\pi_n(\mathcal{B},x_0)$ mapped by the inclusion map to $\pi_n(G,x_0)$ belong to the kernel of $i_\ast$. All the homotopy generators that belong to the submanifold $\mathcal{B}$ are trivial in the relative homotopy group. Intuitively, one can think of $\mathcal{B}$ being reduced to one (base) point. The elements of $Im(j_\ast)\subset \pi_n(G,\mathcal{B})$ can be interpreted as generators of $\pi_n(G,\mathcal{B})$ preserved by contracting $\mathcal{B}$ to a point. These elements are generators of both $\pi_n(G,x_0)$ and $\pi_n(G,\mathcal{B})$. The exactness of the long sequence in $\pi_n(G,\mathcal{B})$ means that the generators of $\pi_n(G,x_0)$ and $\pi_n(G,\mathcal{B})$ are further characterised by the fact that they belong to the kernel of the connection morphism $\delta$ into $\pi_{n-1}(\mathcal{B},x_0)$. \end{rem}

\emph{Example:} Consider $G=T^2$, $\mathcal{B}=S^1\subset T^2$. We have $\pi_1(\mathcal{B})=\mathbb{Z}$,  $\pi_2(\mathcal{B})=0$, $\pi_1(G)=\mathbb{Z}^2$, $\pi_2(G)=0$, so:$$\cdots0 \stackrel{i_\ast}{\longrightarrow} 0 \stackrel{j_\ast}{\longrightarrow} \pi_2(T^2,S^1)\stackrel{\delta}{\longrightarrow}\mathbb{Z} \stackrel{i_\ast}{\longrightarrow} \mathbb{Z}^2\cdots  $$
where we used $\pi_2(S^1,x_0)=\pi_2(T^2,x_0)=0$ in the entries to the far left. Now, we have two different cases.\\
In the first case, $S^1$ is nullotopic in $T^2$ (irrelevant in our case). Therefore, its embedding in $T^2$ maps the generator of $\pi_1(S^1,x_0)$ to a trivial element of $\pi_1(T^2,x_0)$. Hence, $ i^*$ = 0, and the sequence reduces to
$$ 0 \stackrel{j_\ast}{\longrightarrow} \pi_2(T^2,S^1)\stackrel{\delta}{\longrightarrow}\mathbb{Z} \stackrel{i_\ast}{\longrightarrow} 0, $$
so $\delta$ is an isomorphism and $\pi_2(T^2,\mathcal{B})=\mathbb{Z}$.\smallbreak
In the second case, $S^1$ is a generator of $\pi_1(T^2)$ (relevant in our case), so the map $i^*$ is now injective. Its kernel is zero, and, by exactness, the long exact sequence reduces to
$$ 0 \stackrel{j_\ast}{\longrightarrow} \pi_2(T^2,S^1)\stackrel{\delta}{\longrightarrow} 0, $$
so we get $\pi_2(T^2,\mathcal{B})=0$. \smallbreak
We can intuitively understand these results as follows, after accepting $\pi_2(T^2)=0$. Taking the relative homotopy $\pi_2(T^2,S^1)$ means considering the maps from a disc $D^2$ into $T^2$, such that the boundary $\partial D^2$ is mapped into $S^1$. Therefore, there is a crucial difference in the two cases. In the first case, if we cut the torus along $S^1$, it splits into two disconnected components: a torus with a hole and a disc $D\simeq D^2$. The homotopy classes map the disc $D^2$ into the holed torus, and the boundary maps into $S^1$. These are obtainable from the maps of $\pi_2(T^2,x_0)$ restricted to the complement of $D$. Therefore, they just give zero up to homotopy. On the other hand, the maps from $D^2$ to $D$ that send $\partial D^2$ into $\partial D$ are obviously in one-to-one correspondence with omotopy classes from $S^2$ into itself and so, they contribute to the homotopy $\pi_2$ with a summand $\mathbb Z$.\\
In the opposite case, if we cut $T^2$ along a generator of $\pi_1(T^2)$, it remains connected, and the maps from $D^2$ to $T^2$ with the boundary sent into $S^1$ can always be seen as obtained by restricting the maps from $S^2$ to $T^2$ that map an equator to $S^1$. So, they give zero.\\[0.3cm]
This example, based on an abelian Lie group, can be generalised to more general Lie groups. For example, any compact Lie group has the same homology as a product of odd-dimensional spheres. Using this fact, one could extend the above example to such cases, replacing the one-dimensional cycle with any combinations of those spheres. See Theorem \ref{theor:104} for the case of compact connected simple Lie groups.

%%%%%%%%%%%%%%%%%%%%%%%%%%%%%%%%%%%%%%%%%%%%%%%%%%%%%%%%%%%%%%%%%%%%

\section{More on the topology of semidiscrete principal bundles}

Given the relevance of the (co)homology groups $H_n(M)$ of the base manifold, in this appendix, we prove another standard result on the topology of principal bundles in the context of our construction. This result refers to the local properties of the semidiscrete principal bundles and is formulated in terms of the multiplicative definition of simplicial homology.\smallbreak 

Let us write the fibration in the following form: $$F \stackrel{i}{\longrightarrow} P \stackrel{\pi}{\longrightarrow} M $$\noindent We denote by $F_k$, $P_k$ and $M_k$ the sets of k-simplices in each of the simplicial complexes involved in the fibration (we can use an arbitrary triangulation of the fibre $F$). In the circumstances of an induced simplicial complex in $P$, the following short sequence is exact for each $k$. $$0 \longrightarrow F_k   \stackrel{i}{\longrightarrow}P_k \stackrel{\pi}{\longrightarrow} M_k\longrightarrow 0$$ \noindent where ``0'' means ``0-simplex''. The only simplices mapped to 0-simplices  by the inclusion are the 0-simplices  themselves, and the k-simplices of an orbit, mapped by the inclusion in the total space, are then projected to 0-simplices  of the base space. These simple considerations imply the same kind of statement in terms of k-path groups instead of simplicial complexes (this time ``0'' means ``0-path'').

\begin{prop}The following short sequence is exact for each k.
\begin{equation}\label{exact} 0 \longrightarrow P_k(F)   \stackrel{i^*}{\longrightarrow}P_k(P) \stackrel{\pi^*}{\longrightarrow} P_k(M)\longrightarrow 0 \end{equation}\end{prop}

By means of the multiplicative homology construction introduced in Section~1, we recover a basic fact about the topology of fibrations in terms of standard homology groups:

\begin{teor}The following long sequence of homology groups is exact:

$$\ldots\,\,   H_{i+1}(M) \stackrel{j}{\longrightarrow}  H_{i}(F)\stackrel{i^*}{\longrightarrow} H_{i} (P)\stackrel{\pi^*}{\longrightarrow} H_{i}(M) \stackrel{j}{\longrightarrow} H_{i-1}(F) \,\, \ldots$$

\end{teor}

\nit{Sketch of proof} The maps $i^*$ and $\pi^*$ are induced by the inclusion and the projection maps, respectively (acting on equivalence classes). The crucial point is the construction of a map $j$, which associates an element of $H_{i-1}(F)$ to an element of $H_{i}(M)$ (first a (i-1)-cycle in $F$ to an i-cycle in $M$). The existence of such a \emph{connection morphism} is implied by the \emph{snake lemma}. Let us describe how this construction works in this context. Consider a cycle $m\in Z_k(M)$. Since $\pi^*$ is surjective, there exists an element in $P_k(B)$ such that $m=\pi(c)$. Observe that for each realisation of a boundary of a path, both the maps $i^\ast$ and $\pi^\ast$ commute with the boundary operator. In particular for $\partial c\in B_{k-1}(P)$ occurs that $\pi \partial c = \partial (\pi c)=\partial m$, but $m$ by construction is a cycle. The sequence in \eqref{exact} is exact in $P_{k-1}(P)$, for this reason there exist an element $f\in P_{k-1}(F)$ such that $\partial c=i^*(f)$.$$ \begin{array}{ccccc}
                                            &                                  & c                                  & \stackrel{\pi^*}{\longrightarrow} & m=\pi^*(c) \\
                                            &                                  & \,\,\,\downarrow^\partial          &     &  \\
               f                            & \stackrel{i^*}{\longrightarrow}  & \partial c=i^*(f)                  &     &  \\
               \,\,\,\downarrow^\partial    &                                  & \,\,\,\downarrow^\partial          &     &  \\
               \partial f                   &  \stackrel{i^*}{\longrightarrow} &          \emptyset                 &     & \end{array}$$\smallbreak

\noindent Since $i^*(\partial f)=\partial i^*(f)=\partial^2 c=\emptyset$, the exactness of \eqref{exact} in $P_{k-1}(F)$ implies that $f$ is a cycle. The map $j$, defined by $j([m])=[f]$, is obviously a group morphism. We should prove that $j$ is well-defined. In the first place, this means that the equivalence class of $f$ in $H_{k-1}(F)$ is independent of the choice of $c$. The reason for this is illustrated by the following diagram:

$$\begin{array}{ccccc}
        u                  &     \stackrel{i^*}{\longrightarrow} &  c\,i^*(u) & \stackrel{\pi^*}{\longrightarrow} & m=\pi^*(c) \\
\,\,\,\downarrow^\partial  &                                     & \,\,\,\downarrow^\partial &  &  \\
  f\,\partial u            & \stackrel{i^*}{\longrightarrow}     &  \partial c\,\partial(i(u))=i(m\,\partial u)  & &
\end{array}$$\smallbreak

\noindent We take a path $u\in P_{k}(F)$ and see that the class defined by $j([m])$ is independent of the variation of $c$ up to a path in the image of the fibre. More generally, we can ``add'' to $b$ such a contribution, but in general this operation cannot be effortlessly expressed in terms of a product of paths.\smallbreak

Furthermore, if $m$ is a boundary, then also $f$ should be a boundary.  Suppose there exists $e$ such that $m=\partial e$. In this case, the injectivity of $i^\ast$ and the surjectivity of $\pi^\ast$, imply that the class of $m$ is determined by projecting a path $c$, which we have eliminated the contribution of a path $q$ in the fibre i.e. $m = \pi^*(c\,/ (i^*(q)))$. Again, as the product of paths is not abelian, the procedure of ``eliminating the contribution'' of a path cannot, in general, be expressed in terms of the product. However, the operation is well-defined in terms of images and kernels of the maps $i^\ast$ and $\pi^\ast$. So the path $q$ is determined up to a contribution of a cycle by $\partial e=\pi^*(c\,/ (i^*(q)))$.$$\begin{array}{ccccc}  &                                   &                             &                                    &    e                            \\
                        &                                   &                             &                                    &  \,\,\,\downarrow^\partial      \\
  q                     & \stackrel{i^*}{\longrightarrow}  & c                           & \stackrel{\pi^*}{\longrightarrow}   & m = \pi^*(c\,/ (i^*(q)))         \\
 \,\,\,\downarrow^\partial    &                                  &\,\,\,\downarrow^\partial          &                         &   \,\,\,\downarrow^\partial      \\
  \partial q=f          & \stackrel{i^*}{\longrightarrow}  & \partial c=i^*(\partial q)  &   \stackrel{\pi^*}{\longrightarrow} & \emptyset \end{array}$$\smallbreak

\noindent Then we take $f=\partial q$ and apply the considerations of the first point of this proof. We see that the diagram closes correctly, that is again $j[m]=f$. \qed\smallbreak

An analogous result can be proved in more abstract terms of simplicial homotopy. It can be referred to as the classical homotopy lifting property that characterises fibre bundles. The validity of the standard results highlighted in this section supports the self-consistency of our construction.\smallbreak

%%%%%%%%%%%%%%%%%%%%%%%%%%%%%%%%%%%%%%%%%%%%%%%%%%%%

\section{Reduced moments}
Consider a set of variables in $\mathbb R^n$. We say that they are statistically distributed on a $A\subseteq \mathbb R^n$ if there exists a probability measure $\mu$ over $\mathbb R^n$ with support on $A$. Given a multi-index $I=(i_1,\ldots,i_n)\in \mathbb N^n$, with height $|I|=\sum_{j=1}^n i_j$, we call moments of order $k$ the integrals:
\begin{align}
    M^I=\langle x_1^{i_1} \cdots x_n^{i_n} \rangle=\int_{\mathbb R^n} x_1^{i_1} \cdots x_n^{i_n} d\mu(x), 
\end{align}
such that $|I|=k$. In particular, $\langle x_j \rangle =\bar x_j$ is called the mean value of $x_j$. We say that the statistical distribution is centred if $\bar x_j=0$ for all $j$. After defining new variables $y_j$ such that $x_j=y_j+\bar x_j$, we see that the statistical distribution of $y_j$ is centred. Therefore, there is no loss of generality in considering only centred distributions.\smallbreak

Given a statistical distribution (a probability density function), we define the reduced moments as follows. Introduce the partition function
\begin{align}
    Z[t_1,\ldots,t_n]=\langle e^{t_1x_1+\cdots+t_nx_n} \rangle=\int_{\mathbb R^n} e^{t_1x_1+\cdots+t_nx_n} d\mu(x).
\end{align}
Then, the moment with multi-index $I$ is
\begin{align}
    M^I=\partial_{t_1}^{i_1}\cdots \partial_{t_n}^{i_n} Z[t_1,\ldots,t_n]_{t=0}.
\end{align}
We define the associated reduced moments by 
\begin{align}
    M^I_{red}=\langle x_1^{i_1} \cdots x_n^{i_n} \rangle_{red}=\partial_{t_1}^{i_1}\cdots \partial_{t_n}^{i_n} \log Z[t_1,\ldots,t_n]_{t=0}.
\end{align}
A moment $M^I$ is said to be decomposable if it is the product of at least two moments: $M^I=M^{I_1}M^{I_2}$, $|I_j|>0$. A moment $M^I$ is said to be connected if it is not the sum of decomposable moments. It is a standard result of Quantum Field Theory that $\log Z[t_1,\ldots,t_n]$ is the generating function of the connected moments, so that, in other words, we see that the reduced moments are nothing but the connected moments. \\
For convenience, we write the relation among moments and reduced moments up to the fourth order:
\begin{align}
   \langle x_j \rangle_{red} =& \langle x_j \rangle; \\
   \langle x_j x_k \rangle_{red} =& \langle x_j x_k \rangle-\langle x_j \rangle \langle x_k \rangle;\\
   \langle x_j x_k x_l \rangle_{red}=& \langle x_j x_k x_l \rangle-\langle x_j \rangle \langle x_k x_l \rangle-\langle x_k \rangle \langle x_j x_l \rangle
   -\langle x_l \rangle \langle x_j x_k \rangle+\langle x_j \rangle \langle x_k \rangle \langle x_l \rangle;\\
   \langle x_j x_k x_l x_m \rangle_{red}=& \langle x_j x_k x_l x_m \rangle-\langle x_j \rangle \langle x_k x_l x_m \rangle- \langle x_k \rangle \langle x_j x_l x_m \rangle-\langle x_l \rangle \langle x_j x_k x_m \rangle - \langle  x_m \rangle \langle x_j x_k x_l \rangle \cr 
   &- \langle x_j x_k \rangle \langle x_l x_m \rangle -\langle x_j x_l \rangle \langle x_k x_m \rangle-\langle x_j x_m \rangle \langle x_k x_l \rangle \cr
   &+ \langle x_j x_k \rangle \langle x_l \rangle\langle x_m \rangle +\langle x_j x_l \rangle \langle x_k \rangle\langle x_m \rangle +\langle x_j x_m \rangle \langle x_k \rangle\langle x_l \rangle +\langle x_k x_l \rangle \langle x_j \rangle\langle x_m \rangle\cr &+\langle x_k x_m \rangle \langle x_j \rangle\langle x_l \rangle +\langle x_l x_m \rangle \langle x_j \rangle \langle x_k \rangle-\langle x_j \rangle \langle x_k \rangle \langle x_l \rangle \langle x_m \rangle.
\end{align}
In the case of a centred distribution, they simplify to
\begin{align}
   \langle x_j \rangle_{red} =& \langle x_j \rangle; \\
   \langle x_j x_k \rangle_{red} =& \langle x_j x_k \rangle;\\
   \langle x_j x_k x_l \rangle_{red}=& \langle x_j x_k x_l \rangle;\\
   \langle x_j x_k x_l x_m \rangle_{red}=& \langle x_j x_k x_l x_m \rangle - \langle x_j x_k \rangle \langle x_l x_m \rangle -\langle x_j x_l \rangle \langle x_k x_m \rangle-\langle x_j x_m \rangle \langle x_k x_l \rangle.
\end{align}
An interesting particular case is the one of a Gaussian distribution, such that
\begin{align}
    d\mu(x)=((2\pi)^n \det g)^{-\frac 12} e^{-\frac 12 \sum_{ij} g_{ij} x_i x_j},
\end{align}
where $g$ is a positive definite $n\times n$ symmetric matrix with indices $g_{ij}$. In this case, one gets:
\begin{align}
    \log Z[t]=\frac 12 \sum_{ij} g^{ij}t_i t_j,
\end{align}
where $g^{ij}$ are the components of $g^{-1}$. We see that the only non-vanishing reduced moments in this case are the ones of order two:
\begin{align}
    \langle x_j x_k \rangle_{red}=\langle x_j x_k \rangle=g^{jk}.
\end{align}
Among the measures that have a finite Radon-Nikodym derivative w.r.t. the Lebesgue measure, the Gaussian distributions are the only ones with this characteristic.
This can be seen as follows. In this class of measures, we can write
\begin{align}
    d\mu(x)=\rho(x)d^nx,
\end{align}
where $\rho$ is a non-negative function with unitary total integral. Therefore, we see that in this case, the partition function
\begin{align}
    e^{W[t_1,\ldots,t_n]} =Z[t_1,\ldots,t_n]=\int_{\mathbb R^n} e^{t_1x_1+\cdots+t_nx_n} \rho(x) d^n x
\end{align}
is the bilateral multidimensional Laplace transform of the function $\rho$, evaluated in $-t$. As we said, $W$ is the generating function of the connected moments. If we know all non-vanishing reduced moments $M^I_{red}$, with multi-index $I$, we can write
\begin{align}
    W[t_1,\ldots,t_n]=\sum_{I\in \mathbb N^n} M^I_{red} \frac {t^I}{I!},
\end{align}
where we have used the standard notations. 
\begin{align}
    t^I=t_1^{i_1}\cdots t_n^{i_n}, \qquad\ I!=i_1!\cdots i_n!.
\end{align}
Therefore, given $W$, $\rho$ is uniquely determined as the inverse bilateral Laplace transform of $e^{W[-t_1,\ldots,-t_n]}$. For
\begin{align}
    W[t]=\frac 12 \sum_{ij} g^{ij}t_i t_j,
\end{align}
this gives back the Gaussian distribution.

%%%%%%%%%%%%%%%%%%%%%%%%%%%%%%%%%%%%%%%%%%%%%%%%%%%%
\section{Symmetric invariant tensors and ``statistically-motivated'' structure groups}

Assume that the space of internal states is a real vector probability space characterised by some known probability density distribution $f$. Furthermore, we assume that for different observers there is no a priori way of identifying vectors in their internal spaces, but they agree on a number of moments of $f$. These simple assumptions give rise to an interesting class of gauge theories. In fact, the k-th moment of $f$ defined on $\mathbb{R}^n$ is a symmetric tensor of order $k$. A natural choice for a structure group in these circumstances is the Lie group with action on $\mathbb{R}^n$ that preserves a specific set of moments. In this section, we briefly recall the relevant theory and some interesting non-trivial examples of classical Lie groups that stabilise symmetric tensors.\smallbreak  

Here we recall some relations between compact simple Lie groups' cohomology and invariant polynomials. Let G be a compact simple Lie group, $\mathfrak{g}$ its Lie algebra. $G$ is naturally endowed with a left-invariant $\mathfrak{g}$-valued 1-form, the Maurer-Cartan form $J$, which satisfies the Maurer-Cartan equation:
\begin{align}
    dJ+\frac 12[J,J]=0.
\end{align}
Now consider the $(2k-1)$-form 
\begin{align}
    \Omega_{2k-1}:=Tr(J^{\wedge (2k-1)}).
\end{align}
Using the Maurer-Cartan equation, the cyclicity of the trace and the fact that $2k-1$ is odd, we immediately get that $\Omega_{2k-1}$ is closed, $d\Omega_{2k-1}=0$.
For some choices of $k$, the trace may be zero, but when it is not, then it is easy to show that it cannot be exact, so, in this case, it defines a non-trivial element
of the de Rham cohomology of the group. On the other hand, for $T_a$, $a=1,\ldots, n$ a basis of $\mathfrak g$, we can write
\begin{align}
  Tr(J^{\wedge (2k-1)})=J^{a_1}\wedge J^{a_2}\wedge \cdots \wedge J^{a_{2k-3}} \wedge J^{a_{2k-2}} \wedge J^{a_{2k-1}}
  Tr(T_{a_1}T_{a_2}\cdots T^{a_{2k-3}}T_{a_{2k-2}}T_{a_{2k-1}}),
\end{align}
where Einstein's convention on repeated indices has to be intended. Now, $J^{a_{2j-3}} \wedge J^{a_{2j-2}}$, $j=2,\ldots,k$, is antisymmetric under interchanging the indices ${2j-3} \leftrightarrow {2j-2}$. Therefore, we can correspondingly replace $T^{a_{2j-3}}T_{a_{2j-2}}$ with
\begin{align}
    \frac 12 [T^{a_{2j-3}},T_{a_{2j-2}}] =c_{a_{2j-3}a_{2j-2}}^{b} T_b,
\end{align}
where $c_{ab}^d$ are the structure constants. Therefore,
\begin{align}
  Tr(J^{\wedge (2k-1)})=J^{a_1}\wedge \cdots \wedge J^{a_{2k-1}} \frac 1{2^{2k-2}} c_{a_{1}a_{2}}^{b_1} \cdots c_{a_{2k-3}a_{2k-2}}^{b_{k-1}}
  Tr(T_{b_1}\cdots T_{b_{k-1}}T_{a_{2k-1}}).
\end{align}
Finally, we note that the pairs $J_{a_{2j-3}}\wedge J_{a_{2j-2}}$ commute with any other forms, so only the totally symmetric combinations of 
$\{b_1,\ldots,b_{k-1},a_{2k-1}\}$ contribute to the above sum. If $S_n$ is the permutation group of $n$ elements, we can write
\begin{align}
  Tr(J^{\wedge (2k-1)})=J^{a_1}\wedge \cdots \wedge J_{b_{k-1}}\wedge J^{b_{k}}  c_{a_{1}a_{2}}^{b_1} \cdots c_{a_{2k-3}a_{2k-2}}^{b_{k-1}}\frac {2^{-2k+2}}{k!}
  \sum_{\sigma\in S_k} Tr(T_{b_{\sigma(1)}}\cdots T_{b_{\sigma(k-1)}}T_{b_{\sigma(k)}}).
\end{align}
The quantity 
\begin{align}
   \tau^{b_1,\ldots,b_k}:= \frac {2^{-2k+2}}{k!}\sum_{\sigma\in S_k} Tr(T_{b_{\sigma(1)}}\cdots T_{b_{\sigma(k-1)}}T_{b_{\sigma(k)}})
\end{align}
is obviously an invariant symmetric tensor under the adjoint action of $G$ on the algebra. 
This means that the symmetric polynomial over $\mathbb R^n$ defined by
\begin{align}
    P_k(x_1,\ldots,x_n):=\tau^{b_1,\ldots,b_k} x_{b_1}\cdots x_{b_k},
\end{align}
is invariant under the adjoint action of the group. In particular, by taking the infinitesimal adjoint action of the algebra, the invariance takes the form:
\begin{align}
    0=[T_a, P_k(T_1,\ldots,T_n)].
\end{align}
This means that $P_k(T_1,\ldots, T_n)$ is a Casimir operator for the Lie algebra.
We now state a famous theorem due to Hopf, in a particular form of our interest:

\begin{teor}\label{theor:104}
    Let $G$ be a real connected simple compact Lie group of rank $r$. Therefore, its rational de Rham cohomology $H^\bullet(G,\mathbb Q)$ is the same as the de Rham rational cohomology of the product of $r$ odd-dimensional spheres $S^{2d_j-1}$, where $d_j\geq 2$, $j=1,\ldots, r$ are the degrees of the fundamental Casimir polynomials generating the whole ring of symmetric Casimir polynomials.
\end{teor}
The invariant degrees $d_j$ are known for all connected compact Lie groups. If we indicate with $\vec d_G=\{d_1,\ldots d_g\}$ the vector of the fundamental degrees 
for the group $G$, then we have
\begin{align*}
    &\vec d_{SU(N)}=(2,3,\ldots,N), \qquad \vec d_{SO(2N+1)}=\vec d_{Sp(2N)}=(2,4,6,\ldots,2N), \cr 
    &\vec d_{SO(2N)}=(2,4,\ldots,2N-2,N), \qquad \vec d_{G_2}=(2,6), \qquad \vec d_{F_4}=(2,6,8,12), \cr
    &\vec d_{E_6}=(2,5,6,8,9,12), \qquad \vec d_{E_7}=(2,6,8,10,12,14,18), \qquad d_{E_8}=(2,8,12,14,18,20,24,30).
\end{align*}
  
Examples: from the above list, we see that $SO(n)$ always preserves the covariance but never the skewness (unless it is zero); $SU(n)$ may preserve covariance, skewness and kurtosis; $Sp(n)$ may preserve kurtosis but not skewness. Notice that the above list contains only the fundamental degrees. All polynomials that can be algebraically generated by the fundamental ones are still invariant. This means, for example, that $G_2$ may preserve kurtosis despite having no fundamental invariant polynomial of degree 4. However it cannot preserve skewness, since no polynomials of odd degree can be generated by the fundamental polynomials.\smallbreak 

As a concrete example, consider a Gaussian distribution 
\begin{align}
    d\mu(x)=((2\pi)^n \det g)^{-\frac 12} e^{-\frac 12 \sum_{ij} g_{ij} x_i x_j}.
\end{align}
As we have seen in the previous Section, it is centered and the only nonzero reduced moment is $\langle x_i x_j \rangle_{red}=g^{ij}$. This is equivalent to saying that all the higher degree moments are polynomials in the $g^{ij}$, so have even degrees. For example, 
\begin{align}
    \langle x_i x_j x_k x_l \rangle =g^{ij} g^{kl}+g^{ik} g^{jl}+g^{il} g^{jk}.
\end{align}
They can all be computed by using the Wick-Isserlis Theorem. In this case, it is easy to determine the group of transformations that leave all the moments invariant. Indeed, since they are all polynomials in the covariance, these are all the transformations that leave $g^{jk}$ invariant. In matrix form, for $g^{-1}$ the matrix having elements $g^{jk}$, we see that a transformation $T$ leaves $g^{-1}$ invariant if 
\begin{align}
   T^t g^{-1} T=g^{-1}. 
\end{align}
This is the orthogonal group $O(g^{-1})$. Of course, this is the same as $O(g)$. Since $g$ is symmetric and positive definite, there exists an invertible matrix $M$ such that $g=M^t M$. From this, it follows that the matrix $A:=MgM^{-1}$ is orthogonal in the canonical sense: $A^t A=I$ so that $SO(g)\simeq SO(n)$.

\section{Weak gauge invariance}

There are known examples of gauge theories (for example, in the context of supersymmetric gravitational theories) with the property that the action is not manifestly gauge-invariant, but the dynamics is. In our probabilistic setup, one might be interested in creating actions that take into account the probability distribution over the vertex state space. A slight modification of the gauge-invariance requirement gives rise to an interesting and very general setup that evaluates field configurations against the probability distribution $f$ on $V$. 

\begin{defi}We say that an action is weakly gauge invariant if the process of extremisation determines a natural gauge fixing.\label{weak}\end{defi}

In our construction, by assuming this condition, the relevant dynamics of the fields is still gauge-invariant, so weak gauge invariance of the action gives rise to a genuine gauge theory.

\begin{equation}\label{func}\mathscr{G}(\phi, v)=\int_M f(R*|_Xv(X)).\end{equation}In words, we will consider configurations of the gauge field that maximise the probability weight of the holonomy orbit of the material field in each point of $M$.\smallbreak

Again, no particular regularity conditions on the base manifold are required in order to define the integration. The integration denotes just a summation over a number of points and a number of 2-simplices  in the base manifold, so it can be performed even if the dimension of $M$ is not constant.\smallbreak

\begin{teor}The functional $\mathscr{G}$ is weakly gauge invariant. \end{teor}
\nit{Proof} Theorem~\eqref{parinv} implies that, under a gauge transformation $\Phi$, all elements in the holonomy orbit of $v(X)$ are transformed by the same element $\Phi(X)\in G$. Suppose that we maximise $\mathscr{G}$. The first step can be achieved by applying an overall gauge transformation, which places the orbit $R^*|_X\,v(X)$ in a position that maximises the probability. Obviously, this operation is dynamically irrelevant as it does not involve the states of the fields. So we should determine the configurations of $v$ and $\phi^*$, which maximise the probability with this gauge fixing.\qed\smallbreak

\end{appendix}

\bibliographystyle{unsrt}
\bibliography{Biblio1}

\end{document}